\documentclass[12pt]{article}
\usepackage{graphicx,latexsym}

 \def\pd{\partial} \def\pp{\prime} \def\half{\frac{1}{2}} \def\fr{\frac} 

\def\a{\alpha} \def\b{\beta} \def\dl{\delta} \def\s{\sigma} \def\vphi{\varphi} \def\eps{\epsilon} 
 \def\lam{\lambda} \def\Lam{\Lambda} \def\gm{\gamma} \def\Gm{\Gamma}
\def\om{\omega} \def\Om{\Omega} \def\nb{\nabla} \def\sq{\sqrt} 

\def\hg{{\hat g}}  \def\hnb{{\hat \nabla}}  
\def\hR{{\hat R}}   \def\hG{{\hat G}} 

\def\bnb{{\bar \nabla}} \def\bg{{\bar g}} \def\bDelta{{\bar \Delta}} 
\def\bR{{\bar R}}  \def\bG{{\bar G}} 

\def\lang{\langle} \def\rang{\rangle}
\def\barb{{\bar \beta}}

\def\barF{{\bar F}} \def\barK{{\bar K}} \def\barL{{\bar L}} 
\def\QG{{\rm QG}}

\begin{document}

\begin{titlepage}


\vspace{5mm}

\begin{center}
{\large {\bf Two-Loop Quantum Gravity Corrections to The Cosmological Constant in Landau Gauge}}
\end{center}

\vspace{5mm}

\begin{center}
{\sc Ken-ji Hamada{}$^{a,b}$ and Mikoto Matsuda{}$^b$}
\end{center}

\begin{center}
{}$^a$ {\it Institute of Particle and Nuclear Studies, KEK, Tsukuba 305-0801, Japan} \\
{}$^b$ {\it Department of Particle and Nuclear Physics, SOKENDAI (The Graduate University for Advanced Studies), Tsukuba 305-0801, Japan}
\end{center}

\begin{abstract}
The anomalous dimensions of the Planck mass and the cosmological constant are calculated in a renormalizable quantum conformal gravity with a single dimensionless coupling, which is formulated using dimensional regularization on the basis of Hathrell's works for conformal anomalies. The dynamics of the traceless tensor field is handled by the Weyl action, while that of the conformal-factor field is described by the induced Wess-Zumino actions, including the Riegert action as the kinetic term. Loop calculations are carried out in Landau gauge in order to reduce the number of Feynman diagrams as well as to avoid some uncertainty. Especially, we calculate two-loop quantum gravity corrections to the cosmological constant. It suggests that there is a dynamical solution to the cosmological constant problem.
\end{abstract}

\vspace{5mm}

\end{titlepage}

\section{Introduction}
\setcounter{equation}{0}

From the cosmological experiments dramatically advanced in recent years \cite{wmap,planck}, the primordial spectrum of the universe has been indicated to be almost scale invariant. It seems to suggest that conformal invariance is significant to describe the dynamics of the early universe. Actually, most of the fundamental quantum field theories have conformal invariance at very high energies where mass parameters can be neglected. Also in gravity, it seems to be natural that we require conformal invariance at high energies beyond the Planck scale. Under this thought, we consider a conformal gravity that involves matter fields with conformal couplings.

In the early works of fourth-order quantum gravities \cite{stelle,tomboulis,ft,ab}, the nonconformal $R^2$ action was used as the kinetic term of the conformal-factor field. On the other hand, in order to realize conformal invariance at high energies, we have proposed a renormalizable quantum theory of gravity without the $R^2$ action for several years \cite{hamada02,hamada09,hamada14QG}, in which the conformal-factor dynamics is induced quantum mechanically. Its recent developments are the following.

First, we have found that the conformal dynamics of gravity is described by the combined system of the Weyl action and the induced Riegert action \cite{riegert} at the ultraviolet (UV) limit. When we quantize it \cite{am, amm92, amm97, hs, hh, hamada05, hamada12M4, hamada12RxS3}, conformal symmetry develops into a gauge symmetry, called the Becchi-Rouet-Stora-Tyutin (BRST) conformal symmetry, because it arises as a part of diffeomorphism symmetry. Therefore, all theories connected to one another by conformal transformations become gauge equivalent. In this way, we can realize the background-free nature of quantum gravity. The BRST conformal algebra has been constructed, and by solving the BRST invariance condition, it has been shown that physical states are given by real primary scalar fields only \cite{hamada12M4, hamada12RxS3},\footnote{ 
Due to the presence of this symmetry, ghost modes are no longer gauge invariant. This is an alternative approach to the unitarity problem different from that of Tomboulis \cite{tomboulis}, based on the work of Lee and Wick \cite{lw}.} 
which is consistent with scalar-dominated spectra of the early universe.

The second is that the indefiniteness existing in fourth-order gravitational counterterms when using dimensional regularization has been settled at all orders \cite{hamada14CS} through the study of conformal anomalies \cite{cd, ddi, duff, ds} using the Hathrell's renormalization group (RG) equations \cite{bc, hathrellS,hathrellQED,freeman}. On the basis of this study, we have formulated the renormalizable quantum theory of conformal gravity ruled by a single dimensionless gravitational coupling constant ``$t$" that represents a deviation from the system with the BRST conformal invariance \cite{hamada02,hamada09,hamada14QG}.

In this paper, we calculate quantum gravity corrections to the mass parameters by the coupling $t$ using our renormalizable quantum gravity. The paper is organized as follows. After we briefly review recent developments of our quantum conformal gravity in the next section, we derive propagators and interactions used here emphasizing the relationship with conformal anomalies in section 3. In section 4, we calculate some renormalization factors by controlling infrared (IR) divergences specific to fourth-order theories that should be canceled out. The beta function of the coupling $t$ that has a negative value is given here and its physical meaning is explained. The nonrenormalization theorem of the conformal-factor field, which is a characteristic feature of our model, is explicitly shown at $o(t^2)$ in arbitrary gauge. We then calculate the anomalous dimensions of the cosmological constant and the Planck mass.\footnote{ 
Some of these calculations had been carried out using Feynman gauge in the previous work \cite{hamada09}, but they were incomplete. There were some missing diagrams and also errors in the evaluation of two-loop integrals due to the careless treatment of IR divergences.} 
Especially, the two-loop correction to the cosmological constant is calculated in section 5. At that time, we employ Landau gauge to reduce the number of Feynman diagrams considerably and also to obtain physically acceptable results directly \cite{nielsen, fk}, avoiding the uncertainty due to a gauge-parameter dependence that is not so clear yet. Section 6 is devoted to conclusion and discussion, in which we discuss a dynamical solution to the cosmological constant problem.

Our conventions and some gravitational formulas used here are summarized in Appendix A. In Appendix B, the technique for determining the gravitational counterterm at all orders through the RG equation is presented for QCD in curved space as an example. In Appendix C, we calculate the effective potential for the cosmological constant term, and through its study we demonstrate that IR divergences indeed cancel out. The Feynman parameter integrals are evaluated by treating IR divergences carefully, which are given in Appendix D. In Appendix E, we summarize the integrands of Feynman integrals used in calculations of the two-loop cosmological constant corrections.

\section{Renormalizable Quantum Conformal Gravity}
\setcounter{equation}{0}

First of all, we briefly summarize the previous work on the renormalizable quantum conformal gravity with a single dimensionless coupling constant \cite{hamada14QG}, which is formulated using dimensional regularization that preserves diffeomorphism invariance.\footnote{ 
Another advantage of using this regularization is that the result is independent of how to choose the path integral measure because of $\dl^D(0)=\int d^D k =0$. On the other hand, in a four-dimensional regularization, the contribution from the measure such as conformal anomalies is obtained by evaluating the quantity $\dl^4(0)=\lang x| x^\pp \rang|_{x^\pp \to x}$.} 

Through the study of conformal anomalies for quantum field theories with conformal couplings in curved space, we have recently determined the form of the gravitational actions \cite{hamada14CS} by solving Hathrell's RG equations \cite{hathrellS,hathrellQED,freeman}, which are derived from the consideration of correlation functions among normal products such as the energy-momentum tensor and so on \cite{bc}. We then have been able to fix the indefiniteness existing in fourth-order gravitational actions. By employing this action, we define the renormalizable quantum theory of conformal gravity.

The fourth-order action determined in this way is given by two terms\footnote{ 
If we introduce a nonconformally invariant dimensionless coupling, we also have to add the $R^2$ term in addition to these two terms.
} 
\begin{eqnarray}
    S_g  = \int d^D x \sq{g} \left\{  \fr{1}{t_0^2} F_D + b_0 G_D  \right\} ,
        \label{4th order gravity action}
\end{eqnarray}
while the lower-derivative gravitational actions such as the Einstein term are introduced later. The first term is the $D$-dimensional Weyl action defined by
\begin{eqnarray*}
    F_D = C_{\mu\nu\lam\s}^2 = R_{\mu\nu\lam\s}R^{\mu\nu\lam\s}
          - \fr{4}{D - 2}R_{\mu\nu}R^{\mu\nu}  + \fr{2}{(D - 1)(D - 2)}R^2 
\end{eqnarray*}
and $t_0$ is a dynamical coupling constant. The second term that reduces to the Euler density at four dimensions is significant for the conformal-factor dynamics, which is given by \cite{hamada14CS} 
\begin{eqnarray}
    G_D =G_4 + (D-4) \chi(D) H^2,  
       \label{expression of G_D}
\end{eqnarray}
where $G_4$ is the usual Euler combination and $H$ is the rescaled scalar curvature defined by $G_4 = R_{\mu\nu\lam\s}^2 - 4R_{\mu\nu}^2 + R^2$ and $H = R/(D-1)$, respectively. The coefficient $\chi$ is a finite function of $D$ only expanded in series of $D-4$ as
\begin{eqnarray}
   \chi(D) = \sum_{n=1}^\infty \chi_n (D-4)^{n-1} ,
      \label{expansion expression of chi}
\end{eqnarray}
which can be determined order by order. As mentioned below, $b_0$ is not an independent dynamical coupling.

The first several values in (\ref{expansion expression of chi}) have been calculated explicitly \cite{hamada14CS} for QED in curved space \cite{hathrellQED}. In Appendix B, its calculation is generalized to the case of QCD \cite{freeman} with arbitrary gauge group and fermion representation. These theories give the same values as
\begin{eqnarray}
   \chi_1 = \half,  \qquad \chi_2 = \fr{3}{4} .
     \label{values of chi_1 and chi_2}
\end{eqnarray}
For the QED case, the third value has been calculated as $\chi_3=1/3$ using three loop calculations. The first value is also obtained by the conformally coupled scalar theory in curved space \cite{hathrellS}.

Furthermore, we have shown that the conformal anomaly associated with the counterterm (\ref{expression of G_D}) is expressed in the form $E_D = G_D -4\chi \nb^2 H$. Here, it is significant that the familiar ambiguous $\nb^2 R$ term is fixed completely and this combination reduces to $E_4=G_4 -2 \nb^2 R/3$, proposed by Riegert \cite{riegert} in the four-dimensional limit due to $\chi_1=1/2$.

The perturbation in $t_0$ implies that the metric field is expanded about a conformally flat spacetime satisfying $C_{\mu\nu\lam\s}=0$, which is defined by
\begin{eqnarray}
    g_{\mu\nu}  =  e^{2\phi}\bg_{\mu\nu},  \qquad
    \bg_{\mu\nu}  = (\hg e^{t_0 h_0})_{\mu\nu}= \hg_{\mu\nu} + t_0 h_{0 \mu\nu} 
                       + \fr{t_0^2}{2} h_{0\mu}^\lam h_{0\lam\nu} + \cdots ,
            \label{expansion of metric field}
\end{eqnarray}
where $h_{0\mu\nu}= \hg_{\mu\lam}h^\lam_{0\nu}$ and $h^\mu_{0\mu}=0$, and $\hg_{\mu\nu}$ is the background metric. The quantum gravity can be thus described as a quantum field theory defined on the background $\hg_{\mu\nu}$.

The significance of this theory is that the conformal factor $e^{2\phi}$ is treated exactly without introducing its own coupling constant. It ensures the independence under the change of the background $\hg_{\mu\nu} \to  e^{2\s}\hg_{\mu\nu}$, because, as is apparent from (\ref{expansion of metric field}), this change can be absorbed by rewriting the integration variable as $\phi \to \phi - \s$ in quantized gravity. Consequently, we can choose the flat background without affecting the results.

The renormalization factors for the traceless tensor field and the coupling constant are defined as usual by 
\begin{eqnarray}
    h_{0\mu\nu}=Z_h^{1/2}h_{\mu\nu},  \quad t_0 = \mu^{2-D/2} Z_t t  ,
       \label{definitions of renormalization factors}
\end{eqnarray}
where $\mu$ is an arbitrary mass scale to make up for the loss of mass dimensions, and thus the renormalized coupling $t$ becomes dimensionless. On the other hand, the conformal-factor field $\phi$ is not renormalized such that \cite{hamada02}
\begin{eqnarray}
    Z_{\phi}=1 ,
        \label{Z_phi = 1}       
\end{eqnarray}
because there is no coupling constant for this field. This is one of the most significant properties in our renormalization calculations, which reflects the independence of how to choose the background metric as mentioned above.

The renormalization factors of $Z_h$ and $Z_t$ are expanded as usual: 
\begin{eqnarray*}
    \log Z_h = \sum_{n=1}^\infty \fr{f_n}{(D-4)^n} , \qquad
    \log Z_t^{-2} = \sum_{n=1}^\infty \fr{g_n}{(D-4)^n} .
\end{eqnarray*}
Using these factors, we can renormalize UV divergences proportional to the $F_D$ term. The beta function of $\a_t =t^2/4\pi$ is defined by 
\begin{eqnarray}
    \b_t  \equiv  \fr{\mu}{\a_t} \fr{d \a_t}{d\mu} = D-4 + \barb_t ,
       \label{definition of beta function for alpha_t}
\end{eqnarray} 
where $\barb_t = \mu d(\log Z_t^{-2})/d\mu$.

In addition to these renormalization factors, we also introduce the bare parameter $b_0$ to renormalize UV divergences proportional to the $G_D$ term. The nonrenormalization theorem of $\phi$ is related to the geometrical property of $G_D$ (\ref{expression of G_D}) \cite{ds}. Since its volume integral becomes topological at four dimensions, it is not dynamical at the classical level. Therefore, $b_0$ is not an independent dynamical coupling. So we expand the bare parameter $b_0$ in a pure-pole series as
\begin{eqnarray}
  b_0 = \fr{\mu^{D-4}}{(4\pi)^{D/2}} \sum_{n=1}^\infty \fr{b_n}{(D-4)^n} . 
       \label{definition of bare coefficient b_0}
\end{eqnarray}
Since the field-dependent part of the volume integral of $G_D$ starts from the zero, namely $o(D-4)$, the finite field dependence just comes out at the quantum level by canceling out the zero with the UV pole in $b_0$. The finite term in the action induced in this way describes the dynamics of the conformal-factor field.

Here, the residues $b_n ~(n \geq 2)$ depend on the coupling constants only, while the simple-pole residue has a coupling-independent part, which is divided as 
\begin{eqnarray}
    b_1 = b + b_1^\pp , 
       \label{definition fo prime b_1}
\end{eqnarray}
where $b_1^\pp$ is coupling dependent and $b$ is a constant part.

In order to carry out the renormalization systematically incorporating the dynamics induced quantum mechanically, we propose the following procedure. For the moment, $b$ is regarded as a new coupling constant. The effective action is then finite up to the topological term as follows:
\begin{eqnarray*}
    \Gm = \fr{\mu^{D-4}}{(4\pi)^{D/2}} \fr{b-b_c}{D-4} \int d^D x \sq{\hg} \hG_4 + \Gm_{\rm ren}(t,b) ,
\end{eqnarray*} 
where $\Gm_{\rm ren}$ is the renormalized quantity that depends on the coupling constants. The divergent term exists in a curved background only. The constant $b_c$ comes from the sum of direct one-loop calculations \cite{cd, ddi, duff}, which is given by
\begin{eqnarray}
    b_c = \fr{1}{360} \left( N_S + 11 N_F + 62 N_A \right) + \fr{769}{180}  
       \label{value of coefficient b_c}
\end{eqnarray}
for the quantum gravity model coupled with $N_S$ conformally coupled scalars, $N_F$ fermions, and $N_A$ gauge fields. Here, the last term is the sum of $87/20$ and $-7/90$ coming from the gravitational fields $h_{\mu\nu}$ and $\phi$, respectively \cite{ft, amm92, hs}. After the renormalization procedure is carried out, we take $b=b_c$. In this way, we can obtain the finite effective action $\Gm_{\rm ren}(t,b_c)$ whose dynamics is governed by a single gravitational coupling $t$.

From the RG equation $\mu db_0/d\mu=0$, we obtain the following expression:
\begin{eqnarray}
   \mu \fr{db}{d\mu} = (D-4) \barb_b  ,
      \label{RG equation of b}
\end{eqnarray}
where $\barb_b$ is a finite function given by 
\begin{eqnarray*}
    \barb_b = - \left( \fr{\pd b_1}{\pd b} \right)^{-1} 
        \left( b_1  + \a_t \fr{\pd b_1}{\pd \a_t} \right).
\end{eqnarray*}
Here, in order to be able to replace the coupling $b$ to the constant $b_c$ at the end, the condition $\mu db/d\mu = 0$ should be satisfied at four dimensions. Therefore, (\ref{RG equation of b}) ensures the validity of the renormalization procedure proposed above.

From the RG analysis of QED and QCD in curved space, we find that $b_1^\pp$ in (\ref{definition fo prime b_1}) arises at the fourth order of the gauge-coupling constant (see Eq.(\ref{b_1 for QCD in curved space})). From this fact and the similarity between the gauge field and the traceless tensor field ruled by the Weyl action, we can guess that the $\a_t$ dependence of $b_1^\pp$ is also given by 
\begin{eqnarray}
   b_1^\pp = o(\a_t^2) ,
     \label{assumption for b_1^prime}
\end{eqnarray} 
and then we obtain $\barb_b = -b +o(\a_t^2)$. This assumption should be verified through direct two-loop calculations of three-point functions of the traceless tensor field or indirect calculations using the RG equation, but this hard work is not complete yet.

\section{Propagators and Interactions}
\setcounter{equation}{0}

In this paper, we consider the quantum gravity system (\ref{4th order gravity action}) with the Einstein action and the cosmological constant term,
\begin{eqnarray}
    S  = \int d^D x \sq{g} \left\{  \fr{1}{t_0^2} F_D + b_0 G_D  - \fr{M_0^2}{2} R + \Lam_0 \right\} ,
       \label{full gravitational action}
\end{eqnarray}
and calculate the anomalous dimensions of these mass parameters.

Here, we first derive gravitational propagators and interactions used later. The background metric $\hg_{\mu\nu}$ is then chosen to be the flat Euclidean metric $\dl_{\mu\nu}$ and we take a convention that the same lower indices denote contraction in the flat metric.

\subsection{The $F_D$ term}

The Weyl term in $D$ dimensions is expanded as follows:
\begin{eqnarray}
     &&  \fr{1}{t_0^2} \int d^D x \hbox{$\sq{g}$} F_D
         = \fr{1}{t_0^2} \int d^D x \hbox{$\sq{\hg}$} e^{(D-4)\phi}{\bar C}_{\mu\nu\lam\s}^2 
              \nonumber \\
     &&  = \int d^D x \hbox{$\sq{\hg}$} \left[ \fr{1}{t_0^2} {\bar C}_{\mu\nu\lam\s}^2
         + \fr{D-4}{t_0^2} \phi  \,  {\bar C}_{\mu\nu\lam\s}^2 + \cdots \right] .
          \label{expansion of F_D term}
\end{eqnarray}
The first term of the right-hand side gives the propagator and self-interactions of the traceless tensor field. The second and other terms are the induced Wess-Zumino actions associated with the Weyl-squared conformal anomaly, which give new interactions that involve the conformal-factor field.

We first define the gauge-fixed propagator for the traceless tensor field. The kinetic term is given by
\begin{eqnarray*}
     \int d^D x \left\{  \fr{D - 3}{D - 2} \left( h_{0\mu\nu} \pd^4 h_{0\mu\nu}  +  2\chi_{0\mu} \pd^2 \chi_{0\mu} \right)
         -  \fr{D - 3}{D - 1} \chi_{0\mu} \pd_{\mu}\pd_{\nu}\chi_{0\nu}   \right\}
\end{eqnarray*}
where $\chi_{0\mu}=\pd_\nu h_{0\mu\nu}$ and $\pd^2 = \pd_\mu \pd_\mu$. According to the standard procedure of the gauge-fixing, we introduce the following gauge-fixing term \cite{ft}:
\begin{eqnarray*}
    S_{\rm gh + gf}  = \int d^D x \, \dl_{\rm B} \left\{ 
        - i \, \tilde{c}_{0\mu} N_{\mu\nu}  \left(  \chi_{0\nu} - \fr{\zeta_0}{2}B_{0\nu}  \right) \right\}
\end{eqnarray*}
where $\tilde{c}_{0\mu}$ is the antighost and $B_{0\mu}$ is the auxiliary field. $N_{\mu\nu}$ is a symmetric second-order differential operator, which is here defined as
\begin{eqnarray*}
      N_{\mu\nu}=  \fr{2(D-3)}{D-2} \biggl( 
                 -2 \pd^2\dl_{\mu\nu} + \frac{D-2}{D-1} \pd_{\mu}\pd_{\nu} \biggr)   .
\end{eqnarray*}
The BRST transformation denoted by $\dl_{\rm B}$ is defined by introducing the ghost field $c_{0\mu}$ as follows:
\begin{eqnarray*}
    \dl_{\rm B} h_{0\mu\nu}  &=&  \pd_\mu c_{0\nu} \!+\! \pd_\nu c_{0\mu} 
         - \fr{2}{D} \dl_{\mu\nu} \pd_\lam c_{0\lam} \!+\! t_0 c_{0\lam} \pd_\lam h_{0\mu\nu}
                 \nonumber \\
    &&  + \fr{t_0}{2} h_{0\mu\lam} \left( \pd_\nu c_{0\lam} - \pd_\lam c_{0\nu} \right) 
        + \fr{t_0}{2} h_{0\nu\lam} \left( \pd_\mu c_{0\lam} - \pd_\lam c_{0\mu} \right)  + \cdots ,
                 \nonumber \\
    \dl_{\rm B} \phi &=& t_0 c_{0\lam} \pd_\lam \phi + \fr{t_0}{D} \pd_\lam c_{0\lam} ,
                 \nonumber \\
    \dl_{\rm B} c_{0\mu}  &=&  t_0 c_{0\lam} \pd_\lam c_{0\mu},  \qquad
    \dl_{\rm B} \tilde{c}_{0\mu}  = i B_{0\mu},  \qquad   \dl_{\rm B} B_{0\mu} = 0 .
\end{eqnarray*}
The gauge-fixing term and the ghost action are then expressed as
\begin{eqnarray*}
    S_{\rm gh + gf}  =  \int d^D x \left\{  B_{0\mu} N_{\mu\nu} \chi_{0\nu}  - \fr{\zeta_0}{2} B_{0\mu} N_{\mu\nu} B_{0\nu}
           +i \tilde{c}_{0\mu} N_{\mu\nu} \pd_\lam (\dl_{\rm B}h_{0\nu\lam})  \right\} .
\end{eqnarray*}
Here, note that the gauge-fixing term does not depend on the conformal-factor field. Integrating out the auxiliary field, we obtain the following gauge-fixing term:
\begin{eqnarray*}
     S_{\rm gf}  =  \int d^D x  \left\{ \fr{1}{2\zeta_0}  \chi_{0\mu} N_{\mu\nu} \chi_{0\nu} \right\} .
\end{eqnarray*}

The renormalization of the ghost sector is carried out as usual by introducing its own renormalization factor and the gauge parameter is renormalized by $\zeta_0 = Z_h \zeta$.

Let us derive the propagator of the traceless tensor field in arbitrary gauge. The equation of motion is now given by $K^{(\zeta)}_{\mu\nu,\lam\s}(k) h_{\lam\s}(k)=0$ in momentum space, where
\begin{eqnarray*}
    K^{(\zeta)}_{\mu\nu,\lam\s}(k)  &=&
         \fr{2(D \!-\! 3)}{D - 2} \biggl\{ I^{\rm H}_{\mu\nu,\lam\s} k^4 
         + \fr{1 \!-\! \zeta}{\zeta} \biggl[  \half k^2 \bigl( \dl_{\mu\lam} k_\nu k_\s  +  \dl_{\nu\lam} k_\mu k_\s 
                \nonumber \\
    &&  + \dl_{\mu\s} k_\nu k_\lam  +  \dl_{\nu\s} k_\mu k_\lam \bigr) 
        - \fr{1}{D - 1} k^2 \bigl( \dl_{\mu\nu} k_\lam k_\s + \dl_{\lam\s} k_\mu k_\nu \bigr) 
                \nonumber \\
    &&  + \fr{1}{D(D \!-\! 1)} \dl_{\mu\nu} \dl_{\lam\s} k^4  - \fr{D \!-\! 2}{D \!-\! 1} k_\mu k_\nu k_\lam k_\s 
         \biggr] \biggr\} 
\end{eqnarray*}
and
\begin{eqnarray*}
    I^{\rm H}_{\mu\nu, \lam\s} =  \half \left( \dl_{\mu\lam}\dl_{\nu\s}+ \dl_{\mu\s}\dl_{\nu\lam} \right) - \fr{1}{D}\dl_{\mu\nu}\dl_{\lam\s} .
\end{eqnarray*}
By solving the inverse of $K^{(\zeta)}_{\mu\nu,\lam\s}$, we obtain the propagator
\begin{eqnarray}
        \langle h_{\mu\nu}(k) h_{\lam\s}(-k) \rangle = \fr{D-2}{2(D-3)} \frac{1}{k^4} I^{(\zeta)}_{\mu\nu, \lam\s}(k) ,
          \label{propagator of traceless tensor field in arbitrary gauge}
\end{eqnarray}
where 
\begin{eqnarray*}
   I^{(\zeta)}_{\mu\nu, \lam\s}(k)  &=&
         I^{\rm H}_{\mu\nu, \lam\s}  + (\zeta - 1) \Biggl[ \half \biggl( \dl_{\mu\lam} \fr{k_\nu k_\s}{k^2}  
         +  \dl_{\nu\s} \fr{k_\mu k_\lam}{k^2}  +  \dl_{\mu\s} \fr{k_\nu k_\lam}{k^2} 
              \nonumber \\
   &&    +  \dl_{\nu\lam} \fr{k_\mu k_\s}{k^2} \biggr)
         - \fr{1}{D-1} \left( \dl_{\mu\nu} \fr{k_\lam k_\s}{k^2} +  \dl_{\lam\s} \fr{k_\mu k_\nu}{k^2} \right)
          \nonumber \\
   &&    + \fr{1}{D(D-1)} \dl_{\mu\nu} \dl_{\lam\s} 
         - \fr{D-2}{D-1} \fr{k_\mu k_\nu k_\lam k_\s}{k^4}  
        \Biggr] .
\end{eqnarray*}
This tensor satisfies 
\begin{eqnarray*}
   k_\mu I^{(\zeta)}_{\mu\nu, \lam\s}(k) 
   = \zeta \left( \half k_\lam \dl_{\nu\s} + \half k_\s \dl_{\nu\lam} - \fr{1}{D} k_\nu \dl_{\lam\s} \right) , 
\end{eqnarray*}
and thus it becomes transverse when $\zeta=0$. In the following, the choice of $\zeta=0$ is called Landau gauge, while $\zeta =1$ is called Feynman gauge.

Lastly, we write down the three-point interaction coming from the second term of (\ref{expansion of F_D term}), which is given by
\begin{eqnarray}
    S_{{\rm F}[\phi hh]}^{D-4}  &=& 
         (D-4) \int d^D x  \, \phi \, \Biggl\{ \pd_\lam \pd_\s h_{\mu\nu} \pd_\lam \pd_\s h_{\mu\nu} 
         -2 \pd_\nu \pd_\lam h_{\mu\s} \pd_\mu \pd_\lam h_{\nu\s}
                  \nonumber \\
    &&   + \pd_\lam \pd_\s h_{\mu\nu} \pd_\mu \pd_\nu h_{\lam\s}
         - \fr{2}{D-2} \biggl( \half \pd^2 h_{\mu\nu} \pd^2 h_{\mu\nu}
         - \pd^2 h_{\mu\nu} \pd_\mu \chi_\nu
            \nonumber \\
    &&   + \pd_\mu \chi_\nu \pd_\mu \chi_\nu  
         + \pd_\mu  \chi_\nu \pd_\nu  \chi_\mu   -  \chi_\mu \pd^2 \chi_\mu  \biggr)
         + \fr{2}{(D-1)(D-2)} \pd_\mu \chi_\mu \pd_\nu \chi_\nu  \Biggr\} 
                \nonumber \\ 
   &=& (D-4) \int \fr{d^D p d^D q}{(2\pi)^{2D}} 
            \phi(-p-q) h_{\mu\nu}(p) h_{\lam\s}(q) W^3_{\mu\nu,\lam\s}(p,q)  , 
          \label{3-point vertex from Weyl action}
\end{eqnarray}
where $\chi_\mu = \pd_\nu h_{\mu\nu}$. The momentum function $W^3_{\mu\nu,\lam\s}$ is defined through this equation. This interaction is necessary to calculate the two-loop cosmological constant correction in section 5.

\subsection{The $G_D$ term}

Next, we write down the kinetic term and the interactions derived from the $G_D$ action. From the expression of the bare coefficient $b_0$ (\ref{definition of bare coefficient b_0}) and the expansion formula (\ref{expansion of G_D}), this action is expanded as follows: 
\begin{eqnarray}
     &&  b_0  \int  d^D x \sq{g} G_D
         =  \fr{\mu^{D-4}}{(4\pi)^{D/2}} \int d^D x \Biggl\{
          \biggl(  \fr{b_1}{D - 4}+\fr{b_2}{(D - 4)^2} + \cdots \biggr) \bG_4
            \nonumber  \\
     && \quad
         + \biggl( b_1 +\fr{b_2}{D - 4} +\cdots \biggr)
         \biggl( 2 \phi \bDelta_4 \phi +\bG_4 \phi -\fr{2}{3} \bnb^2 \bR \phi + \fr{1}{18}\bR^2 \biggr)
             \nonumber   \\
     &&  \quad
         + \bigl[ (D - 4)b_1 +\cdots \bigr]
         \biggl( \phi^2 \bDelta_4 \phi + \half \bG_4 \phi^2 + 3 \phi \bnb^4 \phi 
             \nonumber \\
     &&  \qquad
         + 4 \phi \bR^{\mu\nu} \bnb_\mu \bnb_\nu \phi   - \fr{14}{9} \phi \bR \bnb^2 \phi 
         + \fr{10}{9} \phi \bnb^\lam \bR \bnb_\lam \phi  +\cdots  \biggr)
             \nonumber \\
     &&  \quad
         + \left[ (D - 4)^2 b_1  +  \cdots \right]  \left[ \fr{1}{3} \phi^3 \bDelta_4 \phi 
         + \left( 4 \chi_3  - \half \right)   \left( \bnb^\mu \phi \bnb_\mu \phi \right)^2 
         + \cdots \right]
             \nonumber \\
     &&  \quad
         + \cdots \Biggr\} .
          \label{expansion of G_D action}
\end{eqnarray}
Here, $\Delta_4$ is the fourth-order differential operator defined by \cite{riegert}
\begin{eqnarray*}
   \Delta_4 = \nb^4 + 2R^{\mu\nu}\nb_\mu \nb_\nu - \fr{2}{3} R \nb^2 + \fr{1}{3}\nb^\mu R \nb_\mu ,
\end{eqnarray*}
where $\sq{g}\Delta_4$ becomes conformally invariant at $D=4$ for a scalar quantity. The pole terms in (\ref{expansion of G_D action}) give the counterterms, while the others are the kinetic and interaction terms induced quantum mechanically. The terms that we do not use in this paper are denoted by the dots here.

The first line of the expansion (\ref{expansion of G_D action}) gives the counterterm subtracting UV divergences proportional to the Euler term $\bG_4$, which determine the residue $b_n$ in (\ref{definition of bare coefficient b_0}). The second line gives the Riegert action \cite{riegert}, which is the Wess-Zumino action associated with the conformal anomaly $E_4$. It includes the bilinear term of the conformal-factor field $\phi$ as
\begin{eqnarray*}
    \fr{\mu^{D-4}}{(4\pi)^{D/2}} 2 b \int d^D x   \phi \pd^4 \phi 
\end{eqnarray*}
at the lowest of the perturbations. Since this term is independent of the coupling $t$, we can use it as the kinetic term of $\phi$ and then the propagator is given by 
\begin{eqnarray}
    \langle \phi(k) \phi(-k) \rangle = \mu^{4-D} \frac{(4\pi)^{D/2}}{4b} \frac{1}{k^4} .
       \label{propagator of conformal-factor field}
\end{eqnarray}
Therefore, quantum corrections from this field are expanded in $1/b$, which corresponds to considering the large-$N$ expansion for the number of matter fields $N_{S,F,A}$ (\ref{value of coefficient b_c}).

The three-point self-interaction is induced in the third line as
\begin{eqnarray}
    S_{{\rm G}[\phi\phi\phi]}^{(D-4)b} = (D-4) b  \fr{\mu^{D-4}}{(4\pi)^{D/2}} \int d^D x \phi^2 \pd^4 \phi .
        \label{three-point vertex of phi}
\end{eqnarray}
Here, note that due to the presence of the $D-4$ factor, the contribution of this interaction to UV divergences appears in two or more loops.

Furthermore, expanding the metric $\bg_{\mu\nu}$ (\ref{expansion of metric field}) in each term by the traceless tensor field, we obtain the interactions between the conformal-factor field and the traceless tensor field. From the $-2\phi \bnb^2 \bR/3$ and $\bR^2/18$ terms in the second line of (\ref{expansion of G_D action}), we obtain two quadratic interactions
\begin{eqnarray}
   S_{{\rm G}[\phi h]}^{bt}  &=& 
       - bt \fr{\mu^{D/2-2}}{(4\pi)^{D/2}} \int d^D x  \frac{2}{3} \pd^2 \phi \pd_{\mu} \chi_\mu ,
                 \nonumber \\
   S_{{\rm G}[hh]}^{b t^2}  &=&
       \fr{bt^2}{(4\pi)^{D/2}} \int d^D x  \fr{1}{18} \pd_{\mu} \chi_\mu \pd_\nu \chi_\nu .
         \label{two-point interactions in Riegert sector}
\end{eqnarray}
Note that in Landau gauge these interactions do not contribute to loop calculations. This is one of the reasons why we employ Landau gauge. We can then considerably reduce the number of Feynman diagrams.

The three-point interaction derived from the $2\phi {\bar \Delta}_4 \phi$ term is given by
\begin{eqnarray}
   S_{{\rm G}[\phi\phi h]}^{bt}   &=&
      b \fr{\mu^{D-4}}{(4\pi)^{D/2}} \int d^D x  \, 2 \phi \bDelta_4 \phi \Bigl|_{o(t)}
                 \nonumber \\   
   &=&  bt \, \fr{\mu^{D/2-2}}{(4\pi)^{D/2}} \int d^D x  \biggl\{  4\pd_{\mu} \phi \pd_{\nu} \pd^2 \phi 
       + \frac{8}{3}\pd_{\mu}\pd_{\lam}\phi \pd_{\nu}\pd_{\lam}\phi 
                 \nonumber \\ 
   &&  -\frac{4}{3}\pd_{\lam}\phi \pd_{\mu}\pd_{\nu}\pd_{\lam} \phi 
       - 4\pd_{\mu}\pd_{\nu}\phi \pd^2 \phi 
       \biggr\} h_{\mu\nu} 
                \nonumber \\ 
   &=& bt \fr{\mu^{D/2-2}}{(4\pi)^{D/2}} \int \fr{d^D p d^D q}{(2\pi)^{2D}} 
            \phi(p)\phi(q) h_{\mu\nu}(-p-q) V^3_{\mu\nu}(p,q)  
         \label{3-point vertex bt V^3}             
\end{eqnarray}
and the four-point interaction is
\begin{eqnarray}
   S_{{\rm G}[\phi\phi hh]}^{b t^2}   &=&
       b \fr{\mu^{D-4}}{(4\pi)^{D/2}} \int d^D x  \, 2 \phi \bDelta_4 \phi \Bigl|_{o(t^2)}
                 \nonumber \\
   &=& \fr{bt^2}{(4\pi)^{D/2}} \int \fr{d^D p d^D q d^D r d^D s}{(2\pi)^{4D}} 
            (2\pi)^D \dl^D(p+q+r+s) 
                 \nonumber \\
   &&  \times  \phi(p)\phi(s) h_{\mu\nu}(q) h_{\lam\s}(r) V^4_{\mu\nu, \lam\s}(q,r;s) .
                  \label{4-point vertex bt^2 V^4}             
\end{eqnarray}
The momentum functions $V^3_{\mu\nu}$ and $V^4_{\mu\nu, \lam\s}$ are defined through these equations. Although the explicit expression of the four-point interaction is very complicated, we can straightforwardly derive it using the formulas given at the end of Appendix A.

Furthermore, in order to calculate the two-loop cosmological constant corrections in section 5, we need the following interactions. The three-point interaction with $bt^2$ coming from $\phi (\bG_4-2\bnb^2 \bR/3)$ in the second line of (\ref{expansion of G_D action}) is given by
\begin{eqnarray}
    S_{{\rm G}[\phi hh]}^{bt^2} 
    &=& b \fr{\mu^{D-4}}{(4\pi)^{D/2}} \int d^D x \, \phi \left( \bG_4 - \fr{2}{3} \bnb^2 \bR \right) \biggr|_{o(t^2)}
                  \nonumber \\
    &=& \fr{ b t^2}{(4\pi)^{D/2}} \int d^D x \, \phi
         \Biggl\{  \pd_\lam \pd_\s h_{\mu\nu} \pd_\lam \pd_\s h_{\mu\nu} -2 \pd_\nu \pd_\lam h_{\mu\s} \pd_\mu \pd_\lam h_{\nu\s} 
                  \nonumber \\
    &&   + \pd_\lam \pd_\s h_{\mu\nu} \pd_\mu \pd_\nu h_{\lam\s}
         - \biggl(  \pd^2 h_{\mu\nu} \pd^2 h_{\mu\nu}   - 4 \pd^2 h_{\mu\nu} \pd_\mu \chi_\nu 
         + 2 \pd_\mu \chi_\nu \pd_\mu \chi_\nu  
                  \nonumber \\
    &&   + 2 \pd_\mu \chi_\nu \pd_\nu \chi_\mu   \biggr)
         + \pd_\mu \chi_\mu \pd_\nu \chi_\nu
         + \fr{1}{3} \biggl[ \half \pd^2 \left( \pd_\lam h_{\mu\nu} \pd_\lam h_{\mu\nu} \right)
         + 2 \pd^2 \left( h_{\mu\nu} \pd_\mu \chi_\nu \right)  
                  \nonumber \\
    &&   + \pd^2 \left( \chi_\mu \chi_\mu \right)  + 2 h_{\mu\nu} \pd_\mu \pd_\nu \pd_\lam \chi_\lam
         + 2 \chi_\mu \pd_\mu \pd_\nu \chi_\nu \biggr]
         \Biggr\} 
                \nonumber \\ 
   &=& \fr{bt^2}{(4\pi)^{D/2}} \int \fr{d^D p d^D q}{(2\pi)^{2D}} 
            \phi(-p-q) h_{\mu\nu}(p) h_{\lam\s}(q) S^3_{\mu\nu,\lam\s}(p,q)   .
         \label{3-point vertex bt^2 S^3}             
\end{eqnarray}
The three-point interaction with $(D-4) bt$ obtained by expanding the terms listed in the third and fourth lines of (\ref{expansion of G_D action}) up to $o(t)$ is given by
\begin{eqnarray}
    S_{{\rm G}[\phi\phi h]}^{(D-4)bt}  
       &=& (D-4) b \fr{\mu^{D-4}}{(4\pi)^{D/2}} \int d^D x \biggl[
             \half \bG_4 \phi^2 + 3 \phi \bnb^4 \phi  + 4 \phi \bR^{\mu\nu} \bnb_\mu \bnb_\nu \phi  
                 \nonumber \\
       &&   - \fr{14}{9} \phi \bR \bnb^2 \phi   + \fr{10}{9} \phi \bnb^\lam \bR \bnb_\lam \phi \biggr] \biggr|_{o(t)}
               \nonumber \\
      &=&  (D-4)bt \fr{\mu^{D/2-2}}{(4\pi)^{D/2}} \int d^D x \, \phi 
           \biggl\{   - 6 h_{\mu\nu} \pd_\mu \pd_\nu \pd^2 \phi    - 6 \chi_\mu \pd_\mu \pd^2 \phi
                  \nonumber \\
      &&  - 2 \pd^2 h_{\mu\nu} \pd_\mu \pd_\nu \phi    + 4 \pd_\mu \chi_\nu \pd_\mu \pd_\nu \phi 
          - \fr{14}{9} \pd_\mu \chi_\mu \pd^2 \phi      
          + \fr{10}{9} \pd_\mu \pd_\nu \chi_\nu \pd_\mu \phi                                                 
          \biggr\} 
                  \nonumber \\
     &=&  (D-4)bt \fr{\mu^{D/2-2}}{(4\pi)^{D/2}} \int \fr{d^D p d^D q}{(2\pi)^{2D}} 
            \phi(p)\phi(q) h_{\mu\nu}(-p-q) T^3_{\mu\nu}(p,q) .  
                  \nonumber \\     
     &&   \label{3-point vertex (D-4)bt T^3}             
\end{eqnarray}
Expanding further up to $o(t^2)$, we obtain the following four-point interaction:
\begin{eqnarray}
    S_{{\rm G}[\phi\phi h h]}^{(D-4)bt^2}  
       &=& (D-4) b \fr{\mu^{D-4}}{(4\pi)^{D/2}} \int d^D x \biggl[
             \half \bG_4 \phi^2 + 3 \phi \bnb^4 \phi  + 4 \phi \bR^{\mu\nu} \bnb_\mu \bnb_\nu \phi  
                 \nonumber \\
    &&   - \fr{14}{9} \phi \bR \bnb^2 \phi   + \fr{10}{9} \phi \bnb^\lam \bR \bnb_\lam \phi \biggr] \biggr|_{o(t^2)}
               \nonumber \\
    &=&  (D-4) \fr{bt^2}{(4\pi)^{D/2}} \int \fr{d^D p d^D q d^D r d^D s}{(2\pi)^{4D}} 
            (2\pi)^D \dl^D(p+q+r+s) 
                 \nonumber \\
    &&   \times  \phi(p)\phi(s) h_{\mu\nu}(q) h_{\lam\s}(r) T^4_{\mu\nu, \lam\s}(q,r;s) ,
         \label{4-point vertex (D-4)bt^2 T^4}             
\end{eqnarray}
where the explicit form of this expansion can be derived using the formulas in Appendix A. The momentum functions $S^3_{\mu\nu,\lam\s}$, $T^3_{\mu\nu}$ and $T^4_{\mu\nu, \lam\s}$ are defined through these interactions.

\subsection{The Mass Parameter Terms}

We here present interactions derived from the Einstein action and the cosmological constant term in the full action $S$ (\ref{full gravitational action}) and define the renormalization factors for the mass parameters. Note that unlike the four-derivative actions, the exponential factor of $\phi$ remains in these actions. Owing to the nonrenormalization theorem (\ref{Z_phi = 1}), the treatment of this factor can be facilitated.

We first expand the Einstein action up to the second order of the coupling constant as
\begin{eqnarray}
    && -\fr{M_0^2}{2} \int d^D x \sq{g} R 
         = -\fr{M_0^2}{2} \int d^D x e^{(D-2)\phi} \Bigl\{ \bR - (D-1)\bnb^2 \phi  \Bigr\}
                \nonumber \\
    && = \fr{3}{2} M_0^2 \int d^D x e^{(D-2)\phi} \Biggl\{
                   \fr{D-1}{3} \pd^2 \phi + \fr{D-2}{3} t_0 h_{0 \mu\nu} \left( -\pd_\mu \pd_\nu \phi +\pd_\mu \phi \pd_\nu \phi \right)
                     \nonumber \\
    &&  \quad
               + \fr{D-1}{6} t_0^2 h_{0 \mu\lam} h_{0 \nu\lam} \pd_\mu \pd_\nu \phi 
               + \fr{D-1}{6} t_0^2 h_{0 \mu\nu} \pd_\mu h_{0 \nu\lam} \pd_\lam \phi
                    \nonumber \\
    &&  \quad
               -\fr{D-3}{6} t_0^2 h_{0 \mu\nu} \chi_{0\mu} \pd_\nu \phi 
               + \fr{t_0^2}{12} \pd_\lam h_{0 \mu\nu} \pd_\lam h_{0 \mu\nu}
               -\fr{t_0^2}{6} \chi_{0 \mu} \chi_{0 \mu} 
                  \Biggr\} .
           \label{expansion of Einstein action}
\end{eqnarray}
The renormalization factor is defined by
\begin{eqnarray}
    M_0^2 = \mu^{D-4} Z_{\rm EH} M^2 .
       \label{definition of Planck mass renormlization factor}
\end{eqnarray}
The anomalous dimension for the Planck mass is then defined by
\begin{eqnarray}
   \gm_{\rm EH} \equiv - \fr{\mu}{M^2} \fr{d M^2}{d\mu} = D-4 + {\bar \gm}_{\rm EH},
       \label{definition of anomalous dimension for Planck mass}
\end{eqnarray}
where ${\bar \gm}_{\rm EH}=\mu d(\log Z_{\rm EH})/d\mu$.

The cosmological constant term is simply written in terms of the exponential factor of the $\phi$ field as
\begin{eqnarray*}
    \Lam_0 \int d^D x \sq{g} = \Lam_0 \int d^D x  e^{D\phi} .
\end{eqnarray*}
The renormalization factor is defined by
\begin{eqnarray}
   \Lam_0 = \mu^{D-4} Z_\Lam \left( \Lam + L_M M^4 \right) ,
      \label{definition of Lambda renormlization factor}
\end{eqnarray}
where $L_M$ is the pure-pole term. The anomalous dimension for the cosmological constant is defined by
\begin{eqnarray}
   \gm_\Lam \equiv - \fr{\mu}{\Lam} \fr{d \Lam}{d\mu} = D-4 + {\bar \gm}_\Lam + \fr{M^4}{\Lam} {\bar \dl}_\Lam,
      \label{definition of anomalous dimension for cosmological constant}
\end{eqnarray}
where ${\bar \gm}_\Lam = \mu d(\log Z_\Lam)/d\mu$ and 
\begin{eqnarray}
   {\bar \dl}_\Lam = \mu \fr{dL_M}{d\mu} -(D-4)L_M + \left( {\bar \gm}_\Lam -2 {\bar \gm}_{\rm EH} \right) L_M .
      \label{definition of delta_Lambda}
\end{eqnarray}

\section{Some Results of Renormalization Factors}
\setcounter{equation}{0}

In this section, we present some results of the renormalization factors for loop diagrams with gravitational internal lines. Some of them have already been calculated elsewhere. We here add new calculations in arbitrary gauge as well.

First, we mention how to treat IR divergences. In fourth-order theories, in general, IR divergences become stronger than those in the usual second-order field theories. Further, since the Einstein term and the cosmological constant term have the exponential factor of $\phi$, these terms cannot be considered as usual mass terms. So, we have to regularize IR divergences by introducing an infinitesimal mass parameter $z$ in the propagators (\ref{propagator of traceless tensor field in arbitrary gauge}) and (\ref{propagator of conformal-factor field}) as
\begin{eqnarray}
   \fr{1}{k^4} \to \fr{1}{k_z^4} = \fr{1}{(k^2+z^2)^2} ,
       \label{IR cutoff in propagator}
\end{eqnarray}
while we do not introduce $z$ in the tensor part $I^{(\zeta)}_{\mu\nu,\lam\s}(k)$ to preserve the transverse and traceless properties. Since this mass parameter violates diffeomorphism invariance, it is a fictitious parameter that should be canceled out at the end.\footnote{
This implies that what is called a massive graviton is not gauge invariant. To begin with, the ordinary particle picture itself is incorrect in a background-free spacetime, as mentioned in section 4.1.}

In the following arguments, we set the dimension as
\begin{eqnarray*}
     D = 4-2\eps  
\end{eqnarray*}
and $1/{\bar \eps}=1/\eps - \gm + \log 4\pi$. In Feynman diagrams, the conformal-factor field $\phi$ and the traceless tensor field $h_{\mu\nu}$ are denoted by a solid line and a spiral line, respectively.

\subsection{The Beta Function}

\begin{figure}[h]
\begin{center}
\includegraphics[scale=1]{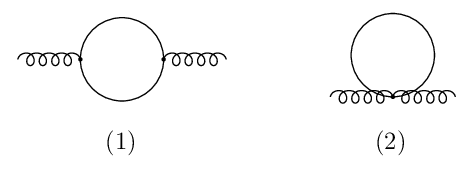}
\end{center}
\vspace{-6mm}
\caption{\label{Z_h one loop by phi}{\small The two-point function of $h_{\mu\nu}$ with a loop of $\phi$.}}
\end{figure}

Let us first calculate the beta function (\ref{definition of beta function for alpha_t}) of $\a_t=t^2/4\pi$. We here calculate the contribution from the two-point function of $h_{\mu\nu}$ with an internal $\phi$-line denoted by Fig.\ref{Z_h one loop by phi}, as an example.

Using the three-point interaction $S_{{\rm G}[\phi\phi h]}^{bt}$ with the momentum function $V^3_{\mu\nu}$ (\ref{3-point vertex bt V^3}), we can calculate the contribution from the diagram (1) in Fig.\ref{Z_h one loop by phi} as  
\begin{eqnarray*}
   \Gm_1^{\rm W} &=& - \fr{\mu^{4-D}}{16} t^2 \int \fr{d^D k}{(2\pi)^D} h_{\mu\nu}(k) h_{\lam\s}(-k) 
             \int \fr{d^D p}{(2\pi)^D} \fr{1}{p_z^4 (p+k)_z^4} 
                  \nonumber \\
         &&  \times 
             V^3_{\mu\nu}(p,-p-k) V^3_{\lam\s}(-p,p+k)
                  \nonumber \\
         &=& \fr{\a_t}{4\pi}  \! \int \!  \fr{d^D k}{(2\pi)^D} h_{\mu\nu}(k) h_{\lam\s}(-k)  
             \biggl\{  \fr{1}{30} \biggl( \half \dl_{\mu\lam}\dl_{\nu\s} k^4  - \dl_{\mu\lam} k_\nu k_\s k^2 
                  \nonumber \\
         &&  + \fr{1}{3} k_\mu k_\nu k_\lam k_\s  \biggr)
             \biggl( \fr{1}{{\bar \eps}} - \log \fr{k^2}{\mu^2} + \fr{229}{60} \biggr)
             - \fr{1}{270} k_\mu k_\nu k_\lam k_\s  \biggr\} ,
\end{eqnarray*}
where there is no $b$ dependence and IR divergences cancel out within this diagram. On the other hand, the tadpole diagram (2) coming from the four-point interaction $S_{{\rm G}[\phi\phi hh]}^{b t^2}$ (\ref{4-point vertex bt^2 V^4}) gives no contributions because the tadpole integral vanishes at the limit $z \to 0$ due to the presence of derivatives on the $\phi$ field in the interaction.

The right-hand side of the above can be combined into the $D$-dimensional Weyl form and thus the effective action from Fig.\ref{Z_h one loop by phi} is given by
\begin{eqnarray*}
   \Gm_1^{\rm W} &=&  \fr{\a_t}{4\pi}  \! \int \!  \fr{d^D k}{(2\pi)^D} h_{\mu\nu}(k) h_{\lam\s}(-k)  
       \biggl\{ \fr{1}{30} \left( \fr{1}{{\bar \eps}} - \log \fr{k^2}{\mu^2} + \fr{289}{60} \right) 
              \nonumber \\
    &&  \times  
       \biggl[ \fr{D-3}{D-2}  \left( \dl_{\mu\lam}\dl_{\nu\s} k^4  - 2 \dl_{\mu\lam} k_\nu k_\s k^2 \right)
       + \fr{D-3}{D-1} k_\mu k_\nu k_\lam k_\s  \biggr] \biggr\} .
\end{eqnarray*}
This divergence can be removed by the field renormalization factor $Z_h$ defined in (\ref{definitions of renormalization factors}) such that $Z_h -1$ is taken to be $- (1/30) (\a_t/4\pi\eps)$. Since this diagram is gauge invariant, it has a relationship with the renormalization factor $Z_t$ (\ref{definitions of renormalization factors}) such as $Z_t Z_h^{1/2}=1$. Thus, we obtain the contribution to $Z_t -1$ from Fig.\ref{Z_h one loop by phi} to be $(1/60)(\a_t/4\pi\eps)$. This result is consistent with the previous calculation using the DeWitt-Schwinger method in four dimensions \cite{amm92}.

In general, the renormalization factor for the coupling constant is given by \cite{cd, ddi, duff}
\begin{eqnarray*}
    Z_t = 1-  \left[ \fr{1}{480} \left( N_S + 6 N_F + 12 N_A \right) + \fr{197}{120} \right]  
                \fr{\a_t}{4\pi} \fr{1}{\eps}   + o(\a_t^2) .
\end{eqnarray*}
For the contribution from the traceless tensor field, we here quote the result \cite{ft, amm92, hs} obtained by using the background field method \cite{abbott} as follows. Introducing the background traceless tensor field as $\hg_{\mu\nu}=(e^{t{\hat h}})_{\mu\nu}$ and calculating the two-point function of the background field ${\hat h}_{\mu\nu}$, we obtain the contribution $-199/120$ for $Z_t$ using the relation $Z_t Z_{{\hat h}}^{1/2}=1$ ensured by the gauge invariance of the background field, where $Z_{{\hat h}}$ is the renormalization factor of the background field ${\hat h}_{\mu\nu}$. The sum of this value and $1/60$ from the conformal-factor field calculated above gives the last term at $o(\a_t)$.

Hence, the beta function (\ref{definition of beta function for alpha_t}) has the negative value as follows:
\begin{eqnarray*}
   \b_t \equiv \fr{\mu}{\a_t} \fr{d\a_t}{d\mu} 
        = - \left[ \fr{1}{120} \left( N_S + 6 N_F + 12 N_A \right) + \fr{197}{30} \right] \fr{\a_t}{4\pi} + o(\a_t^2) .
\end{eqnarray*}
The coupling $\a_t$ thus indicates the asymptotic freedom, which ensures that we develop the perturbation theory about conformally flat spacetime.

Note that the asymptotic limit here does not mean the realization of a picture in which free gravitons are propagating in the flat spacetime because the conformal factor is still nonperturbative and so the spacetime totally fluctuates quantum mechanically. And also, it indicates that scalarlike fluctuations by the conformal factor are much more dominant than tensor fluctuations at very high energies.

\subsection{The Nonrenormalization Theorem}

\begin{figure}[h]
\begin{center}
\includegraphics[scale=1]{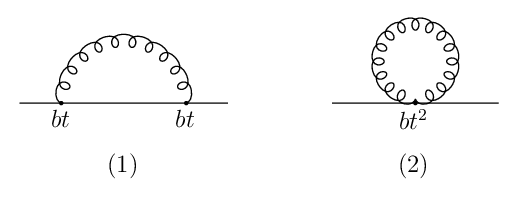}
\end{center}
\vspace{-8mm}
\caption{\label{Z_phi one loop}{\small The two-point function of $\phi$ at $o(\a_t)$.}}
\end{figure}

Here, we show the nonrenormalization theorem (\ref{Z_phi = 1}) at $o(\a_t)$ \cite{hamada02} in arbitrary gauge. We calculate the two-point function of $\phi$ depicted in Fig.\ref{Z_phi one loop} and show that all divergences cancel out.

The contribution from the diagram (1) in Fig.\ref{Z_phi one loop} is calculated using the three-point interaction $S_{{\rm G}[\phi\phi h]}^{bt}$ (\ref{3-point vertex bt V^3}) as
\begin{eqnarray*}
  \Gm^{\rm R}_1 &=& - \half \fr{b t^2}{(4\pi)^{D/2}} \fr{D-2}{2(D-3)} \int \fr{d^D k}{(2\pi)^D} \phi(k) \phi(-k) 
             \nonumber \\
    && \times
        \int \fr{d^D p}{(2\pi)^D} \fr{1}{p_z^4 (p-k)_z^4} I^{(\zeta)}_{\mu\nu,\lam\s}(p) 
        V^3_{\mu\nu}(p-k,k) V^3_{\lam\s} (p-k,k) .
\end{eqnarray*}
The explicit expression of the integrand is given in Appendix E. Evaluating IR divergences at $z \ll 1$, we obtain
\begin{eqnarray*}
  \Gm^{\rm R}_1  &=&   \fr{\mu^{D-4}}{(4\pi)^{D/2}} 2b \int \fr{d^D k}{(2\pi)^D} \phi(k) \phi(-k) k^4 
           \nonumber \\
  && \times  \fr{\a_t}{4\pi}  \left\{ 
        - \fr{5}{3} \left( \fr{1}{\bar{\eps}} - \log \fr{z^2}{\mu^2} \right) - \fr{43}{18}
         + \zeta \left[ - \fr{4}{3} \left( \fr{1}{\bar{\eps}} - \log \fr{z^2}{\mu^2} \right) - \fr{10}{9} \right]
        \right\} .
\end{eqnarray*} 
Here, the nonlocal term $\log (k^2/\mu^2)$ does not appear, which cancels out.

The tadpole diagram (2) in Fig.\ref{Z_phi one loop} can be calculated using the four-point interaction $S_{{\rm G}[\phi\phi hh]}^{b t^2}$ (\ref{4-point vertex bt^2 V^4}), which gives 
\begin{eqnarray*}
   \Gm^{\rm R}_2 &=&   \fr{\mu^{D-4}}{(4\pi)^{D/2}} 2b \int \fr{d^D k}{(2\pi)^D} \phi(k) \phi(-k) k^4 
           \nonumber \\
  && \times \fr{\a_t}{4\pi} \left\{  \fr{5}{3} \left( \fr{1}{\bar{\eps}}  -\log \fr{z^2}{\mu^2} \right) + \fr{11}{36} 
    + \zeta \left[ \fr{4}{3} \left( \fr{1}{\bar{\eps}} -\log \fr{z^2}{\mu^2} \right) + \fr{13}{9} \right] \right\} .
\end{eqnarray*}
This contribution comes from the terms without derivatives on $h_{\mu\nu}$ in (\ref{4-point vertex bt^2 V^4}), because if there are derivatives on $h_{\mu\nu}$ it gives a vanishing contribution at $z \to 0$ for such a one-loop tadpole diagram.\footnote{ 
Note that in two-loop calculations that involve such a tadpole diagram discussed in the next section, there are nonvanishing contributions from the interaction terms with derivatives on $h_{\mu\nu}$.} 

Now, combining the contributions from two diagrams (1) and (2) in Fig.\ref{Z_phi one loop}, we find that both UV divergences and IR divergences indeed cancel out and we obtain 
\begin{eqnarray*}
   \Gm^{\rm R}_{1+2}  =  2b \fr{\mu^{D-4}}{(4\pi)^{D/2}} \fr{\a_t}{4\pi} \int \fr{d^D k}{(2\pi)^D} \phi(k) \phi(-k) k^4 
     \left[  - \fr{25}{12} + \fr{1}{3} \zeta  \right] .
\end{eqnarray*}
Thus, we can see that the nonrenormalization theorem (\ref{Z_phi = 1}) holds at $o(\a_t)$.

The other nontrivial checks of the nonrenormalization theorem (\ref{Z_phi = 1}) have been given in the quantum gravity model coupled to QED up to $o(\a^3)$ and $o(\a^3/b)$ with an internal $\phi$-line \cite{hamada02} using Hathrell's result \cite{hathrellQED}, where $\a$ is the fine structure constant of QED.

\subsection{Renormalizations of The Mass Parameters}

\begin{figure}[h]
\begin{center}
\includegraphics[scale=1]{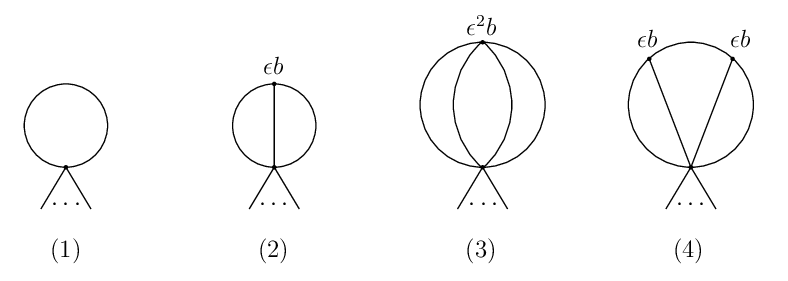}
\end{center}
\vspace{-8mm}
\caption{\label{Z_Lam at 3 loop}{\small The diagrams for $Z_\Lam$ up to $o(1/b^3)$.}}
\end{figure}

The renormalization of the cosmological constant has been carried out up to $o(1/b^3)$. The diagrams up to three loops that yield simple poles are given in Fig.\ref{Z_Lam at 3 loop}. These diagrams are evaluated with particular attention to the dependence on the mass scale $z$ and then extract UV divergences only. On the other hand, IR divergences are ignored here, which should cancel out after all. How IR divergences disappear in the effective cosmological constant term is demonstrated at the one-loop level in Appendix C.

The result for the renormalization factor $Z_\Lam$ defined in (\ref{definition of Lambda renormlization factor}) is given by \cite{hamada14QG}
\begin{eqnarray*}
   Z_\Lam = 1 - \left[ \fr{2}{b} + \fr{2}{b^2} + \fr{10}{3} \fr{1}{b^3} \right] \fr{1}{\eps} ,
\end{eqnarray*}
where the three-point self-interaction of $\phi$ (\ref{three-point vertex of phi}) contributes to the diagrams (2) and (4) in Fig.\ref{Z_Lam at 3 loop} and the four-point self-interactions of $\phi$ in the fifth line of (\ref{expansion of G_D action}) contribute to the diagram (3) in Fig.\ref{Z_Lam at 3 loop}.\footnote{ 
Although the second one in two four-point interactions depends on the coefficient $\chi_3$, the result becomes independent of it.} 
Using the equation $\mu db/d\mu = 2\eps [b + o(\a_t^2)]$ (\ref{RG equation of b}), we obtain 
\begin{eqnarray}
   {\bar \gm}_\Lam = \fr{4}{b} + \fr{8}{b^2} + \fr{20}{b^3} .
      \label{anomalous dimension of Lambda at t=0}
\end{eqnarray}
By replacing $b$ with the constant $b_c$ at $D=4$ at last, we obtain the expression of the anomalous dimension. This value agrees with the first three terms of the $1/b_c$-expansion of the exact solution $\gm_\Lam = 2b_c (1-\sq{1-4/b_c})-4$ derived from the BRST conformal algebra in four dimensions \cite{hamada12M4, hamada12RxS3}. This result shows that our gravitational action $G_D$ with the values (\ref{values of chi_1 and chi_2}) is correct also in quantum gravity beyond the curved-space argument given in \cite{hamada14CS} and Appendix B.

In Landau gauge, there are no corrections of $o(\a_t)$ to the cosmological constant. The next-order loop correction is given at $o(\a_t/b)$, which is discussed in the next section.

\begin{figure}[h]
\begin{center}
\includegraphics[scale=1]{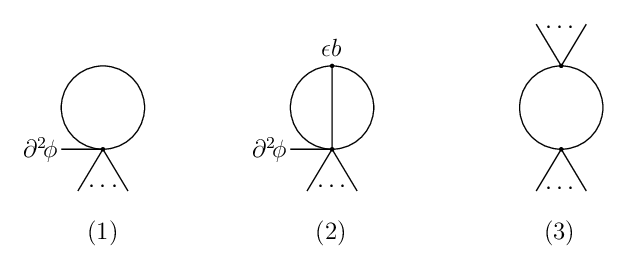}
\end{center}
\vspace{-8mm}
\caption{\label{Z_EH and L_M}{\small The first two are the diagrams for $Z_{\rm EH}$ and the last one is for the pole-term $L_M$.}}
\end{figure}

The Einstein term is also evaluated in the same way. The diagrams for $Z_{\rm EH}$ (\ref{definition of Planck mass renormlization factor}) up to $o(1/b^2)$ are given by (1) and (2) in Fig.\ref{Z_EH and L_M}. On the other hand, (3) in Fig.\ref{Z_EH and L_M} gives the pure-pole term $L_M$ in the renormalization of the cosmological constant term (\ref{definition of Lambda renormlization factor}). The results of these factors are \cite{hamada09}
\begin{eqnarray*}
   Z_{\rm EH} = 1 - \left[ \fr{1}{2b} + \fr{1}{4b^2} \right] \fr{1}{\eps}, \qquad 
   L_M = \fr{9}{16}\fr{(4\pi)^2}{b^2} \fr{1}{\eps} .
\end{eqnarray*}
From $Z_{\rm EH}$, we obtain the anomalous dimension (\ref{definition of anomalous dimension for Planck mass}) as ${\bar \gm}_{\rm EH}= 1/b+1/b^2$, which also agrees with the exact solution $\gm_{\rm EH} = 2b_c (1-\sq{1-2/b_c})-2$. The pole term $L_M$ gives the contribution to ${\bar \dl}_\Lam$ (\ref{definition of delta_Lambda}) in the mass-dependent part of (\ref{definition of anomalous dimension for cosmological constant}), which is
\begin{eqnarray}
    {\bar \dl}_\Lam = - \fr{9(4\pi)^2}{8b^2} .
      \label{result of delta_Lambda}
\end{eqnarray}

\begin{figure}[h]
\begin{center}
\includegraphics[scale=0.9]{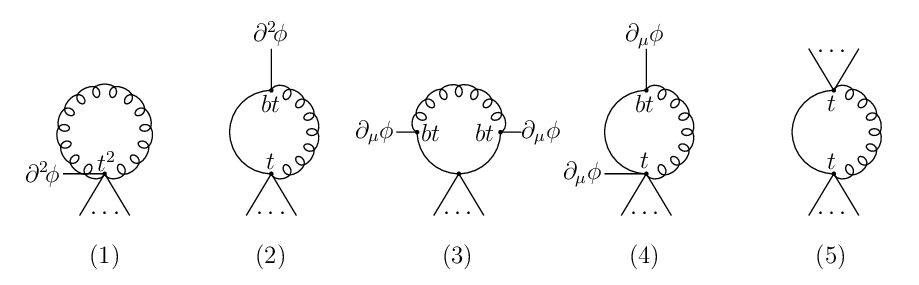}
\end{center}
\vspace{-8mm}
\caption{\label{loop corrections to Planck mass at o(alpha_t)}{\small The first four are the diagrams for $Z_{\rm EH}$ at $o(\a_t)$ and the last one is for $L_M$ at $o(\a_t/b)$.}}
\end{figure}

The potentially divergent $o(\a_t)$ loop diagrams in Landau gauge with the interactions (\ref{expansion of Einstein action}) are depicted in Fig.\ref{loop corrections to Planck mass at o(alpha_t)}, in which the first four diagrams contribute to $Z_{\rm EH}$. However, the three diagrams (2), (3) and (4) have no UV divergences in Landau gauge. The last diagram (5) that contributes to $L_M$ also has no UV divergences. Thus, only (1) in Fig.\ref{loop corrections to Planck mass at o(alpha_t)} gives the contribution such that
\begin{eqnarray*}
     Z_{\rm EH} = 1 - \fr{5}{8} \fr{\a_t}{4\pi} \fr{1}{\eps} .
\end{eqnarray*}
Combining with the coupling-independent part, we obtain the anomalous dimension
\begin{eqnarray}
    \gm_{\rm EH} = \fr{1}{b} + \fr{1}{b^2} + \fr{5}{4} \fr{\a_t}{4\pi} 
       \label{result for Planck mass anomalous dimension}
\end{eqnarray} 
with $b=b_c$.

\section{Two-Loop Corrections to The Cosmological Constant in Landau Gauge}
\setcounter{equation}{0}

\begin{figure}[h]
\begin{center}
\includegraphics[scale=1]{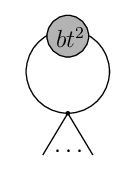}
\end{center}
\vspace{-8mm}
\caption{\label{two loop for Lambda}{\small The two-loop correction to $Z_\Lam$ at $o(\a_t/b)$ in Landau gauge, where the gray circle is depicted in Fig.\ref{bt^2 correction in Landau gauge}}}
\end{figure}

\begin{figure}[h]
\begin{center}
\includegraphics[scale=0.7]{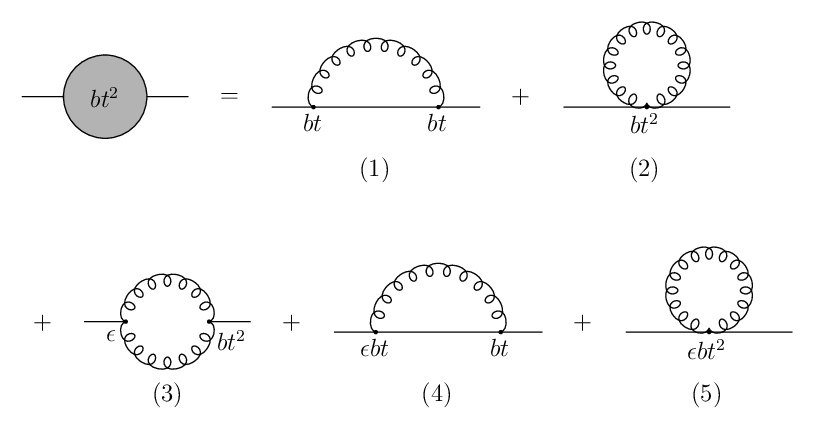}
\end{center}
\vspace{-8mm}
\caption{\label{bt^2 correction in Landau gauge}{\small The subdiagrams for $Z_\Lam$ at $o(\a_t/b)$.}}
\end{figure}

Now, we can calculate two-loop quantum gravity corrections to the cosmological constant, which are given at $o(\a_t/b)$ in Landau gauge. The Feynman diagram is given by Fig.\ref{two loop for Lambda} with the subdiagrams in Fig.\ref{bt^2 correction in Landau gauge}.

The integral expression of the effective action for one-loop subdiagram (a) in Fig.\ref{bt^2 correction in Landau gauge} denoted by $\Gm^{\rm R}_a \,(a=1, \cdots, 5)$ is given in Appendix E. The momentum integrations for $\Gm^{\rm R}_1$ and $\Gm^{\rm R}_2$ have been performed in section 4, which yields UV divergences, while $\Gm^{\rm R}_3$, $\Gm^{\rm R}_4$ and $\Gm^{\rm R}_5$ become finite because of the $D-4$ factor in the associated interactions.

The two-loop cosmological constant correction with the subdiagram (a) in Fig.\ref{bt^2 correction in Landau gauge} is denoted by $\Gm^\Lam_a$. It is calculated using the integrand $F_a$ given in Appendix E to define $\Gm^{\rm R}_a$. The momentum integration is performed using the Feynman integral formulas given in Appendix D. For each diagram, the calculation is done in arbitrary gauge and then we take Landau gauge. Each result is given as follows.

The contribution from the two-loop diagram of Fig.\ref{two loop for Lambda} with the subdiagram (1) in Fig.\ref{bt^2 correction in Landau gauge} is given by
\begin{eqnarray}
  \Gm^\Lam_1 &=&  \fr{t^2}{b} \mu^{4-D} (4\pi)^{D/2} \fr{D^2(D-2)}{64(D-3)} \Lam \int d^D x e^{D\phi}
       \int \fr{d^D p d^D q}{(2\pi)^{2D}} \fr{F_1(p,q)}{p_z^8 q_z^4 (q-p)_z^4}
             \nonumber \\
    &=&  \fr{t^2}{b} (4\pi)^{D/2-4} \mu^{4-D} (z^2)^{D-4} \Lam \int d^D x e^{D\phi}
              \nonumber \\
     &&  \times  \left[ \fr{40}{3(D-4)^2} + \fr{6}{D-4} + \zeta \left( \fr{32}{3(D-4)^2} + \fr{4}{D-4} \right) \right] ,
        \label{two loop result 1}
\end{eqnarray}
where the integrand $F_1$ is obtained by contracting two $V^3_{\mu\nu}$ in (\ref{3-point vertex bt V^3}), which is given by (\ref{F_1}).

The contribution from the two-loop diagram with (2) is given by\footnote{ 
Note that in this calculation, the product of the integrals such as $\int d^D p d^D q p^2 q^6/p_z^4 q_z^4$ yields a simple pole, though the single tadpole integral $\int d^D p p^2/p_z^4$ vanishes at $z \to 0$.} 
\begin{eqnarray}
  \Gm^\Lam_2 &=& - \fr{t^2}{b} \mu^{4-D} (4\pi)^{D/2} \fr{D^2(D-2)}{32(D-3)} \Lam \int d^D x e^{D\phi}
       \int \fr{d^D p d^D q}{(2\pi)^{2D}} \fr{F_2(p,q)}{p_z^8 q_z^4}
             \nonumber \\
    &=& - \fr{t^2}{b} (4\pi)^{D/2-4} \mu^{4-D} (z^2)^{D-4} \Lam \int d^D x e^{D\phi}
              \nonumber \\
     &&  \times  \left[ \fr{40}{3(D-4)^2} + \fr{124}{9(D-4)} + \zeta \left( \fr{32}{3(D-4)^2} + \fr{47}{9(D-4)} \right) \right] ,
        \label{two loop result 2}
\end{eqnarray}
where the integrand $F_2$ is obtained by contracting $V^4_{\mu\nu,\lam\s}$ in (\ref{4-point vertex bt^2 V^4}), which is given by (\ref{F_2}). Consequently, the double poles in $\Gm^\Lam_1$ (\ref{two loop result 1}) and $\Gm^\Lam_2$ (\ref{two loop result 2}) cancel out.

The contribution from the two-loop diagram with (3) is given by
\begin{eqnarray}
  \Gm^\Lam_3 &=&  \fr{t^2}{b} \mu^{4-D} (4\pi)^{D/2} \fr{D^2(D-2)^2(D-4)}{32(D-3)^2} \Lam \int d^D x e^{D\phi}
       \int \fr{d^D p d^D q}{(2\pi)^{2D}} \fr{F_3(p,q)}{p_z^8 q_z^4 (q-p)_z^4}
             \nonumber \\
    &=& \fr{t^2}{b} (4\pi)^{D/2-4} \mu^{4-D} (z^2)^{D-4} \fr{25}{3(D-4)} \Lam \int d^D x e^{D\phi} ,
        \label{two loop result 3}
\end{eqnarray}
where the integrand $F_3$ is obtained by contracting $W^3_{\mu\nu,\lam\s}$ in (\ref{3-point vertex from Weyl action}) and $S^3_{\mu\nu,\lam\s}$ in (\ref{3-point vertex bt^2 S^3}), which is given by (\ref{F_3}). The result is independent of $\zeta$ due to the property of the interaction (\ref{3-point vertex from Weyl action}).

The contribution from the two-loop diagram with (4) is given by
\begin{eqnarray}
  \Gm^\Lam_4 &=&  \fr{t^2}{b} \mu^{4-D} (4\pi)^{D/2} \fr{D^2(D-2)(D-4)}{32(D-3)} \Lam \int d^D x e^{D\phi}
       \int \fr{d^D p d^D q}{(2\pi)^{2D}} \fr{F_4(p,q)}{p_z^8 q_z^4 (q-p)_z^4}
             \nonumber \\
    &=& \fr{t^2}{b} (4\pi)^{D/2-4} \mu^{4-D} (z^2)^{D-4} \left[ \fr{110}{3(D-4)} + \zeta \fr{94}{3(D-4)} \right]
         \Lam \int d^D x e^{D\phi} ,
        \label{two loop result 4}
\end{eqnarray}
where the integrand $F_4$ is obtained by contracting $T^3_{\mu\nu}$ in (\ref{3-point vertex (D-4)bt T^3}) and $V^3_{\mu\nu}$ in (\ref{3-point vertex bt V^3}), which is given by (\ref{F_4}).

The contribution from the two-loop diagram with (5) is given by\footnote{ 
The double pole in $\Gm^\Lam_2$ originates from the interaction terms without derivatives on $h_{\mu\nu}$ that come from the $2\phi \bnb^4 \phi$ part in (\ref{4-point vertex bt^2 V^4}), and also the simple pole of $\Gm^\Lam_5$ originates from such interaction terms in the $(D-4)3\phi \bnb^4 \phi$ part of (\ref{4-point vertex (D-4)bt^2 T^4}). Consequently, the relationship $\Gm^\Lam_5|_{\hbox{simple pole}} = (3/2)(D-4) \times \Gm^\Lam_2 |_{\hbox{double pole}}$ is satisfied.} 
\begin{eqnarray}
  \Gm^\Lam_5 &=&  - \fr{t^2}{b} \mu^{4-D} (4\pi)^{D/2} \fr{D^2(D-2)(D-4)}{32(D-3)} \Lam \int d^D x e^{D\phi}
       \int \fr{d^D p d^D q}{(2\pi)^{2D}} \fr{F_5(p,q)}{p_z^8 q_z^4}
             \nonumber \\
    &=& - \fr{t^2}{b} (4\pi)^{D/2-4} \mu^{4-D} (z^2)^{D-4} \left[ \fr{20}{D-4} + \zeta \fr{16}{D-4} \right] 
         \Lam \int d^D x e^{D\phi} ,
        \label{two loop result 5}
\end{eqnarray}
where the integrand $F_5$ is obtained by contracting $T^4_{\mu\nu,\lam\s}$ in (\ref{4-point vertex (D-4)bt^2 T^4}), which is given by (\ref{F_5}).

Combining the contributions from (\ref{two loop result 1}) to (\ref{two loop result 5}), we finally obtain the following result in Landau gauge:
\begin{eqnarray*}
    \Gm^\Lam |_{\zeta=0} = \fr{t^2}{b} (4\pi)^{D/2-4} \mu^{4-D} (z^2)^{D-4} \fr{155}{9(D-4)} \Lam \int d^D x e^{D\phi} .
\end{eqnarray*}
Thus, the renormalization factor to remove this UV divergence is given by
\begin{eqnarray*}
     Z_\Lam  = 1 + \fr{t^2}{b(4\pi)^2} \fr{155}{18} \fr{1}{\eps}
\end{eqnarray*}
Therefore, the anomalous dimension (\ref{definition of anomalous dimension for cosmological constant}), including (\ref{anomalous dimension of Lambda at t=0}) and (\ref{result of delta_Lambda}) in the previous section, is given by
\begin{eqnarray}
     \gm_\Lam = \fr{4}{b} + \fr{8}{b^2} + \fr{20}{b^3} - \fr{9(4\pi)^2}{8 b^2}\fr{M^4}{\Lam} - \fr{310}{9 b} \fr{\a_t}{4\pi} 
        \label{final expression of anomalous dimension for cosmological constant}
\end{eqnarray}
with $b=b_c$.

Finally, we discuss the validity of the choice of Landau gauge. The anomalous dimension in Landau gauge (\ref{final expression of anomalous dimension for cosmological constant}) vanishes when we take the large $b$ limit. It is quite an acceptable result because in this limit the $\phi$ field becomes classical and thus the anomalous contributions from this field should disappear.

On the other hand, when we calculate the anomalous dimension in arbitrary gauge, there is a correction of $o(\a_t)$ proportional to $\zeta$ because the interaction $S_{{\rm G}[\phi h]}^{bt}$ in (\ref{two-point interactions in Riegert sector}) is then enabled, which does not vanish at the large $b$ limit. Furthermore, this interaction might cause the gauge-parameter dependence of $o(\a_t)$ in the coefficient $b_1^\pp$ supposed to be (\ref{assumption for b_1^prime}). If so, it induces new interactions that give a contribution of $o(\a_t)$ to the cosmological constant.

In this way, in arbitrary gauge, unnatural behavior in the anomalous dimension occurs. It may be because the interaction (\ref{two-point interactions in Riegert sector}) gives the contribution of a positive power of $b$ in loop corrections. It can be seen by rescaling the conformal-factor field as $\phi \to \phi/\sq{b}$ to remove the $b$ dependence in the kinetic term. The interaction term is then expanded in a negative power of $\sq{b}$, apart from the interactions in (\ref{two-point interactions in Riegert sector}) that have a positive power of $\sq{b}$. As a result, there arise the anomalous dimensions with a positive power of $b$ in arbitrary gauge.

One of the reasons why such a behavior occurs is that there exists the dimensionless product of the bare couplings like $(D-4)(4\pi)^{D/2} b_0 t_0^2 = b t^2 + o(t^4)$ in our renormalization procedure so that any functions of it become independent of $\mu$. So, we expect that the positive $b$ dependence in the renormalization factor can be factored out as a function of such a dimensionless product, which does not contribute to the anomalous dimension.

In any case, all such unnatural behaviors will vanish in Landau gauge. Of course, physical quantities should be gauge independent.\footnote{
Since we treat the constant part $b$ in (\ref{definition fo prime b_1}) as a coupling constant in order to incorporate the induced conformal-factor dynamics systematically, which temporarily breaks diffeomorphism invariance, the $\zeta$-independence of $b_1^\pp$ is not guaranteed. Therefore, as long as we use this trick, we may have to take the Landau gauge that is a fixed point of the RG flow for $\zeta$.} 
Since our formalism using dimensional regularization respects diffeomorphism invariance, we think that such an uncertainty will be resolved in the future.

\section{Conclusion and Discussion}
\setcounter{equation}{0}

We studied the renormalizable quantum conformal gravity with a single dimensionless coupling $t$ formulated using dimensional regularization.  The coupling $t$ introduced in front of the Weyl action handles the dynamics of the traceless tensor field, while the dynamics of the conformal-factor field is ruled by the $1/b_c$ expansion of the coefficient (\ref{value of coefficient b_c}) in front of the Riegert action that is induced from the gravitational counterterm $G_D$ (\ref{expression of G_D}) determined through the analysis of conformal anomalies using Hathrell's RG equations.

After carrying out some calculations of renormalization factors, including various consistency checks of our renormalization procedure, we have calculated the anomalous dimensions of the Planck mass and the cosmological constant at the order of $\a_t=t^2/4\pi$. We then employed Landau gauge not only to reduce the number of Feynman diagrams but also to avoid the indefiniteness as mentioned in the latter half of section 5. The results are given by (\ref{result for Planck mass anomalous dimension}) and (\ref{final expression of anomalous dimension for cosmological constant}), respectively. For the cosmological constant, such a correction appears at the order of $\a_t/b_c$ through two-loop diagrams.

The $\a_t$ correction to the anomalous dimension of the Planck mass is positive, while that of the cosmological constant becomes negative. This suggests that the Einstein term dominates the cosmological constant term in the low energy region. It might give a dynamical solution of the cosmological constant problem. The low energy effective theory of gravity valid below the IR energy scale $\Lam_\QG$, indicated from the negative beta function, would be given by expansion in derivatives starting around the Einstein action \cite{hhy06}.\footnote{ 
In the case of QCD, the dynamics of gauge fields disappears below the dynamical QCD scale, and the meson and baryon become dynamical fields. In quantum gravity, although the dynamics of fourth-order conformal gravity disappears below $\Lam_\QG$, the metric tensor still remains as the dynamical variable in Einstein gravity, which would be given by a composite field in which the conformal factor and the traceless tensor field are tightly binding.} 

The IR scale $\Lam_\QG$ separates the background-free quantum gravity phase and our classical universe where gravitons and elementary particles are propagating. If its value is taken to be about $10^{17}$GeV below the Planck mass scale, we can construct an inflationary scenario that ignites at the Planck scale and eventually ends at the $\Lam_\QG$ scale \cite{hy, hhy06, hhy10}.

\appendix 

\section{Useful Formulas for Gravitational Fields}
\setcounter{equation}{0}

The Christoffel symbol and the Riemann curvature tensor are defined by
\begin{eqnarray*}
    \Gm^\lam_{\mu\nu} &=&  \half g^{\lam\s} \left( \pd_\mu g_{\nu\s}  + \pd_\nu g_{\mu\s} - \pd_\s g_{\mu\nu} \right),
             \nonumber \\
    R^{\lam}_{~\mu\s\nu} &=&  \pd_{\s}\Gm^{\lam}_{~\mu\nu}- \pd_{\nu}\Gm^{\lam}_{~\mu\s}
                    + \Gm^\lam_{\rho\s} \Gm^\rho_{\mu\nu}  - \Gm^\lam_{\rho\nu}\Gm^\rho_{\mu\s} .
\end{eqnarray*}
The Ricci tensor and the Ricci scalar are defined by $R_{\mu\nu} = R^{\lam}_{~\mu\lam\nu}$ and $R=R^\mu_{~\mu}$, respectively. The covariant derivative then satisfies the following commutation relation:
\begin{eqnarray*}
   \left[ \nb_\mu, \nb_\nu \right] A_{\lam_1 \cdots \lam_n}
     = \sum_{j=1}^n R_{\mu \nu \lam_j}^{~~~~\s_j} A_{\lam_1 \cdots \s_j \cdots \lam_n} .
\end{eqnarray*}

The Weyl tensor is defined by
\begin{eqnarray*}
    C_{\mu\nu\lam\s} &=&  R_{\mu\nu\lam\s}  -  \fr{1}{D-2} \left(  g_{\mu\lam}R_{\nu\s}  -  g_{\mu\s}R_{\nu\lam} 
         -  g_{\nu\lam}R_{\mu\s}  +  g_{\nu\s}R_{\mu\lam}   \right) 
                \nonumber \\
    &&   +  \fr{1}{(D-1)(D-2)}  \left(  g_{\mu\lam} g_{\nu\s}  -  g_{\mu\s} g_{\nu\lam}  \right)  R .
\end{eqnarray*}
It satisfies the traceless conditions $C^\mu_{~\mu\lam\s}=C^\mu_{~\nu\mu\s}=0$. The number of independent components is $(D-3)D(D+1)(D+2)/12$. In three dimensions it vanishes identically and in four dimensions it has ten components.

Let us decompose the metric field as $g_{\mu\nu}=e^{2\phi}\bg_{\mu\nu}$. The curvatures are then expressed as
\begin{eqnarray*}
   \Gm^{\lam}_{~\mu\nu} &=& {\bar \Gm}^{\lam}_{~\mu\nu} +\bg^{\lam}_{~\mu}\bnb_{\nu}\phi 
       +\bg^{\lam}_{~\nu}\bnb_{\mu}\phi -\bg_{\mu\nu} \bnb^{\lam}\phi,
           \nonumber \\
   R^{\lam}_{~\mu\s\nu}  &=&  \bR^{\lam}_{~\mu\s\nu} + \bg^{\lam}_{~\nu}\bDelta_{\mu\s}   
       -\bg^{\lam}_{~\s}\bDelta_{\mu\nu}+\bg_{\mu\s}\bDelta^{\lam}_{~\nu}
       -\bg_{\mu\nu}\bDelta^{\lam}_{~\s} 
               \nonumber  \\
    && + \bigl( \bg^{\lam}_{~\nu}\bg_{\mu\s}    -\bg^{\lam}_{~\s}\bg_{\mu\nu} \bigr) 
               \bnb_{\rho}\phi\bnb^{\rho}\phi, 
              \nonumber \\ 
    R_{\mu\nu}  &=&  \bR_{\mu\nu}-(D-2)\bDelta_{\mu\nu} 
        -\bg_{\mu\nu} \left\{ \bnb^2 \phi   +(D-2)\bnb_{\lam}\phi \bnb^{\lam}\phi \right\},
               \nonumber  \\
    R  &=&  e^{-2\phi} \left\{ \bR -2(D-1)\bnb^2 \phi 
        -(D-1)(D-2)\bnb_{\lam}\phi \bnb^{\lam}\phi \right\} 
               \nonumber  \\
    C^\lam_{~\mu\s\nu}  &=&  {\bar C}^\lam_{~\mu\s\nu}  ,
\end{eqnarray*}
where $\bDelta_{\mu\nu}=\bnb_{\mu}\bnb_{\nu}\phi-\bnb_{\mu}\phi\bnb_{\nu}\phi$. The quantities with the bar are defined by using the metric $\bg_{\mu\nu}$. Thus, the square of the Weyl tensor is expressed as $\sq{g}C_{\mu\nu\lam\s}^2= \sq{\bg}e^{(D-4)\phi}{\bar C}_{\mu\nu\lam\s}^2$, and the Euler density is
\begin{eqnarray}
   \sq{g} G_4  = \sq{\hg}  e^{(D - 4)\phi}  \left[ \bG_4 + (D - 3) \bnb_\mu J^\mu 
                  + (D \!-\! 3)(D - 4) K \right] ,
           \label{mode decomposition of G_4}
\end{eqnarray}
where
\begin{eqnarray*}
    J^\mu  &=&
         8 \bR^{\mu\nu} \bnb_\nu \phi -4 \bR \bnb^\mu \phi 
         + 4(D-2) \bigl( \bnb^\mu \phi \bnb^2 \phi  - \bnb^\mu \bnb^\nu \phi \bnb_\nu \phi 
             \nonumber \\
       &&  + \bnb^\mu \phi \bnb_\lam \phi \bnb^\lam \phi \bigr) ,
             \nonumber \\
    K  &=&  4 \bR^{\mu\nu} \bnb_\mu \phi \bnb_\nu \phi  - 2 \bR \bnb_\lam \phi \bnb^\lam \phi 
         + 4 (D-2) \bnb^2 \phi \bnb_\lam \phi \bnb^\lam \phi 
             \nonumber \\
       &&         + (D-1)(D-2) \left( \bnb_\lam \phi \bnb^\lam \phi \right)^2 .
\end{eqnarray*}
Therefore, the $\phi$ dependence of $\sq{g}G_4$ becomes the divergence form in four dimensions.

\subsection{Expansion of the $G_D$ action}

The volume integral of $G_4$ is expanded in a power series of $D-4$ as   
\begin{eqnarray*}
    && \int d^D x \sq{g}  G_4  = \int  d^D x \sq{\bg} e^{(D-4)\phi}  \left[
           \bG_4 + (D - 3) \bnb_\mu J^\mu + (D - 3)(D - 4) K  \right]
             \nonumber \\ 
    && = \sum_{n=0}^\infty \fr{(D - 4)^n}{n!}  \int  d^D x \sq{\bg} \left\{
           \phi^n \bG_4 + (D - 3) \left( \phi^n \bnb_\mu J^\mu + n \phi^{n-1} K \right) \right\}
             \nonumber \\ 
    && = \sum_{n=0}^{\infty} \fr{(D - 4)^n}{n!}  \int d^D x \sq{\hg} \,  \biggl\{ 
           \phi^n \bG_4 +4(D - 3)\phi^n \bR^{\mu\nu} \bnb_\mu \bnb_\nu \phi 
             \nonumber \\ 
    && \quad
          -2(D - 3) \phi^n \bR \bnb^2 \phi -2(D - 2)(D - 3)(D - 4) \phi^n \bnb^2 \phi 
          \bnb^\lam \phi \bnb_\lam \phi 
             \nonumber  \\
    && \quad
         -(D - 2)(D - 3)^2(D - 4) \phi^n (\bnb^\lam \phi \bnb_\lam \phi )^2 
          \biggr\} 
\end{eqnarray*}
and the square of $H=R/(D-1)$ multiplied by $D-4$ is expanded as  
\begin{eqnarray*}
    && (D - 4) \int d^D x \sq{g}  H^2  
            \nonumber \\
    && = \sum_{n=0}^{\infty} \fr{(D - 4)^n}{n!}
         \int d^D x \sq{\hg}  \biggl\{  \fr{(D - 4)}{(D - 1)^2} \phi^n \bR^2  
         - \fr{2(D - 6)}{D - 1} \phi^n \bR \bnb^2 \phi  
             \nonumber \\ 
    && \quad      
         + \fr{2(D - 2)}{D - 1} \phi^n \bnb^\lam \bR \bnb_\lam \phi  
         +4 \phi^n \bnb^4 \phi
         +8(D - 4) \phi^n \bnb^2 \phi \bnb^\lam \phi \bnb_\lam \phi 
             \nonumber \\ 
    && \quad      
         +(D - 2)^2(D - 4) \phi^n (\bnb^\lam \phi \bnb_\lam \phi )^2 
          \biggr\} .
\end{eqnarray*}

Combining these expressions and $\chi(D)=1/2+3(D-4)/4+\chi_3(D-4)^2+\chi_4 (D-4)^3 +\cdots$, we can expand the volume integral of $G_D$ as    
\begin{eqnarray}
   && \int d^D x \sq{g} G_D =  \int  d^D x \sq{g} \left[ G_4 + (D - 4)\chi(D)H^2 \right] 
          \nonumber \\
   && = \sum_{n=0}^{\infty} \fr{(D - 4)^n}{n!}  \int   d^D x \sq{\hg}
         \biggl\{  \phi^n \bG_4 + \fr{D - 4}{(D - 1)^2}\chi(D) \phi^n \bR^2
            \nonumber \\
   && \quad
       +4(D - 3)\phi^n \bR^{\mu\nu} \bnb_\mu \bnb_\nu \phi
       - 2\left[ D - 3 +\fr{D - 6}{D - 1}\chi(D) \right] \phi^n \bR\bnb^2 \phi 
            \nonumber  \\ 
   && \quad
       + \fr{2(D - 2)}{D - 1}\chi(D) \phi^n \bnb^\lam \bR \bnb_\lam \phi
       + 4\chi(D) \phi^n \bnb^4 \phi
            \nonumber  \\ 
   && \quad   
       +2(D - 4)\left[ -(D - 2)(D - 3)+4\chi(D) \right] \phi^n \bnb^2 \phi \bnb_\lam \phi \bnb^\lam \phi   
             \nonumber\\ 
   && \quad
       +(D - 2)(D - 4) \left[ -(D - 3)^2 +(D - 2)\chi(D) \right] 
       \phi^n \left( \bnb_\lam \phi \bnb^\lam \phi \right)^2
             \biggr\} 
                \nonumber \\ 
   && = \int  d^D x \sq{\hg} \Biggl\{ 
       \bG_4 + (D - 4) \biggl( 2\phi \bDelta_4 \phi +\bG_4 \phi -\fr{2}{3} \bR \bnb^2 \phi + \fr{1}{18}\bR^2 \biggr) 
              \nonumber \\ 
   && \quad
       + (D - 4)^2 \biggl( \phi^2 \bDelta_4 \phi + \half \bG_4 \phi^2 
                 + 3\phi \bnb^4 \phi  +4 \phi \bR^{\mu\nu} \bnb_\mu \bnb_\nu \phi  
              \nonumber  \\ 
   && \quad         
        -\fr{14}{9} \phi\bR \bnb^2  \phi + \fr{10}{9} \phi \bnb^\lam \bR \bnb_\lam \phi 
        -\fr{7}{9} \bR \bnb^2 \phi  + \fr{1}{18} \bR^2 \phi +\fr{5}{108}\bR^2 \biggr)  
              \nonumber \\ 
   && \quad
        + (D - 4)^3 \Biggl[
        \fr{1}{3} \phi^3 \bDelta_4 \phi  + \fr{1}{6} \bG_4 \phi^3  
        + \left( 4\chi_3 -\half \right) \left( \bnb_\lam \phi \bnb^\lam \phi \right)^2 
                      \nonumber \\
   && \quad 
       + \left( 8 \chi_3 -2 \right) \bnb^2 \phi \bnb^\lam \phi \bnb_\lam \phi  
       + \fr{3}{2} \phi^2 \bnb^4 \phi 
       + 2 \phi^2 \bR^{\mu\nu} \bnb_\mu \bnb_\nu \phi 
       -\fr{7}{9} \phi^2 \bR \bnb^2 \phi
              \nonumber \\
   && \quad    
       + \fr{5}{9} \phi^2 \bnb^\lam \bR \bnb_\lam \phi + \fr{1}{36} \bR^2 \phi^2 
       + 4\chi_3 \phi \bnb^4 \phi 
       + \left( \fr{4}{3}\chi_3 -\fr{35}{54} \right) \phi \bR \bnb^2 \phi 
            \nonumber \\
   && \quad
       + \left( \fr{4}{3}\chi_3 + \fr{7}{54} \right) \phi \bnb^\lam \bR \bnb_\lam \phi
       + \fr{5}{108} \bR^2 \phi               
       + \left( -\fr{4}{3}\chi_3 + \fr{7}{27} \right) \bR \bnb^2 \phi
            \nonumber  \\ 
   && \quad  
       + \left( \fr{1}{9}\chi_3 -\fr{1}{27} \right) \bR^2 
       \Biggr] 
       + o((D - 4)^4)  \Biggr\} ,
      \label{expansion of G_D}
\end{eqnarray}
where the dependence on the coefficient $\chi_4$ arises from $o((D-4)^4)$.

\subsection{Expansions in Traceless Tensor Fields}

Let $\bg_{\mu\nu}=( \hg e^h )_{\mu\nu}$ using the traceless tensor $h^\mu_{~\nu}$. The curvature quantities with the bar are then expanded up to $o(h^2)$ as follows:
\begin{eqnarray*}
   \bar{\Gm}^\lam_{\mu\nu}  &=&  \hat{\Gm}^\lam_{\mu\nu}  + \hnb_{(\mu}h^\lam_{~\nu)} - \half \hnb^\lam h_{\mu\nu}
          + \half \hnb_{(\mu} (h^2)^\lam_{~\nu)} - \fr{1}{4} \hnb^\lam (h^2)_{\mu\nu}
              \nonumber \\
      &&  - h^\lam_{~\s} \hnb_{(\mu} h^\s_{~\nu)} + \half h^\lam_{~\s}\hnb^\s h_{\mu\nu},
              \nonumber \\
   \bR  &=&  \hR -\hR_{\mu\nu} h^{\mu\nu} + \hnb_\mu \hnb_\nu h^{\mu\nu} 
          + \half \hR^\s_{~\mu\lam\nu} h^\lam_{~\s} h^{\mu\nu} 
          - \fr{1}{4} \hnb^\lam h^\mu_{~\nu} \hnb_\lam h^\nu_{~\mu}
                \nonumber \\
      &&  + \half \hnb_\nu h^\nu_{~\mu} \hnb_\lam h^{\lam\mu}
          - \hnb_\mu ( h^\mu_{~\nu} \hnb^\lam h^\nu_{~\lam}) ,
                \nonumber \\
   \bR_{\mu\nu}  &=&  \hR_{\mu\nu}  -\hR^\s_{~\mu\lam\nu} h^\lam_{~\s}
          + \hR^\lam_{(\mu}h_{\nu)\lam} + \hnb_{(\mu} \hnb^\lam h_{\nu)\lam} 
          - \half \hnb^2 h_{\mu\nu}
                \nonumber \\
      &&  - \half h^\lam_{(\mu} \hnb^2 h_{\nu)\lam} 
          - \half \hnb^\lam h^\s_{~\mu} \hnb_\s h_{\nu\lam}
          - \fr{1}{4} \hnb_\mu h^\lam_{~\s} \hnb_\nu h^\s_{~\lam}
                \nonumber \\
      &&  - \half \hnb_\lam ( h^\lam_{~\s} \hnb_{(\mu} h^\s_{~\nu)})
          + \half \hnb_\lam ( h^\s_{(\mu} \hnb_{\nu)} h^\lam_{~\s})
          + \half \hnb_\lam ( h^\lam_{~\s} \hnb^\s h_{\mu\nu}) .
\end{eqnarray*}
Here, the contraction is taken by the background metric $\hg_{\mu\nu}$ and the traceless condition is $h^\mu_{~\mu}=\hg^{\mu\nu}h_{\mu\nu}=0$. The symmetric product is defined by $a_{(\mu}b_{\nu)}=(a_\mu b_\nu + a_\nu b_\mu)/2$.

When we employ the flat background $\hg_{\mu\nu}=\dl_{\mu\nu}$, the expansions of the squared curvatures with the bar and so on are given up to $o(h^2)$ by
\begin{eqnarray*}
   \bR^{\mu\nu\lam\s} \bR_{\mu\nu\lam\s}  &=&  \pd_\lam \pd_\s h_{\mu\nu} \pd_\lam \pd_\s h_{\mu\nu} 
        -2 \pd_\nu \pd_\lam h_{\mu\s} \pd_\mu \pd_\lam h_{\nu\s}
        + \pd_\lam \pd_\s h_{\mu\nu} \pd_\mu \pd_\nu h_{\lam\s} ,
            \nonumber \\
   \bR^{\mu\nu} \bR_{\mu\nu}   &=&  \half \pd_\mu \chi_\nu \pd_\mu \chi_\nu    
        - \pd^2 h_{\mu\nu} \pd_\mu \chi_\nu   + \half \pd_\mu  \chi_\nu \pd_\nu  \chi_\mu  
        + \fr{1}{4} \pd^2 h_{\mu\nu} \pd^2 h_{\mu\nu} ,
            \nonumber \\
   \bR^2   &=&  \pd_\mu \chi_\mu \pd_\nu \chi_\nu ,
            \nonumber \\
   \bnb^2 \bR   &=&
        \pd^2 \pd_\mu \chi_\mu - \fr{1}{4} \pd^2 \left( \pd_\lam h_{\mu\nu} \pd_\lam h_{\mu\nu} \right)
        - \half \pd^2 \left( \chi_\mu \chi_\mu \right) 
        - \pd^2 \left( h_{\mu\nu} \pd_\mu \chi_\nu \right) 
            \nonumber \\
     && - h_{\mu\nu} \pd_\mu \pd_\nu \pd_\lam \chi_\lam  
        - \chi_\mu \pd_\mu \pd_\nu \chi_\nu ,
\end{eqnarray*}
where $\chi_\mu = \pd_\nu h_{\mu\nu}$. The same lower indices denote contraction in the flat metric. The Euler density with the bar at $o(h^2)$ can be written in the divergence form
\begin{eqnarray}
   \bG_4  &=& \pd_\s L_\s 
      \label{total divergence formula of barG_4}
\end{eqnarray}
in any dimensions, where 
\begin{eqnarray*}
   L_\s  &=&   \pd_\lam h_{\mu\nu} \pd_\lam \pd_\s h_{\mu\nu} - \pd_\s h_{\mu\nu} \pd^2 h_{\mu\nu}
        - 2 \pd_\lam h_{\mu\nu} \pd_\lam \pd_\mu h_{\nu\s}  
        - 2 \pd_\lam h_{\nu\s} \pd_\lam \chi_\nu  
            \nonumber \\
     && + 4 \pd_\s h_{\mu\nu} \pd_\mu \chi_\nu   
        + \pd_\lam h_{\mu\nu} \pd_\mu \pd_\nu h_{\lam\s}
        - \pd_\lam h_{\nu\s} \pd_\nu \chi_\lam
        - \chi_\lam \pd_\lam \chi_\s  + \chi_\s \pd_\lam \chi_\lam .
\end{eqnarray*}

The quantities with the bar including the $\phi$ field are expanded as
\begin{eqnarray*}
  \bnb^2 \phi &=& \pd^2 \phi  - \chi_\mu \pd_\mu \phi - h_{\mu\nu} \pd_\mu \pd_\nu \phi 
                   + \half h_{\mu\lam} \pd_\mu h_{\nu\lam} \pd_\nu \phi 
                   + \half h_{\mu\lam} \chi_\lam \pd_\mu \phi
                      \nonumber \\
               &&    + \half h_{\mu\lam} h_{\nu\lam} \pd_\mu \pd_\nu \phi ,
                     \nonumber \\
  \bnb^4 \phi  &=& \pd^4 \phi  - \pd^2 ( h_{\mu\nu} \pd_\mu \pd_\nu \phi + \chi_\mu \pd_\mu \phi) 
         - h_{\mu\nu} \pd_\mu \pd_\nu \pd^2 \phi - \chi_\mu \pd_\mu \pd^2 \phi  
               \nonumber \\
     &&  + \pd^2 \biggl( \half h_{\mu\lam} h_{\nu\lam} \pd_\mu \pd_\nu \phi  + \half h_{\mu\nu} \pd_\mu h_{\nu\lam} \pd_\lam \phi
         + \half h_{\mu\nu} \chi_\mu \pd_\nu \phi \biggr)
               \nonumber \\
     &&  + h_{\mu\nu} \biggl( \pd_\mu \pd_\nu h_{\lam\s} \pd_\lam \pd_\s \phi   + 2 \pd_\mu h_{\lam\s} \pd_\nu \pd_\lam \pd_\s \phi 
         + h_{\lam\s} \pd_\mu \pd_\nu \pd_\lam \pd_\s \phi 
             \nonumber \\
     &&  + \half \pd_\mu h_{\nu\lam} \pd_\lam \pd^2 \phi   + \half \chi_\mu  \pd_\nu \pd^2 \phi 
         + \pd_\mu \pd_\nu \chi_\lam \pd_\lam \phi    + 2 \pd_\mu \chi_\lam \pd_\nu \pd_\lam \phi
             \nonumber \\
     &&  + 2 \chi_\lam \pd_\mu \pd_\nu  \pd_\lam \phi \biggr)
         + \chi_\mu \pd_\mu \chi_\nu \pd_\nu \phi   + \chi_\mu \chi_\nu \pd_\mu \pd_\nu \phi  
         + \chi_\mu \pd_\mu h_{\nu\lam} \pd_\nu \pd_\lam \phi  
             \nonumber \\
    &&   + \half h_{\mu\lam} h_{\nu \lam} \pd_\mu \pd_\nu \pd^2 \phi 
\end{eqnarray*}
and
\begin{eqnarray*}
  \bR^{\mu\nu} \bnb_\mu \bnb_\nu \phi    
    &=& \pd_\mu \chi_\nu \pd_\mu \pd_\nu \phi - \half \pd^2 h_{\mu\nu} \pd_\mu \pd_\nu \phi 
        - \half \pd_\lam  h_{\mu\nu} \pd_\lam \chi_\mu  \pd_\nu \phi   
             \nonumber \\
    &&  - \half \pd_\lam h_{\mu\nu} \pd_\mu \chi_\lam \pd_\nu \phi   + \half \pd_\lam h_{\mu\nu} \pd_\mu \chi_\nu \pd_\lam \phi   
        + \half \pd^2 h_{\mu\nu} \pd_\mu h_{\nu\lam} \pd_\lam \phi
             \nonumber \\
    &&  - \fr{1}{4} \pd^2 h_{\mu\nu} \pd_\lam h_{\mu\nu} \pd_\lam \phi    - h_{\mu\nu} \pd_\lam \chi_\mu \pd_\nu \pd_\lam \phi
        - h_{\mu\nu} \pd_\mu \chi_\lam \pd_\nu \pd_\lam \phi    
             \nonumber \\
    &&  + \half h_{\mu\nu} \pd^2 h_{\mu\lam} \pd_\nu \pd_\lam \phi  - \half \pd_\lam h_{\mu\nu} \pd_\mu h_{\lam\s} \pd_\nu \pd_\s \phi   
            \nonumber \\
    &&  - \fr{1}{4} \pd_\lam h_{\mu\nu} \pd_\s h_{\mu\nu} \pd_\lam \pd_\s \phi
        - \half \pd_\lam ( h_{\lam\s} \pd_\mu h_{\nu\s} ) \pd_\mu \pd_\nu \phi  
            \nonumber \\
    &&  + \half \pd_\lam ( h_{\mu\s} \pd_\nu h_{\lam\s}) \pd_\mu \pd_\nu \phi
        + \half \pd_\lam ( h_{\lam\s} \pd_\s h_{\mu\nu}) \pd_\mu \pd_\nu \phi  ,
            \nonumber \\
  \bR \bnb^2 \phi  &=& \pd_\mu \chi_\mu \pd^2 \phi   
        - \chi_\mu \biggl( \pd_\nu \chi_\nu \pd_\mu \phi + \half \chi_\mu \pd^2 \phi \biggr)
        - \fr{1}{4} \pd_\lam h_{\mu\nu} \pd_\lam h_{\mu\nu} \pd^2 \phi 
            \nonumber \\
    &&  - h_{\mu\nu} \biggl( \pd_\mu \chi_\nu \pd^2 \phi  + \pd_\lam \chi_\lam \pd_\mu \pd_\nu \phi  \biggr) ,
            \nonumber \\
   \bnb^\mu \bR \bnb_\mu \phi  &=& \pd_\mu \pd_\nu \chi_\nu \pd_\mu \phi    
        - \half \pd_\lam \pd_\s h_{\mu\nu} \pd_\lam h_{\mu\nu} \pd_\s \phi  - \half \pd_\mu ( \chi_\nu \chi_\nu ) \pd_\mu \phi
            \nonumber \\
    &&  - \pd_\lam ( h_{\mu\nu} \pd_\mu \chi_\nu ) \pd_\lam \phi    - h_{\mu\nu} \pd_\mu \pd_\lam \chi_\lam \pd_\nu \phi .
\end{eqnarray*}

\section{Determination of Gravitational Counterterms for QCD in Curved Space}
\setcounter{equation}{0}

We here briefly review recent achievements of how to determine the forms of gravitational counterterms in dimensional regularization.  In the previous paper \cite{hamada14CS}, we have determined them based on QED in curved space \cite{hathrellQED} as a prototype of conformally coupled quantum field theory. We here advance the argument to non-Abelian gauge theories in curved space \cite{freeman}, including fermions with an arbitrary representation.

The QCD action in curved space is defined by
\begin{eqnarray*}
   S &=& \int d^D x \sq{g} \biggl\{ \fr{1}{g_0^2} \left[ \fr{1}{4} F_{0\mu\nu}^a F_0^{a \mu\nu} 
                  + \fr{1}{2\xi_0} \left( \nb^\mu A^a_{0\mu} \right)^2 \right] 
                  + i {\bar \psi}_0 \gm^\mu D_\mu \psi_0 
            \nonumber \\
     &&  - i \pd^\mu {\tilde c}^a_0 \left( \pd_\mu c^a_0 + f^{abc} A^b_{0\mu} c^c_0 \right)  
         + a_0 F_D + b_0 G_4 + c_0 H^2 \biggr\} ,
\end{eqnarray*}
where $F^a_{0\mu\nu}=\pd_\mu A^a_{0\nu} - \pd_\nu A^a_{0\nu} + f^{abc} A^b_{0\mu} A^b_{0\nu}$ and $D_\mu = \pd_\mu + \om_{\mu\a\b}\Sigma_{\a\b}/2 + A^a_{0\mu} T^a$. $T^a$ is the generator of the Lie group, which   satisfies $[ T^a , T^b] = f^{abc} T^c$. 
The spin connection with Euclidean indices denoted by $\a$ and $\b$ here is defined using the vielbein $e^\mu_\a$ as $\om_{\mu\a\b}=e^\nu_\a (\pd_\mu e_{\nu\b} -\Gm^\lam_{\mu\nu} e_{\lam\b})$. The gamma matrix can be described by $\gm^\mu = e^\mu_\a \gm_\a$ and $\{ \gm_\a, \gm_\b \}=-2 \dl_{\a\b}$. The Lorentz generator is then given by $\Sigma_{\a\b}= - [\gm_\a, \gm_\b]/4$.

Here, for later convenience, we use the convention that the gauge coupling is factored out, and thus the field strength, the fermion and ghost actions do not manifestly depend on the coupling. The renormalization factors are then defined by $g_0 = \mu^{2-D/2} Z_g g$, $A_{0\mu}^a = \mu^{2-D/2} Z_A^{1/2} A_\mu^a$, $\psi_0 = Z_2^{1/2} \psi$ and $\xi_0 = Z_A Z_g^{-2} \xi$. Using $\a_g =g^2/4\pi$, the beta function is defined by $\b_g = (\mu/\a_g)d\a_g/d\mu =D-4 + \barb_g$ and the anomalous dimensions of the fields are $\gm_A = \mu d(\log Z_A)/d\mu$ and $\gm_2 = \mu d(\log Z_2)/d\mu$.

For the moment, we consider three gravitational counterterms with the bare couplings $a_0$, $b_0$ and $c_0$. The end of this appendix is to see that the last two are related to each other through the RG equations at all orders and thus we can combine them into the form $G_D$ (\ref{expression of G_D}).

The bare gravitational couplings are defined by $a_0 = \mu^{D-4}(a+L_a)$ with the pole term expanded as $L_a =\sum_{n=1}^\infty a_n /(D-4)^n$ and similar equations for $b_0$ and $c_0$.\footnote{ 
In quantum conformal gravity, we set $a_0=1/t_0^2$ and $b_0$ is taken to be the pure-pole term without the constant $b$ as (\ref{definition of bare coefficient b_0}), while $c_0$ is expressed by $b_0$ as discussed here.} 
The beta function of these couplings is defined by $\b_a = \mu da/d\mu = -(D-4)a + \barb_a$ and so on. The RG equations are $\barb_a = - \pd ( \a_g a_1)/\pd \a_g$ and $\pd(\a_g a_{n+1})/\pd \a_g + \barb_g \a_g \pd a_n/\pd \a_g = 0$ and similar equations for $b_n$ and $c_n$.

In the following, we essentially use some normal products denoted by the symbol $[ ~~]$. The equation of motion operators defined by $E_{0A}=(A_{0\mu}^a/\sq{g}) \dl S/\dl A_{0\mu}^a$ and $E_{0\psi}= ( {\bar \psi}_0 \dl S/\dl {\bar \psi}_0 + \psi_0 \dl S/\dl \psi_0 )/\sq{g}$ are the simplest normal products. It is because $\lang E_{0A,\psi} \prod A \prod \psi \rang$ becomes finite for any renormalized correlation function denoted by $\lang \prod A \prod \psi \rang$, which can be easily shown by carrying out the partial integral for each field variable in the path integral. Thus, we can write them as $E_{0A}=[E_A]$ and $E_{0\psi}=[E_\psi]$.

The normal product for the square of the gauge field strength is given by
\begin{eqnarray}
   \fr{1}{4g^2} \left[ F^a_{\mu\nu} F^{a \mu\nu} \right] 
    &=& \fr{D-4}{\b_g} \fr{1}{4g_0^2} F^a_{0\mu\nu} F^{a \mu\nu} - \fr{{\bar \gm}_A}{2\b_g} E_{0A} - \fr{{\bar \gm}_2}{2\b_g} E_{0\psi} 
                 \nonumber \\
    && + \fr{D-4}{\b_g} \mu^{D-4} \Biggl[ \left( L_a + \fr{{\bar \b}_a}{D-4} \right) F_D + \left( L_b + \fr{{\bar \b}_b}{D-4} \right) G_4
                 \nonumber \\
    &&  + \left( L_c + \fr{{\bar \b}_c}{D-4} \right) H^2 - \fr{4(\s+L_\s)}{D-4} \nb^2 H  \Biggr] ,
        \label{expression of normal product [F^2]}
\end{eqnarray}
where the anomalous dimensions with the bar are defined by ${\bar \gm}_A = \gm_A + [\gm_A-(D-4)] \xi\pd(\log Z_A)/\pd \xi$ and ${\bar \gm}_2 = \gm_2 + [\gm_A-(D-4)] \xi\pd(\log Z_2)/\pd \xi$. Note that $1/\b_g$ has poles and so $(D-4)/\b_g = 1 + \sum_{n=1}^\infty (-\barb_g)^n/(D-4)^n$.

In order to see that (\ref{expression of normal product [F^2]}) is a normal product, according to the technique developed in \cite{hathrellQED, freeman}, we consider a renormalized correlation function $(\a_g \pd/\pd \a_g - \xi \pd/\pd \xi) \lang \prod A \prod \psi \rang$. This finite correlation function can be expressed using (\ref{expression of normal product [F^2]}) in the form $\lang \int [ F^a_{\mu\nu} F^{a \mu\nu} ] \prod A \prod \psi \rang/4g^2$, up to the terms of gauge-fixing origin that becomes BRST trivial in physical correlation functions without ghost fields. In this way, we can determine the form of the normal product (\ref{expression of normal product [F^2]}), apart from the last divergence term. The last term can be determined by imposing a further condition such that the last three terms are finally combined into the form $E_D$ given in section 2.

Note that since the differential operators $\a_g \pd/\pd \a_g$ and $\xi \pd/\pd \xi$ do not act on the bare fields that are integration variables, they pass through the bare field strength, the fermion and ghost bare actions in our convention. It simplifies the calculation significantly.

The energy-momentum tensor is defined by $\theta^{\mu\nu} = (2/\sq{g}) \dl S/\dl g_{\mu\nu}$ and its trace is denoted by $\theta= \theta^{\mu}_{~\mu}=\dl S/\dl \Om$. This is also the normal product because $\lang \theta^{\mu\nu}\prod A \prod \psi \rang$ that is obtained by the variation of a renormalized correlation function is trivially finite. As a convention, we do not use the symbol $[~]$ for the energy-momentum tensor.

\paragraph{Two-Point Functions}
Since the partition function is finite, its gravitational variations are also finite. Therefore, carrying out the variation two times, we obtain the condition $\lang \theta(x) \theta(y) \rang - \lang \dl \theta(x)/\dl \Om(y) \rang = {\rm finite}$. From this, in momentum space, we obtain $\lang \theta(p) \theta(-p) \rang_{\rm flat} - 8 c_0 p^4 = {\rm finite}$. For later convenience, we introduce the variable ${\bar \theta} = \theta -(D-1)[E_\psi]/2$. Since the two-point function with $[E_\psi]$ vanishes,\footnote{ 
\label{vanishing of two-point function with E_psi} Since one-point functions are dimensionally regularized to zero for first- and second-order massless theories in flat space, $\lang [E_\psi(x)] P(y) \rang_{\rm flat} = \lang \dl P(y)/\dl \chi(x) \rang_{\rm flat} = 0$ is satisfied for a polynomial composite $P(y)$ in the fields ${\bar \psi}(y)$ and $\psi(y)$, where $\dl/\dl \chi = ( {\bar \psi}_0 \dl /\dl {\bar \psi}_0 + \psi_0 \dl /\dl \psi_0 )/\sq{g}$.} 
we obtain 
\begin{eqnarray*}
        \lang {\bar \theta}(p) {\bar \theta}(-p) \rang_{\rm flat} - 8p^4 \mu^{D-4} L_c = {\rm finite} .
\end{eqnarray*}

Next, we introduce the composite operator in flat space
\begin{eqnarray*}
   \{A^2\} \equiv \fr{D-4}{\b_g} \fr{1}{4 g_0^2} F^a_{0\mu\nu} F_0^{a\mu\nu} 
           = \fr{1}{4 g^2} \left[ F^a_{\mu\nu} F^{a\mu\nu} \right] 
                + \fr{1}{2\b_g} \left( {\bar \gm}_A [E_A] + {\bar \gm}_2 [E_\psi] \right) ,
\end{eqnarray*}
and consider its two-point function defined by $\Gm_{AA}(p^2) = \left\lang \{A^2(p)\} \{A^2(-p)\} \right\rang_{\rm flat}$ in momentum space. Although $\{A^2\}$ is not a normal product because $1/\b_g$ yields poles, the contributions from these terms with poles vanish due to the property of the equation of motion operators.  Therefore, $\Gm_{AA}$ is given by the two-point function of the normal product $[F^a_{\mu\nu}F^{a\mu\nu}]$.   Since correlation functions among normal products do not yield nonlocal poles in general, it can be expressed in the following form:
\begin{eqnarray}
     \Gm_{AA}(p^2) - p^4 \mu^{D-4} \left( \fr{D-4}{\b} \right)^2 L_x = {\rm finite} ,
       \label{defining equation of L_x}
\end{eqnarray}
where $L_x=\sum_{n=1}^\infty x_n/(D-4)^n$ is a new pure-pole term defined by this equation.

Since ${\bar \theta}|_{\rm flat} = \b_g \{A^2\}$ up to the term of gauge-fixing origin, $\b_g^2 \Gm_{AA} = \lang {\bar \theta} {\bar \theta} \rang_{\rm flat}$ is satisfied. So, we obtain the relation
\begin{eqnarray}
      (D-4)^2 L_x -8 L_c = {\rm finite} .
       \label{relation between L_x and L_c}
\end{eqnarray}
This implies that the simple-pole residue $c_1$ of $L_c$ can be determined from the residue $x_3$ of $L_x$ as $c_1 = x_3/8$.

We next consider the RG equation that relates $x_3$ with $x_1$. In order to derive it, we use the fact that if $F$ is a finite quantity, $\b_g^{-n} \mu d (\b_g^n F)/d\mu$ is also finite in spite of the presence of the pole factor $\b_g^{-n}$ because of $\b_g^{-n}\mu d\b_g^n/d\mu = n \a_g \pd \barb_g/\pd \a_g$. Applying this fact for $n=2$ to the finite equation (\ref{defining equation of L_x}) as $F$, we obtain the RG equation
\begin{eqnarray}
    \fr{1}{\b_g^2} \mu \fr{d}{d\mu} \left\{ \mu^{D-4} (D-4)^2 L_x \right\} = {\rm finite} ,
       \label{RG equation for L_x}
\end{eqnarray}
where we use the fact that $\mu d(\b_g \{A^2\})/d\mu = 0$ because $\b_g \{A^2\}$ can be described in terms of bare quantities. Expanding this equation in Laurent series and extracting finiteness conditions, we can derive the relation among the residues $x_n$. Solving it, we obtain 
\begin{eqnarray}
    x_3(\a_g)  = - \fr{\barb_g(\a_g)}{\a_g} \int^{\a_g}_0 d\a \left\{ \a^2 x_1(\a) \fr{\pd}{\pd \a} 
          \left( \fr{\barb_g(\a)}{\a} \right) \right\}  .
        \label{equation of x_3 from x_1}
\end{eqnarray}
As shown later, the lowest term of $x_1$ is $o(1)$ and thus $x_3$ and $c_1$ start from $o(\a_g^3)$.

\paragraph{Three-Point Functions}

We also consider the three-point function of $\theta$. In terms of ${\bar \theta}$, it is expressed in flat space as
\begin{eqnarray*}
    && \lang {\bar \theta}(x) {\bar \theta}(y) {\bar \theta}(z) \rang_{\rm flat} 
       - \lang {\bar \theta}(x) {\bar \theta}_2(y,z) \rang_{\rm flat}  
       - \lang {\bar \theta}(y) {\bar \theta}_2(z,x) \rang_{\rm flat} 
       - \lang {\bar \theta}(z) {\bar \theta}_2(x,y) \rang_{\rm flat}
             \nonumber \\
    && + \biggl\lang \fr{\dl^3 S}{\dl \Om(x) \dl \Om(y) \dl \Om(z)} \biggr\rang_{\rm flat} = {\rm finite} ,
\end{eqnarray*}
where ${\bar \theta}_2(x,y) = \dl {\bar \theta}(x)/\dl \Om(y) - (D-1) \dl {\bar \theta}(x)/2\dl \chi(y)$, where $\dl/\dl \chi$ is defined in footnote \ref{vanishing of two-point function with E_psi}.

The three-point function of $\{A^2\}$ is denoted by $\Gm_{AAA}$ as before. Since ${\bar \theta}|_{\rm flat}=\b_g \{A^2\}$ and ${\bar \theta}_2(x,y)|_{\rm flat}=-4\b_g \{A^2\} \dl^D(x-y)+8c_0 \pd^4 \dl^D (x-y)$, the finiteness condition above can be written in momentum space as
\begin{eqnarray*}
    && \b_g^3 \Gm_{AAA}(p_x^2,p_y^2,p_z^2) + 4\b_g^2 \left\{ \Gm_{AA}(p_x^2) + \Gm_{AA}(p_y^2) +\Gm_{AA}(p_z^2) \right\}
            \nonumber \\
    && + b_0 B(p^2_x,p^2_y,p^2_z) + c_0 C(p^2_x,p^2_y,p^2_z) = {\rm finite} ,
\end{eqnarray*}
where the last two functions are given by $B= -2(D-2)(D-3)(D-4) [ p_x^4 + p_y^4 + p_z^4  -2\left(p_x^2 p_y^2 + p_y^2 p_z^2 + p_z^2 p_x^2 \right) ]$ and $ C = -4 [ (D+2) ( p_x^4 + p_y^4 + p_z^4 ) +4 (p_x^2 p_y^2 + p_y^2 p_z^2 + p_z^2 p_x^2 ) ]$.

Consider the special cases that some momenta are taken to be on shell. Combining (\ref{defining equation of L_x}) and (\ref{relation between L_x and L_c}), we obtain
\begin{eqnarray}
    && \b_g^3 \Gm_{AAA}(p^2,0,0) - p^4 \mu^{D-4} \left[ 2(D-2)(D-3)(D-4)L_b + 4(D-6)L_c \right] 
         \nonumber \\
    && = {\rm finite} 
      \label{equations for beta^3 x Gamma_AAA}
\end{eqnarray}
and $\b_g^3 \Gm_{AAA}(p^2,p^2,0) - 8(D-4)p^4 \mu^{D-4} L_c ={\rm finite}$.

In general, $\Gm_{AAA}$ has the following form:
\begin{eqnarray}
  && \Gm_{AAA}(p_x^2,p_y^2,p_z^2) - \sum {\rm poles} \times \left\{ \Gm_{AA}(p_x^2) + \Gm_{AA}(p_y^2) + \Gm_{AA}(p_z^2) \right\} 
       \nonumber \\
  && - \mu^{D-4} \sum {\rm poles} \times \left\{ \hbox{terms in } p_i^2 p_j^2 \right\} = {\rm finite} .
      \label{general expression of Gm_AAA}
\end{eqnarray}
This expression cannot be obtained by dividing (\ref{equations for beta^3 x Gamma_AAA}) by $1/\b_g^3$ because $1/\b_g$ has poles.\footnote{ 
Since three-point functions with the equation of motion operators do not vanish, the terms with $1/\b_g$ in $\{A^2\}$ produce nonlocal poles. Thus, unlike $\Gm_{AA}$, $\Gm_{AAA}$ has nonlocal poles. The second term in (\ref{general expression of Gm_AAA}) plays a role in canceling out such nonlocal poles. } 
In order to determine the pure-pole factor in front of $\Gm_{AA}$, we consider the equation obtained by applying $\a_g \pd/\pd \a_g$ to (\ref{defining equation of L_x}), which yields the equation for $\Gm_{AAA}(p^2,p^2,0)$ because of $\a_g \pd S/\pd \a_g |_{\rm flat} = - \int d^D x \{ A^2 \}$ up to the gauge-fixing term origin and $\a_g \pd \{ A^2 \}/\pd \a_g = - [1 + (\a_g^2/\b_g)\pd (\barb_g/\a_g)/\pd \a_g] \{A^2\}$. The pole factor can be extracted from this equation and fixed to be $(\a_g^2/\b_g) \pd(\barb_g/\a_g)/\pd \a_g$. Therefore, $\Gm_{AAA}(p^2,0,0)$ has the following form:
\begin{eqnarray}
    \Gm_{AAA}(p^2,0,0) - \fr{\a_g^2}{\b_g} \fr{\pd}{\pd \a_g} \left( \fr{\barb_g}{\a_g} \right) \Gm_{AA}(p^2)
     - p^4 \mu^{D-4} \left( \fr{D-4}{\b_g} \right)^3 L_y = {\rm finite} .
     \label{defining equation of L_y}
\end{eqnarray}
Here, the last pure-pole term $L_y = \sum_{n=1}^\infty y_n/(D-4)^n$ cannot be deduced from the argument above, which is defined through this equation.

Multiplying (\ref{defining equation of L_y}) by $\b_g^3$, we obtain another equation of $\b_g^3 \Gm_{AAA}$, independent of (\ref{equations for beta^3 x Gamma_AAA}). By eliminating $\b_g^3 \Gm_{AAA}$ from these equations, we obtain the following pole relation:
\begin{eqnarray}
   && 2(D-2)(D-3)(D-4)L_b + 4 \left[ D-6 -2\a_g^2 \fr{\pd}{\pd \a_g} \left( \fr{\barb_g}{\a_g} \right) \right] L_c - (D-4)^3 L_y 
             \nonumber \\
   && = {\rm finite} .
       \label{RG equation of L_b}
\end{eqnarray}
where (\ref{defining equation of L_x}) and (\ref{relation between L_x and L_c}) are used.

The RG equation for $L_y$ can be derived as similar to the derivation of (\ref{RG equation for L_x}). Consider the equation obtained by applying $\b_g^{-3} \mu d/d\mu$ to (\ref{defining equation of L_y}) multiplied by $\b_g^3$. Since that $\mu d(\b_g^3\Gm_{AAA})/d\mu = \mu d(\b_g^2\Gm_{AA})/d\mu = 0$, we obtain the following RG equation:
\begin{eqnarray}
   \left( \fr{D-4}{\b_g} \right)^3  \left[ (D-4)L_y + \b_g \a_g \fr{\pd}{\pd \a_g}L_y \right]
    + \a_g^2 \fr{\pd^2 \barb_g}{\pd \a_g^2} \left( \fr{D-4}{\b_g} \right)^2 L_x = {\rm finite} ,
        \label{RG equation of L_y}
\end{eqnarray}
where (\ref{defining equation of L_x}) is used.

\paragraph{Gravitational Counterterms}

The four RG equations (\ref{relation between L_x and L_c}), (\ref{RG equation for L_x}), (\ref{RG equation of L_b}) and (\ref{RG equation of L_y}) imply that the pole terms $L_b$ and $L_c$ are related to each other at all orders. Thus, we can combine these two terms into $G_D$ (\ref{expression of G_D}) as introduced in section 2. The coupling constant $c$ in $c_0$ is then removed. By solving the RG equations, we can determine the function $\chi(D)$ order by order when it is expanded as (\ref{expansion expression of chi}) \cite{hamada14CS}.

Since the derived RG equations have the same forms as those obtained for the QED in curved space, we can solve them in the same way as in the QED case. The information needed to solve the RG equations is the simple pole of $L_x$ and $L_y$ and the QCD beta function expanded as
\begin{eqnarray*}
   x_1 &=& X_1 + X_2 \a_g +o(\a_g^2),  \qquad
   y_1 = Y_1 + Y_2 \a_g  +o(\a_g^2) ,
         \nonumber \\
   \barb_g &=& \b_1 \a_g + \b_2 \a_g^2 + o(\a_g^3) .
\end{eqnarray*}
The solution for the first three terms of $\chi$ is then given by
\begin{eqnarray*}
      &&  \chi_1 = \half , \qquad  \chi_2 = 1 -\fr{Y_1}{4X_1},   
             \nonumber \\
      &&    \chi_3 = \fr{1}{8} \left( 2- \fr{Y_1}{X_1} \right) \left( 3 -\fr{Y_1}{X_1} \right) 
             - \fr{1}{6} \fr{\b_2}{\b_1^2} \left( 1 - \fr{Y_1}{X_1} \right) 
             + \fr{1}{6} \fr{X_2}{\b_1 X_1} \left( \fr{Y_2}{X_2} - \fr{3}{2} \fr{Y_1}{X_1} \right) .
\end{eqnarray*}

The explicit values of the coefficients $X_1$ and $Y_1$ are obtained from the one-loop calculations of $\Gm_{AA}$ and $\Gm_{AAA}$, respectively. They are given by
\begin{eqnarray*}
    \Gm_{AA}(p^2) = - \fr{r}{2} \fr{\mu^{D-4} }{(4\pi)^2} p^4 \fr{1}{D-4}  , \qquad
    \Gm_{AAA}(p^2,0,0) = - \fr{r}{2} \fr{\mu^{D-4} }{(4\pi)^2} p^4 \fr{1}{D-4} .
\end{eqnarray*}
where $r$ is the dimension of the Lie group. From these, we obtain $X_1=Y_1=-r/2(4\pi)^2$ and thus $\chi_2$ is determined to be $3/4$.

In this way, we can see that at least $\chi_1$ and $\chi_2$ are the universal coefficients independent of the gauge group and the fermion representation. At present, it is not clear whether the coefficient $\chi_n ~(n \geq 3)$ has a universal value independent of the theories or not. In any case, $\chi_n$ can be determined at all orders.

Finally, we calculate the explicit value of $b_1^\pp$, which is the coupling-dependent part of $b_1$.\footnote{ 
Note that the coupling-independent part cannot be determined from the RG equation.} 
From $\chi_1=1/2$, we obtain the relation $b_2 = 2c_1 + o(\a_g^4)$. The residue $c_1 = x_3/8$ is obtained from $x_1$ through (\ref{equation of x_3 from x_1}). Since $x_1=X_1+o(\a_g)$, we obtain $c_1= - \b_1 \b_2 X_1 \a_g^3/96 + o(\a_g^4)$ and therefore $b_2$. Further, using the RG equation among $b_n$, we obtain
\begin{eqnarray}
    b_1^\pp = \fr{\b_2 X_1}{24} \a_g^2 + o(\a_g^3) .
       \label{b_1 for QCD in curved space}
\end{eqnarray} 
Thus, the coupling dependent part of the residue $b_1$ starts from $o(\a_g^2)$.

\section{Effective Potential and How To Handle IR Divergences}
\setcounter{equation}{0}

In this appendix, we calculate the one-loop effective potential for the cosmological constant term, and then we demonstrate that IR divergences indeed cancel out \cite{hamada09}.

\begin{figure}[h]
\begin{center}
\includegraphics[scale=0.6]{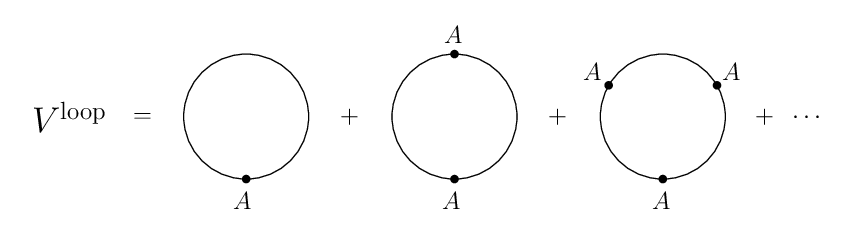}
\end{center}
\vspace{-8mm}
\caption{\label{effective cosmological potential}{\small One-loop effective potential for the cosmological constant term.}}
\end{figure}

We here introduce the background $\s$ and expand the $\phi$ field about it as follows: $\phi = \s + \sq{(4\pi)^{D/2}/4b \mu^{D-4}} \, \vphi$. We then expand the action up to the second order of $\vphi$, which is given by
\begin{eqnarray*}
     S |_{\vphi^2} = \int d^D x \left\{ \half \vphi \pd^4 \vphi + \fr{D^2(4\pi)^{D/2}}{8b} \Lam e^{D\s} \vphi^2 
               -\fr{1}{{\bar \eps}} \fr{2}{b} \mu^{D-4} \Lam e^{D\s}  \right\} .
\end{eqnarray*}
The last term is the counterterm to remove UV divergences. The kinetic term is given by ${\cal D}= k^4 + A$ in momentum space, where $A = (D^2 (4\pi)^{D/2}/4b) \, \Lam e^{D\s}$, and ${\cal D}_0=k^4$ for the free field. The one-loop effective potential depicted in Fig.\ref{effective cosmological potential} is then given by
\begin{eqnarray*}
     V^{\rm loop}  &=& - \log \left[ \det \left( {\cal D}_0^{-1} {\cal D} \right) \right]^{-\half}
                   \nonumber \\
           &=& \fr{\mu^{4 - D}}{2} \int \fr{d^D k}{(2\pi)^D} \log \left( 1 + \fr{A}{k_z^4} \right)
            =  \fr{\mu^{4 - D}}{2}  \sum_{n=1}^\infty \fr{(-1)^{n-1}}{n} A^n I_n ,
\end{eqnarray*}
where the IR cutoff is introduced as (\ref{IR cutoff in propagator}) and $I_n$ is the tadpole-type integral defined by 
\begin{eqnarray*}
    I_n = \int \fr{d^D k}{(2\pi)^D} \fr{1}{(k_z^4)^n} . 
\end{eqnarray*}
The integral $I_1$ has both UV and IR divergences. For $n \geq 2$, the integral has IR divergences only. They are given by
\begin{eqnarray*}
    I_1 =  \fr{1}{(4\pi)^2} \left( \fr{1}{{\bar \eps}} - \log z^2 \right), \quad
    I_{n (\geq 2)} =  \fr{1}{(4\pi)^2} \fr{z^{2(2-2n)}}{(2n - 1)(2n - 2)}  .
\end{eqnarray*}

After UV divergences are removed and $D=4$ is taken, we obtain the one-loop correction to the effective potential as
\begin{eqnarray*}
   V^{\rm loop}  &=&  \fr{1}{(4\pi)^2} \left\{ - \fr{A}{2} \log \fr{z^2}{\mu^2}  + z^4 \sum_{n=2}^\infty 
         \fr{(-1)^{n-1} A^n z^{-4n}}{2n(2n - 1)(2n - 2)} \right\}
             \nonumber \\
   &=&   \fr{1}{(4\pi)^2} \biggl\{ - \fr{A}{2} \log z^2 
         + \fr{1}{4} \left( z^4 -A \right) \log \left( 1+ \fr{A}{z^4} \right) 
             \nonumber \\
    &&  \qquad\quad 
         - z^2 \sq{A} \arctan \left( \fr{\sq{A}}{z^2} \right) + \fr{3}{4} A \biggr\} .
\end{eqnarray*}
Here, the sum of the series is calculated as follows. Let $f(x)=\sum_{n=2}^\infty (-1)^{n-1} x^{2n}/2n(2n-1)(2n-2)$ and so the series part is denoted by $z^4 f(\sq{A}/z^2)$. We also consider the series defined by $h(x)=\pd^2/\pd x^2 \{ f(x)/x \}=\sum_{n=3}^\infty (-1)^{n-1}x^{2n-3}/2n$, which can be easily evaluated as $h(x)=[ \log(1+x^2)-x^2]/2x^3$. The former series is then obtained by $f(x)=x \int^x_0 du \int^u_0 dv h(v)$.

Now, we can take the $z \to 0$ limit and obtain the finite expression $V^{\rm loop}= A\{3-\log(A/\mu^4)\}/4(4\pi)^2$. Adding the classical part $\Lam e^{4\s}$ and taking $b=b_c$, we finally obtain the effective potential 
\begin{eqnarray*}
    V = \Lam e^{4\s} \left\{ 1 + \fr{1}{b_c} \left[ 3 - \log \left( \fr{64\pi^2}{\mu^4} \fr{\Lam}{b_c} \right) \right] \right\} .
\end{eqnarray*}
In this way, we can demonstrate that IR divergences cancel out.

\section{Feynman Integral Formulas}
\setcounter{equation}{0}

We here summarize the integral formulas used in one- and two-loop calculations, which are evaluated by paying particular attention to IR divergences.  

\subsection{One-Loop Integral Formulas}

In one-loop calculations, we need the following integral:
\begin{eqnarray}
   I^{(n)}_\a &=& \int \fr{d^D k}{(2\pi)^D} \fr{(k^2)^\a (k \cdot l)^n}{(k^2+z^2)^2 [(k-l)^2 +z^2]^2}
          \nonumber \\
    &=& \Gm(4) \int^1_0 dx \, x(1-x) \int \fr{d^D p}{(2\pi)^D} \fr{(p+xl)^{2\a} ( p \cdot l +x l^2)^n}{(p^2 +K)^4}  ,   
         \label{definition of I integral}              
\end{eqnarray}
where $z$ is an IR cutoff (\ref{IR cutoff in propagator}) and
\begin{eqnarray*}
    K = z^2 + x(1-x) l^2 .
\end{eqnarray*}
In Landau gauge,  we also need the integral of the form
\begin{eqnarray}
   J^{(n)}_\b &=& \int \fr{d^D k}{(2\pi)^D} \fr{(k \cdot l)^n}{(k^2 +z^2)^2 [(k-l)^2 + z^2]^2 (k^2)^\b}
          \nonumber \\
    &=& \fr{\Gm(\b+4)}{\Gm(\b)} \int^1_0 dx \, x (1-x)^{\b+1} \int^1_0 dy \, y (1-y)^{\b-1} 
           \nonumber \\     
     && \times   
         \int \fr{d^D p}{(2\pi)^D} \fr{( p \cdot l + x l^2)^n}{(p^2 + L)^{\b+4}} ,      
     \label{definition of J integral}        
\end{eqnarray}
where 
\begin{eqnarray*}
    L = (x + y - xy) z^2 + x(1-x) l^2 .
\end{eqnarray*}
Note that there is no $z$ dependence in $1/(k^2)^\b$, and therefore $J^{(0)}_2$ is not defined here because there are IR divergences that cannot be handled by the cutoff $z$, but this integral is not necessary in our calculations.

In order to evaluate (\ref{definition of I integral}), we expand the numerator in powers of $p$. We then find that the following integral appears: 
\begin{eqnarray*}
  F_{n,m} = (4\pi)^2 \Gm(4) \int \fr{d^D p}{(2\pi)^D} \fr{(p^2)^n (p \cdot l)^{2m}}{(p^2+K)^4}
          = (l^2)^{\fr{D}{2} + n + 2m -4} \barF_{n,m} ,
\end{eqnarray*}
where $\barF_{n,m}$ is a dimensionless quantity defined by
\begin{eqnarray*}
   \barF_{n,m} = \fr{1}{(4\pi)^{D/2-2}} \fr{\Gm(m+\half)\Gm(n+m+\fr{D}{2})\Gm(4-n-m-\fr{D}{2})}{\Gm(\half)\Gm(m+\fr{D}{2})} 
                  \barK^{ \fr{D}{2}+n+m-4} 
\end{eqnarray*}
with 
\begin{eqnarray}
    \barK = \fr{K}{l^2} = w^2 +x(1-x),  \qquad w^2 = \fr{z^2}{l^2} .
      \label{expression of barK}
\end{eqnarray}
This function satisfies the relation
\begin{eqnarray*}
   \barF_{n-k,k} = \fr{\Gm(\fr{D}{2})\Gm(k+\half)}{\Gm(k+\fr{D}{2})\Gm(\half)} \barF_{n,0} .
\end{eqnarray*}

In order to evaluate the $p$-integral in (\ref{definition of J integral}), we also need the following integral:
\begin{eqnarray*}
  R_{m;\b} &=& (4\pi)^2 \fr{\Gm(\b+4)}{\Gm(\b)} \int \fr{d^D p}{(2\pi)^D} \fr{ (p \cdot l)^{2m}}{(p^2+L)^{\b+4}}
          = (l^2)^{\fr{D}{2} + 2m -\b -4} \bR_{m;\b} , 
\end{eqnarray*}
where the dimensionless part is defined by
\begin{eqnarray*}
   \bR_{m;\b} &=& \fr{1}{(4\pi)^{D/2-2}} \fr{\Gm(m+\half)\Gm(\b+4-m-\fr{D}{2})}{\Gm(\half)\Gm(\b)} 
                  \barL^{ \fr{D}{2}+m-\b-4} ,
\end{eqnarray*}
with
\begin{eqnarray}
   \barL = \fr{L}{l^2} = (x + y - xy)w^2 +x(1-x) .
      \label{expression of barL}
\end{eqnarray}

The integrals $I^{(n)}_\a$(\ref{definition of I integral}) and $J^{(n)}_\a$(\ref{definition of J integral}) are then given by the linear combinations of the parameter integrals of these functions defined by
\begin{eqnarray*}
   \left[ x^a \barF_{n,m} \right] &=& \int^1_0 dx \, x^{a+1} (1-x) \, \barF_{n,m} ,
           \nonumber \\ 
   \left[ x^a \bR_{m;\b} \right] &=& \int^1_0 dx \, x^{a+1} (1-x)^{\b+1} \int^1_0 dy \, y (1-y)^{\b-1} \, \bR_{m;\b} .
\end{eqnarray*}

These parameter integrals have to be evaluated with attention to IR divergences. We here summarize the results used in this paper. First, the integrals $\left[x^a \barF_{n,0} \right] ~(n \geq 1)$ that have no IR divergences are evaluated at $w^2 =0$ and we obtain 
\begin{eqnarray*}
   \left[ \barF_{4,0} \right] &=&  \fr{3}{7 {\bar \eps}} + \fr{729}{980} ,  \qquad
   \left[ \barF_{3,0} \right] = - \fr{4}{5 {\bar \eps}} - \fr{89}{75} ,
            \nonumber \\
   \left[ x \barF_{3,0} \right] &=& - \fr{2}{5 {\bar \eps}} - \fr{89}{150} ,  \qquad
   \left[ x^2 \barF_{3,0} \right] = - \fr{8}{35 {\bar \eps}} - \fr{1276}{3675} ,
            \nonumber \\
   \left[ \barF_{2,0} \right] &=& \fr{1}{{\bar \eps}} + \fr{5}{6} , \qquad
   \left[ x \barF_{2,0} \right] =  \fr{1}{2 {\bar \eps}} + \fr{5}{12} ,
            \nonumber \\
   \left[ x^2 \barF_{2,0} \right] &=& \fr{3}{10 {\bar \eps}} + \fr{27}{100} , \qquad
   \left[ x^3 \barF_{2,0} \right] = \fr{1}{5 {\bar \eps}} + \fr{59}{300} , 
            \nonumber \\
   \left[ x^4 \barF_{2,0} \right] &=& \fr{1}{7 {\bar \eps}} + \fr{449}{2940} , \qquad
   \left[ \barF_{1,0} \right] = 2 ,            
            \nonumber \\
     \left[ x \barF_{1,0} \right] &=& 1 ,  \qquad
   \left[ x^2 \barF_{1,0} \right] = \fr{2}{3} , \qquad
   \left[ x^3 \barF_{1,0} \right] = \half ,  
               \nonumber \\
   \left[ x^4 \barF_{1,0} \right] &=& \fr{2}{5} , \qquad 
   \left[ x^5 \barF_{1,0} \right] = \fr{1}{3} , \qquad 
   \left[ x^6 \barF_{1,0} \right] = \fr{2}{7}  .   
\end{eqnarray*}
The integrals with IR divergences only are evaluated at $D=4$ and $w^2 \ll 1$, which are given by 
\begin{eqnarray*}
   \left[ \barF_{0,0} \right] &=& - 2 \log w^2 - 2,  \qquad
   \left[ x \barF_{0,0} \right] = - \log w^2 - 1 , 
       \nonumber \\
   \left[ x^2 \barF_{0,0} \right] &=& - \log w^2 - 2,  \qquad
   \left[ x^3 \barF_{0,0} \right] = - \log w^2 - \fr{5}{2} , 
       \nonumber \\
   \left[ x^4 \barF_{0,0} \right] &=& - \log w^2 - \fr{17}{6},  \qquad
   \left[ x^5 \barF_{0,0} \right] = - \log w^2 - \fr{37}{12} , 
       \nonumber \\
   \left[ x^6 \barF_{0,0} \right] &=& - \log w^2 - \fr{197}{60},  \qquad
   \left[ x^7 \barF_{0,0} \right] = - \log w^2 - \fr{69}{20} , 
       \nonumber \\
   \left[ x^8 \barF_{0,0} \right] &=& - \log w^2 - \fr{503}{140} .
\end{eqnarray*}

In the same way, we calculate $\left[ x^a \bR_{m;\b} \right]$. For $\b=1$, we obtain
\begin{eqnarray*}
   \left[ \bR_{0;1} \right] &=& \fr{1}{w^2} - 2 \log w^2 - \fr{11}{2},  \qquad
   \left[ x \bR_{0;1} \right] = - 2 \log w^2 - \fr{5}{2} , 
       \nonumber \\
   \left[ x^2 \bR_{0;1} \right] &=& - \log w^2 - \fr{3}{2},  \qquad
   \left[ x^3 \bR_{0;1} \right] = - \log w^2 - \fr{5}{2} , 
       \nonumber \\
   \left[ x^4 \bR_{0;1} \right] &=& - \log w^2 - 3 ,  \qquad
   \left[ x^5 \bR_{0;1} \right] = - \log w^2 - \fr{10}{3} , 
       \nonumber \\
   \left[ x^6 \bR_{0;1} \right] &=& - \log w^2 - \fr{43}{12} ,
       \nonumber \\
   \left[ \bR_{1,1} \right] &=&  - \fr{1}{4} \log w^2 - \fr{1}{8},  \qquad
   \left[ x \bR_{1;1} \right] =  \fr{1}{4} ,  \qquad
   \left[ x^2 \bR_{1;1} \right] =  \fr{1}{8}, 
       \nonumber \\
   \left[ x^3 \bR_{1;1} \right] &=&  \fr{1}{12} ,  \qquad
   \left[ x^4 \bR_{1;1} \right] = \fr{1}{16} , 
       \nonumber \\
   \left[ \bR_{2;1} \right] &=&  \fr{3}{16},  \qquad
   \left[ x \bR_{2;1} \right] =  \fr{1}{16} , \qquad
   \left[ x^2 \bR_{2;1} \right] =  \fr{1}{32}, 
       \nonumber \\
   \left[ \bR_{3;1} \right] &=&  \fr{5}{64} \fr{1}{{\bar \eps}} + \fr{25}{192} , \qquad
   \left[ x \bR_{3;1} \right] =  \fr{1}{32} \fr{1}{{\bar \eps}} + \fr{47}{960} 
\end{eqnarray*}
and for $\b=2$ we obtain\footnote{ 
As mentioned before, $\left[ \bR_{0;2} \right]$ and so $J^{(0)}_2$ are not defined here.} 
\begin{eqnarray*}
   \left[ x \bR_{0;2} \right] &=& \fr{1}{w^2} - 2 \log w^2 - \fr{41}{6} , \qquad
   \left[ x^2 \bR_{0;2} \right] = - 2 \log w^2 - \fr{17}{6}, 
       \nonumber \\
   \left[ x^3 \bR_{0;2} \right] &=& - \log w^2 - \fr{11}{6} , \qquad
   \left[ x^4 \bR_{0;2} \right] = - \log w^2 - \fr{17}{6} , 
       \nonumber \\
   \left[ x^5 \bR_{0;2} \right] &=& - \log w^2 - \fr{10}{3} , \qquad
   \left[ x^6 \bR_{0;2} \right] = - \log w^2 - \fr{11}{3} ,
        \nonumber \\
   \left[ \bR_{1;2} \right] &=&   \fr{1}{4} \log w^2 - \fr{2}{3},  \qquad
   \left[ x \bR_{1;2} \right] =  - \fr{1}{6} \log w^2 - \fr{1}{9} ,  \qquad
       \nonumber \\
   \left[ x^2 \bR_{1;2} \right] &=&  \fr{1}{6}, \qquad
   \left[ x^3 \bR_{1;2} \right] =  \fr{1}{12} , \qquad
   \left[ x^4 \bR_{1;2} \right] = \fr{1}{18} ,  
      \nonumber \\
   \left[ \bR_{2;2} \right] &=& - \fr{1}{8} \log w^2 - \fr{7}{48},  \qquad
   \left[ x \bR_{2;2} \right] =  \fr{1}{16} , \qquad
   \left[ x^2 \bR_{2;2} \right] =  \fr{1}{48}, 
         \nonumber \\
   \left[ \bR_{3;2} \right] &=&  \fr{5}{48} .
\end{eqnarray*}

Using these quantities, we can calculate the integrals of (\ref{definition of I integral}) and (\ref{definition of J integral}). For $I^{(n)}_\a$, we obtain
\begin{eqnarray*}
    I^{(0)}_0 &=& \fr{1}{(4\pi)^2} \fr{1}{l^4} \left( 2 \log l^2 - 2 \log z^2 -2 \right), \quad
    I^{(0)}_1 = \fr{1}{(4\pi)^2} \fr{1}{l^2} \left( \log l^2 - \log z^2 \right), 
          \nonumber \\
    I^{(0)}_2 &=& \fr{1}{(4\pi)^2} \left( \fr{1}{\bar{\eps}} - \log z^2 \right),  \quad
    I^{(0)}_3 = \fr{1}{(4\pi)^2} l^2 \left( \fr{1}{\bar{\eps}} - \log z^2 \right), 
          \nonumber \\
    I^{(0)}_4 &=& \fr{1}{(4\pi)^2} l^4 \left( \fr{1}{\bar{\eps}} - \log z^2 \right),
          \nonumber \\
    I^{(1)}_0 &=& \fr{1}{(4\pi)^2} \fr{1}{l^2} \left(  \log l^2 - \log z^2 - 1  \right), \quad
    I^{(1)}_1 = \fr{1}{(4\pi)^2} \left(  \log l^2 - \log z^2 - 1  \right), 
          \nonumber \\
    I^{(1)}_2 &=& \fr{1}{(4\pi)^2} l^2 \left(  \fr{1}{\bar{\eps}} - \log z^2 \right),  \quad
    I^{(1)}_3 = \fr{1}{(4\pi)^2} l^4 \left(  \fr{1}{\bar{\eps}} - \log z^2 \right), 
          \nonumber \\
    I^{(2)}_0 &=& \fr{1}{(4\pi)^2} \left(  \log l^2 - \log z^2 - \fr{3}{2}  \right),
          \nonumber \\
    I^{(2)}_1 &=& \fr{1}{(4\pi)^2} l^2 \left( \fr{1}{4} \fr{1}{\bar{\eps}} +  \fr{3}{4} \log l^2  - \log z^2 -1 \right), \quad
    I^{(2)}_2 = \fr{1}{(4\pi)^2} l^4 \left(  \fr{1}{\bar{\eps}} - \log z^2 \right), 
          \nonumber \\
    I^{(3)}_0 &=& \fr{1}{(4\pi)^2} l^2 \left(   \log l^2 - \log z^2 - \fr{7}{4}  \right), 
          \nonumber \\
    I^{(3)}_1 &=& \fr{1}{(4\pi)^2} l^4 \left(  \fr{1}{2} \fr{1}{\bar{\eps}} +  \fr{1}{2} \log l^2  - \log z^2 - \fr{3}{4} \right),
          \nonumber \\
    I^{(4)}_0 &=& \fr{1}{(4\pi)^2} l^4 \left(  \fr{1}{8} \fr{1}{\bar{\eps}} + \fr{7}{8} \log l^2  - \log z^2 - \fr{13}{8} \right) . 
\end{eqnarray*}

For $J^{(n)}_\b$ with $\b=1$, we obtain
\begin{eqnarray*}
    J^{(0)}_1 &=& \fr{1}{(4\pi)^2} \fr{1}{l^6} \left( 2 \log l^2 - 2 \log z^2 - \fr{11}{2} + \fr{l^2}{z^2} \right), 
          \nonumber \\
    J^{(1)}_1 &=& \fr{1}{(4\pi)^2} \fr{1}{l^4} \left( 2 \log l^2 - 2 \log z^2  - \fr{5}{2} \right), 
          \nonumber \\
    J^{(2)}_1 &=& \fr{1}{(4\pi)^2} \fr{1}{l^2} \left( \fr{5}{4} \log l^2 - \fr{5}{4} \log z^2  - \fr{13}{8} \right), \quad
    J^{(3)}_1 = \fr{1}{(4\pi)^2} \left( \log l^2  -  \log z^2 - \fr{7}{4} \right),
          \nonumber \\
    J^{(4)}_1 &=& \fr{1}{(4\pi)^2} l^2 \left(  \log l^2 - \log z^2 - \fr{33}{16}  \right), \quad
    J^{(5)}_1 = \fr{1}{(4\pi)^2} l^4 \left( \log l^2 - \log z^2  - \fr{35}{16} \right),
          \nonumber \\
    J^{(6)}_1 &=& \fr{1}{(4\pi)^2} l^6 \left(  \fr{5}{64} \fr{1}{\bar{\eps}} + \fr{59}{64} \log l^2 - \log z^2 - \fr{131}{64} \right) 
\end{eqnarray*}
and for $\b=2$ we obtain
\begin{eqnarray*}
    J^{(1)}_2 &=& \fr{1}{(4\pi)^2} \fr{1}{l^6} \left( 2 \log l^2 - 2 \log z^2  - \fr{41}{6} + \fr{l^2}{z^2} \right),
          \nonumber \\
    J^{(2)}_2 &=& \fr{1}{(4\pi)^2} \fr{1}{l^4} \left( 2 \log l^2 - 2 \log z^2  - \fr{7}{2} + \fr{1}{4} \fr{l^2}{z^2} \right),
          \nonumber \\
    J^{(3)}_2 &=& \fr{1}{(4\pi)^2} \fr{1}{l^2} \left( \fr{3}{2} \log l^2  -  \fr{3}{2} \log z^2 - \fr{13}{6} \right), \quad
    J^{(4)}_2 = \fr{1}{(4\pi)^2} \left(  \fr{9}{8} \log l^2 - \fr{9}{8} \log z^2 - \fr{95}{48}  \right), 
          \nonumber \\
    J^{(5)}_2 &=& \fr{1}{(4\pi)^2} l^2 \left( \log l^2 - \log z^2  - \fr{35}{16} \right), \quad
    J^{(6)}_2 = \fr{1}{(4\pi)^2} l^4 \left(  \log l^2 - \log z^2 - \fr{29}{12} \right) .
\end{eqnarray*}

\subsection{Two-Loop Integral Formulas}

Next, we present the integral formulas to calculate the two-loop cosmological constant corrections depicted in Fig.\ref{two loop for Lambda}. They are given by the two-loop integral involving $I^{(n)}_\a$ and $J^{(n)}_\a$ as follows:
\begin{eqnarray}
    \Lam \left[ (l^2)^{4-n-\a} I^{(n)}_\a \right] &=& \int \fr{d^D l}{(2\pi)^D} \fr{1}{(l^2 +z^2)^4} (l^2)^{4-n-\a} I^{(n)}_\a ,
           \nonumber \\
    \Lam \left[ (l^2)^{4-n+\b} J^{(n)}_\b \right] &=& \int \fr{d^D l}{(2\pi)^D} \fr{1}{(l^2 + z^2)^4} (l^2)^{4-n+\b} J^{(n)}_\b .
       \label{definition of two-loop Lambda integral}
\end{eqnarray}
These integrals can be written in the linear combination of the two-loop integrals defined by
\begin{eqnarray*}
   L\left[x^a \barF_{n,m}\right] &=& (4\pi)^2 \int \fr{d^D l}{(2\pi)^D} \fr{1}{(l^2 + z^2)^4} (l^2)^{2-\eps} \left[ x^a \barF_{n,m} \right] ,
            \nonumber \\
   L\left[x^a \bR_{m;\b}\right] &=& (4\pi)^2 \int \fr{d^D l}{(2\pi)^D} \fr{1}{(l^2 + z^2)^4} (l^2)^{2-\eps} \left[ x^a \bR_{m;\b} \right] .
\end{eqnarray*}

In the following, we extract UV divergences only, and pure IR divergences are disregarded though its evaluation is significant here.

\subsubsection{Two-Loop Integral with $\barF_{n,m}$}

We first evaluate the two-loop integral $L\left[x^a \barF_{n,m}\right]$ whose integrand $[x^a \barF_{n,m}]$ does not have IR divergences. In this case, it can be easily evaluated as follows:
\begin{eqnarray*}
  L\left[x^a \barF_{n,m}\right] &=& \left[ x^a \barF_{n,m} \right] (4\pi)^2 \int \fr{d^D l}{(2\pi)^D} \fr{1}{(l^2 + z^2)^4} (l^2)^{2-\eps}
            \nonumber \\
         &=& \left[ x^a \barF_{n,m} \right] \fr{1}{(4\pi)^{D/2-2}} \fr{\Gm(4-2\eps)\Gm(2\eps)}{\Gm(2-\eps)\Gm(4)} (z^2)^{-2\eps}
             \nonumber \\
         &=& \left[ x^a \barF_{n,m} \right] \left( \half \fr{1}{\bar{\eps}} - \fr{4}{3} \right) z^{-4\eps} .
\end{eqnarray*}
We then obtain the following results:
\begin{eqnarray*}
  L\left[\barF_{4,0}\right] &=& z^{-4\eps} \left( \fr{3}{14 \bar{\eps}^2} -  \fr{391}{1960 \bar{\eps}} \right) ,  \qquad
  L\left[\barF_{3,0}\right] =  z^{-4\eps} \left( - \fr{2}{5 \bar{\eps}^2} +  \fr{71}{150 \bar{\eps}} \right) , 
        \nonumber \\      
  L\left[ x \barF_{3,0}\right] &=&  z^{-4\eps} \left( - \fr{1}{5 \bar{\eps}^2} +  \fr{71}{300 \bar{\eps}} \right) ,  \qquad
  L\left[ x^2 \barF_{3,0}\right]  =  z^{-4\eps} \left( - \fr{4}{35 \bar{\eps}^2} +  \fr{482}{3675 \bar{\eps}} \right) , 
        \nonumber \\
  L\left[ \barF_{2,0}\right] &=& z^{-4\eps} \left(  \fr{1}{2\bar{\eps}^2} - \fr{11}{12 \bar{\eps}} \right) , \qquad
  L\left[ x \barF_{2,0}\right] = z^{-4\eps} \left( \fr{1}{4 \bar{\eps}^2} - \fr{11}{24 \bar{\eps}} \right) , 
        \nonumber \\      
  L\left[ x^2 \barF_{2,0}\right] &=& z^{-4\eps} \left( \fr{3}{20 \bar{\eps}^2} - \fr{53}{200 \bar{\eps}} \right) , \qquad
  L\left[ x^3 \barF_{2,0}\right]  = z^{-4\eps} \left( \fr{1}{10 \bar{\eps}^2} - \fr{101}{600 \bar{\eps}} \right) , 
        \nonumber \\        
  L\left[ x^4 \barF_{2,0}\right] &=& z^{-4\eps} \left( \fr{1}{14 \bar{\eps}^2} - \fr{671}{5880 \bar{\eps}} \right) , \qquad
  L\left[ \barF_{1,0}\right] =  z^{-4\eps} \fr{1}{\bar{\eps}} , 
        \nonumber \\        
  L\left[ x \barF_{1,0}\right]  &=& z^{-4\eps} \fr{1}{2 \bar{\eps}}  ,   \qquad
  L\left[ x^2 \barF_{1,0}\right] = z^{-4\eps} \fr{1}{3 \bar{\eps}} ,   \qquad
  L\left[ x^3 \barF_{1,0}\right] = z^{-4\eps} \fr{1}{4 \bar{\eps}}  ,  
        \nonumber \\
  L\left[ x^4 \barF_{1,0}\right] &=& z^{-4\eps} \fr{1}{5 \bar{\eps}}  , \qquad
  L\left[ x^5 \barF_{1,0}\right]  = z^{-4\eps} \fr{1}{6 \bar{\eps}}  ,  \qquad
  L\left[ x^6 \barF_{1,0}\right]  = z^{-4\eps} \fr{1}{7 \bar{\eps}}  .
\end{eqnarray*}
Noting $z^{-4\eps}= 1-2\eps \log z^2 + o(\eps^2)$, we can find that IR divergences appear. Pure IR divergences are ignored here, while we leave the mixed divergences of the form $(1/\eps)\log z^2$, which should cancel out in the end.

The evaluation of the integral $L [x^a \barF_{0,0}]$ whose integrand $[x^a \barF_{0,0}]$ has IR divergences is carried out as follows. From $\barK=K/l^2$ (\ref{expression of barK}), this integral can be written in the form
\begin{eqnarray*}
    L \left[x^a \barF_{0,0} \right]  &=& 
      (4 \pi)^{4-\fr{D}{2}} \Gm \left( 4- \fr{D}{2} \right) \int^1_0 dx (1-x) x^{a+1} 
      \int \fr{d^D l}{(2\pi)^D} \fr{(l^2)^{2-\eps}}{(l^2 + z^2)^4} \left( \fr{l^2}{K} \right)^{4-\fr{D}{2}}
          \nonumber \\
    &=& (4 \pi)^{4-\fr{D}{2}} \Gm \left( 4- \fr{D}{2} \right) \int^1_0 dx (1-x) x^{1+a} 
           \nonumber \\
    &&  \times
        \int \fr{d^D l}{(2\pi)^D} \fr{1}{[z^2 + x(1-x)l^2]^{2+\eps}} \left( \fr{l^2}{l^2 + z^2} \right)^4 .
\end{eqnarray*}
The last term is now expanded as
\begin{eqnarray}
    \left( \fr{l^2}{l^2 + z^2} \right)^4 = \sum_{s=0}^4 (-1)^s {}_4 C_s \left( \fr{z^2}{l^2 + z^2} \right)^s .
      \label{momentum power expansion of last term of integrand}
\end{eqnarray}
The strength of UV divergences then becomes weak in accordance with the increasing of $s$. In the two-loop integral, we neglect the terms that are not related to UV divergences. In this case, UV divergences appear at $s=0$ only and thus we obtain
\begin{eqnarray*}
    L \left[x^a \barF_{0,0} \right]  &=& 
      (4 \pi)^{4-\fr{D}{2}} \Gm (2+\eps) \int^1_0 dx x^{a-1-\eps} (1-x)^{-1-\eps} 
            \int \fr{d^D l}{(2\pi)^D} \fr{1}{\left[ l^2 + \fr{z^2}{x(1-x)} \right]^{2+\eps}} 
           \nonumber \\
      &=&  ( z^2 )^{-2\eps} (4 \pi)^{4-D} \Gm(2\eps) \int^1_0 dx x^{a-1+\eps} (1-x)^{-1 + \eps}
           \nonumber \\
      &=& z^{-4\eps} (4\pi)^{2\eps} \fr{\Gm(2\eps)\Gm(\eps)\Gm(a+\eps)}{\Gm(a+2\eps)} .
\end{eqnarray*}
From this expression, we obtain
\begin{eqnarray*}
    L \left[ \barF_{0,0} \right]  &=& z^{-4\eps} \fr{1}{\bar{\eps}^2} , \quad
    L \left[ x \barF_{0,0} \right]  = z^{-4\eps} \fr{1}{2 \bar{\eps}^2} , \quad
    L \left[ x^2 \barF_{0,0} \right]  = z^{-4\eps} \left( \fr{1}{2 \bar{\eps}^2} - \fr{1}{2 \bar{\eps}} \right), 
               \nonumber \\
    L \left[ x^3 \barF_{0,0} \right]  &=& z^{-4\eps} \left( \fr{1}{2 \bar{\eps}^2} - \fr{3}{4 \bar{\eps}} \right), \qquad
    L \left[ x^4 \barF_{0,0} \right]   = z^{-4\eps} \left( \fr{1}{2 \bar{\eps}^2} - \fr{11}{12 \bar{\eps}} \right), 
               \nonumber \\
    L \left[ x^5 \barF_{0,0} \right]  &=& z^{-4\eps} \left( \fr{1}{2 \bar{\eps}^2} - \fr{25}{24 \bar{\eps}} \right), \qquad
    L \left[ x^6 \barF_{0,0} \right]   = z^{-4\eps} \left( \fr{1}{2 \bar{\eps}^2} - \fr{137}{120 \bar{\eps}} \right), 
               \nonumber \\
    L \left[ x^7 \barF_{0,0} \right]  &=& z^{-4\eps} \left( \fr{1}{2 \bar{\eps}^2} - \fr{49}{40 \bar{\eps}} \right), \qquad
    L \left[ x^8 \barF_{0,0} \right]   = z^{-4\eps} \left( \fr{1}{2 \bar{\eps}^2} - \fr{363}{280 \bar{\eps}} \right) .
\end{eqnarray*}
Here, note that even though the integrand $[x^a\bar F_{0,0}]$ has no UV divergences, the two-loop integral $L[x^a\bar F_{0,0}]$ has a double pole.

\subsubsection{Two-Loop Integrals with $\bR_{m;\b}$}

Let us evaluate the two-loop integral $L[x^a \bR_{m;\b}]$. We first evaluate the integral whose integrand $[x^a \bR_{m;\b}]$ does not have IR divergences, which are $m=2,3$ for $\b=1$ and $m=3$ for $\b=2$. In this case, we can easily calculate the integral as before. For $\b=1$ and $m=2,3$, we obtain
\begin{eqnarray*}
  L\left[ \bR_{2;1}\right] &=& z^{-4\eps} \fr{3}{32 \bar{\eps}}  ,   \quad
  L\left[ x \bR_{2;1}\right] = z^{-4\eps} \fr{1}{32 \bar{\eps}}  ,  \quad
  L\left[ x^2 \bR_{2;1}\right] = z^{-4\eps} \fr{1}{64 \bar{\eps}}  ,    
            \nonumber \\        
  L\left[ \bR_{3;1}\right] &=& z^{-4\eps} \left( \fr{5}{128 \bar{\eps}^2} - \fr{5}{128 \bar{\eps}}  \right),   \qquad
  L\left[ x \bR_{3;1}\right] = z^{-4\eps} \left( \fr{1}{64 \bar{\eps}^2} - \fr{11}{640 \bar{\eps}}  \right) .
\end{eqnarray*}
and for $\b=2$ and $m=3$, we obtain
\begin{eqnarray*}
  L\left[ \bR_{3;2}\right] = z^{-4\eps} \fr{5}{96 \bar{\eps}} .
\end{eqnarray*}

For the cases when the integrand has IR divergences, which are $m=0,1$ for $\b=1$ and $m=0,1,2$ for $\b=2$, UV divergences are evaluated as follows. From $\barL=L/l^2$ (\ref{expression of barL}), they can be written in the form
\begin{eqnarray}
    && L\left[x^a \bR_{m;\b} \right] 
                   \nonumber \\
    && = (4 \pi)^{4-\fr{D}{2}} \fr{\Gm(m+\half )\Gm ( 4+\b -m -\fr{D}{2})}{\Gm(\half)\Gm(\b)} \int^1_0 dx x^{a+1}(1-x)^{\b+1} 
                   \nonumber \\
    &&   \quad  \times    
            \int^1_0 dy y (1-y)^{\b-1} 
            \int \fr{d^D l}{(2\pi)^D} \fr{(l^2)^{2-\eps}}{(l^2 + z^2)^4} \barL^{\fr{D}{2}+m-4-\b}
                   \nonumber \\
    && = (4 \pi)^{4-\fr{D}{2}} \fr{\Gm(m+\half )\Gm ( 4+\b -m -\fr{D}{2})}{\Gm(\half)\Gm(\b)} \int^1_0 dx x^{a+1} (1-x)^{\b+1} 
         \int^1_0 dy y (1-y)^{\b-1} 
                   \nonumber \\
    &&   \quad \times
            \int \fr{d^D l}{(2\pi)^D} \fr{(l^2)^{\b-m}}{[(x+y-xy)z^2 + x(1-x)l^2]^{2+\b-m+\eps}} \left( \fr{l^2}{l^2 + z^2} \right)^4  ,
               \label{general formula for L[xaR_{m;b}]} 
\end{eqnarray}
where the last term is expanded as (\ref{momentum power expansion of last term of integrand}).

\paragraph{\underline{$\b=1$, $m=0$}}
We first evaluate the integral of $m=0$ for $\b=1$, which has UV divergences for $s=0,1$ in the expansion (\ref{momentum power expansion of last term of integrand}) to the last term. The integration for $s=0$ can be performed as 
\begin{eqnarray*}
    L\left[x^a \bR_{0;1} \right] \Bigl|_{s=0} &=&
    (4\pi)^{4-\fr{D}{2}} \Gm \left( 5 -\fr{D}{2} \right) \int^1_0 dx x^{a-2-\eps} (1-x)^{-1-\eps} \int^1_0 dy y
             \nonumber \\
    && \times  \int \fr{d^D l}{(2\pi)^D} \fr{l^2}{ \left[ l^2 + \fr{(x+y-xy)z^2}{x(1-x)} \right]^{3+\eps}}
             \nonumber \\
    &=&  (z^2)^{-2\eps} (4\pi)^{2\eps} \fr{D}{2} \Gm(2\eps) \int^1_0 dx x^{a-2+\eps} (1-x)^{-1+\eps} 
              \nonumber \\
    &&  \times   \int^1_0 dy y (x+y-xy)^{-2\eps} .
\end{eqnarray*}
Changing the variables as $x=1-u$ and $y=1-v$, we rewrite the right-hand side as 
\begin{eqnarray*}
    (z^2)^{-2\eps} (4\pi)^{2\eps} \fr{D}{2} \Gm(2\eps) \int^1_0 du (1-u)^{a-2+\eps} u^{-1+\eps}  \int^1_0 dv (1-v) (1-uv)^{-2\eps}
\end{eqnarray*}
Here, using the expansion formula
\begin{eqnarray}
   (1-uv)^{-b} = \sum_{r=0}^\infty \fr{\Gm(b+r)}{\Gm(b)} \fr{(uv)^r}{r!} ,
         \label{series expansion}
\end{eqnarray}
we can perform the parameter integrals and thus we obtain 
\begin{eqnarray*}
   (z^2)^{-2\eps} (4\pi)^{2\eps} \fr{D}{2} \Gm(a-1+\eps) \sum_{r=0}^\infty \fr{\Gm(r+2\eps)\Gm(r+\eps)}{(r+2)!\Gm(r+a-1+2\eps)} .
\end{eqnarray*}
The UV divergence in the series part now arises from the $r=0$ term only, while the sum of $r \geq 1$ becomes finite. Therefore, we separate the series into $r=0$ and $r \geq 1$, and the latter will be evaluated by taking $\eps=0$.

We next evaluate the $s=1$ part, which is given by
\begin{eqnarray*}
    L\left[x^a \bR_{0;1} \right] \Bigl|_{s=1} 
    &=& (-4) (z^2)^{-2\eps} (4\pi)^{2\eps} \Gm(1+2\eps) \int^1_0 du (1-u)^{a-1+\eps} u^\eps  
              \nonumber \\
    &&  \times   \int^1_0 dv (1-v) (1-uv)^{-1-2\eps}
              \nonumber \\
     &=& -4 (z^2)^{-2\eps} (4\pi)^{2\eps} \Gm(a+\eps) \sum_{r=0}^\infty \fr{\Gm(r+1+2\eps)\Gm(r+1+\eps)}{(r+2)! \Gm(r+a+1+2\eps)} .
\end{eqnarray*}
The series part becomes finite and so UV divergences appear in the overall $\Gm(a+\eps)$ factor when $a=0$ only.

We can see that the $s \geq 2$ parts do not have UV divergences. We evaluate the $l$-integral by introducing the Feynman parameter further as follows:
\begin{eqnarray*}
    L\left[x^a \bR_{0;1} \right] \Bigl|_s 
    &=& (-)^s {}_4 C_s (z^2)^{-2\eps} (4\pi)^{2\eps} \fr{D}{2} \fr{\Gm(s+2\eps)}{\Gm(s)} 
         \int^1_0 dx x^{a+s-2+\eps} (1-x)^{s-1+\eps}  
              \nonumber \\
    &&  \times   \int^1_0 dy y \int^1_0 dt \fr{(1-t)^{s-1} t^{2+\eps}}{[(1-t)x(1-x)+t(x+y-xy)]^{s+2\eps}} .
\end{eqnarray*}
The parameter integral can be performed at $\eps=0$ for $s \geq 2$ and thus it does not produce UV divergences. For example, we obtain the finite values $L\left[\bR_{0;1} \right] \Bigl|_{s=2}=1.72$ and $L\left[x \bR_{0;1} \right] \Bigl|_{s=2}=0.34$.

Combining the results of $s=0,1$ and extracting the UV divergences, we obtain
\begin{eqnarray*}
    L\left[x^a \bR_{0;1} \right]
    &=&  (z^2)^{-2\eps} (4\pi)^{2\eps} \biggl\{ (2 -\eps) \Gm(a-1+\eps) 
          \biggl[  \half \fr{\Gm(2\eps)\Gm(\eps)}{\Gm(a-1+2\eps)} 
            \nonumber \\
    &&   + \sum_{r=1}^\infty \fr{\Gm(r)^2}{(r+2)! \Gm(r+a-1)}  \biggr]
         - 4 \Gm(a+\eps) \sum_{r=0}^\infty \fr{\Gm(r+1)^2}{(r+2)! \Gm(r+a+1)}  \biggr\} .
\end{eqnarray*}
The explicit values are given by
\begin{eqnarray*}
    L \left[ \bR_{0;1} \right]  &=& z^{-4\eps} \left( \fr{1}{\bar{\eps}^2} - \fr{6}{\bar{\eps}} \right), \qquad
    L \left[ x \bR_{0;1} \right]   = z^{-4\eps} \fr{1}{\bar{\eps}^2} , 
               \nonumber \\
    L \left[ x^2 \bR_{0;1} \right]  &=& z^{-4\eps} \left( \fr{1}{2 \bar{\eps}^2} - \fr{1}{4 \bar{\eps}} \right), \qquad
    L \left[ x^3 \bR_{0;1} \right]   = z^{-4\eps} \left( \fr{1}{2 \bar{\eps}^2} - \fr{3}{4 \bar{\eps}} \right) 
               \nonumber \\
    L \left[ x^4 \bR_{0;1} \right]  &=& z^{-4\eps} \left( \fr{1}{2 \bar{\eps}^2} - \fr{1}{\bar{\eps}} \right), \qquad
    L \left[ x^5 \bR_{0;1} \right]   = z^{-4\eps} \left( \fr{1}{2 \bar{\eps}^2} - \fr{7}{6 \bar{\eps}} \right) 
               \nonumber \\
    L \left[ x^6 \bR_{0;1} \right]  &=& z^{-4\eps} \left( \fr{1}{2 \bar{\eps}^2} - \fr{31}{24 \bar{\eps}} \right) .
\end{eqnarray*}

\paragraph{\underline{$\b=1$, $m=1$}}
In this case, the integral has UV divergences for the $s=0$ part only, which can be performed as follows:
\begin{eqnarray*}
    L\left[x^a \bR_{1;1} \right] \Bigl|_{s=0} &=&
    (4\pi)^{4-\fr{D}{2}} \half \Gm \left( 4 -\fr{D}{2} \right) \int^1_0 dx x^{a-1-\eps} (1-x)^{-\eps} \int^1_0 dy y
             \nonumber \\
    && \times  \int \fr{d^D l}{(2\pi)^D} \fr{1}{ \left[ l^2 + \fr{(x+y-xy)z^2}{x(1-x)} \right]^{2+\eps}}
             \nonumber \\
    &=&  (z^2)^{-2\eps} (4\pi)^{2\eps} \half \Gm(2\eps) \int^1_0 dx x^{a-1+\eps} (1-x)^\eps 
              \nonumber \\
    &&  \times   \int^1_0 dy y (x+y-xy)^{-2\eps} .
\end{eqnarray*}
Changing the variables as before, we perform the parameter integrals using expansion formula (\ref{series expansion}) so that we obtain 
\begin{eqnarray*}
    (z^2)^{-2\eps} (4\pi)^{2\eps} \half \Gm(a+\eps) \sum_{r=0}^\infty \fr{\Gm(r+2\eps)\Gm(r+1+\eps)}{(r+2)! \Gm(r+a+1+2\eps)} .
\end{eqnarray*}
The series part has UV divergences at $r=0$ only and so the sum of $r \geq 1$ can be evaluated by taking $\eps=0$. We thus obtain
\begin{eqnarray*}
    L\left[x^a \bR_{1;1} \right]
    = (z^2)^{-2\eps} (4\pi)^{2\eps} \half \Gm(a+\eps) \biggl[ \half \fr{\Gm(2\eps)\Gm(1+\eps)}{\Gm(a+1+2\eps)}
       + \sum_{r=1}^\infty \fr{\Gm(r)\Gm(r+1)}{(r+2)! \Gm(r+a+1)}  \biggr] .
\end{eqnarray*}
The explicit values are given by
\begin{eqnarray*}
    L \left[ \bR_{1;1} \right]  &=& z^{-4\eps} \left( \fr{1}{8 \bar{\eps}^2} + \fr{1}{8 \bar{\eps}} \right), \quad
    L \left[ x \bR_{1;1} \right]   = z^{-4\eps} \fr{1}{8 \bar{\eps}} ,  \quad
    L \left[ x^2 \bR_{1;1} \right]  = z^{-4\eps} \fr{1}{16 \bar{\eps}} , 
               \nonumber \\
    L \left[ x^3 \bR_{1;1} \right]  &=& z^{-4\eps} \fr{1}{24 \bar{\eps}} , \qquad
    L \left[ x^4 \bR_{1;1} \right]   =  z^{-4\eps}  \fr{1}{32 \bar{\eps}} .
\end{eqnarray*}

\paragraph{\underline{$\b=2$, $m=0$ ($a \geq 1$)}}

In this case, UV divergences arise from the $s=0, 1$ parts of (\ref{general formula for L[xaR_{m;b}]}) with (\ref{momentum power expansion of last term of integrand}) only.  These can be calculated as before and we obtain
\begin{eqnarray*}
   (z^2)^{-2\eps} (4\pi)^{2\eps} \fr{D(D+2)}{4} \Gm(a-2+\eps) 
   \sum_{r=0}^\infty \fr{(r+1)\Gm(r+2\eps)\Gm(r+\eps)}{(r+3)!\Gm(r+a-2+2\eps)} 
\end{eqnarray*}
for the $s=0$ part and 
\begin{eqnarray*}
     (z^2)^{-2\eps} (4\pi)^{2\eps} (-2 D) \Gm(a-1+\eps) 
     \sum_{r=0}^\infty \fr{(r+1)\Gm(r+1+2\eps)\Gm(r+1+\eps)}{(r+3)! \Gm(r+a+2\eps)} 
\end{eqnarray*}
for the $s=1$ part.

Combining the results of $s=0,1$, we obtain UV divergences for $a \geq 1$ as
\begin{eqnarray*}
    L\left[x^a \bR_{0;2} \right]
    &=&  (z^2)^{-2\eps} (4\pi)^{2\eps} \Biggl\{ (2 - \eps)(3 - \eps) \Gm(a-2+\eps) 
          \biggl[  \fr{1}{6} \fr{\Gm(2\eps)\Gm(\eps)}{\Gm(a-2+2\eps)} 
              \nonumber \\
    &&   + \sum_{r=1}^\infty \fr{(r+1)\Gm(r)^2}{(r+3)! \Gm(r+a-2)}  \biggr]
              \nonumber \\
    &&   - 2(4-2\eps) \Gm(a-1+\eps) \sum_{r=0}^\infty \fr{(r+1)\Gm(r+1)^2}{(r+3)! \Gm(r+a)}  \Biggr\} 
\end{eqnarray*}
and the explicit values are
\begin{eqnarray*}
    L \left[ x\bR_{0;2} \right]  &=& z^{-4\eps} \left( \fr{1}{\bar{\eps}^2} - \fr{7}{\bar{\eps}} \right), \qquad
    L \left[ x^2 \bR_{0;2} \right]  = z^{-4\eps} \fr{1}{\bar{\eps}^2} ,
               \nonumber \\
    L \left[ x^3 \bR_{0;2} \right]  &=& z^{-4\eps} \left( \fr{1}{2 \bar{\eps}^2} - \fr{5}{12 \bar{\eps}} \right), \qquad
    L \left[ x^4 \bR_{0;2} \right]   = z^{-4\eps} \left( \fr{1}{2 \bar{\eps}^2} - \fr{11}{12 \bar{\eps}} \right) ,
               \nonumber \\
    L \left[ x^5 \bR_{0;2} \right]  &=& z^{-4\eps} \left( \fr{1}{2 \bar{\eps}^2} - \fr{7}{6 \bar{\eps}} \right), \qquad
    L \left[ x^6 \bR_{0;2} \right]   = z^{-4\eps} \left( \fr{1}{2 \bar{\eps}^2} - \fr{4}{3 \bar{\eps}} \right) .
\end{eqnarray*}

\paragraph{\underline{$\b=2$, $m=1$}}

The UV divergence arises from the $s=0, 1$ parts of (\ref{general formula for L[xaR_{m;b}]}). The $s=0$ part is calculated as
\begin{eqnarray*}
   (z^2)^{-2\eps} (4\pi)^{2\eps} \fr{D}{4} \Gm(a-1+\eps) 
   \sum_{r=0}^\infty \fr{(r+1)\Gm(r+2\eps)\Gm(r+1+\eps)}{(r+3)!\Gm(r+a+2\eps)} 
\end{eqnarray*}
and the $s=1$ part is
\begin{eqnarray*}
     -2 (z^2)^{-2\eps} (4\pi)^{2\eps} \Gm(a+\eps) 
     \sum_{r=0}^\infty \fr{(r+1)\Gm(r+1+2\eps)\Gm(r+2+\eps)}{(r+3)! \Gm(r+a+2+2\eps)} .
\end{eqnarray*}
Thus, we obtain
\begin{eqnarray*}
    L\left[x^a \bR_{1;2} \right]
    &=&  (z^2)^{-2\eps} (4\pi)^{2\eps} \Biggl\{ \left( 1 - \fr{\eps}{2} \right) \Gm(a-1+\eps) 
          \biggl[  \fr{1}{6} \fr{\Gm(2\eps)\Gm(1+\eps)}{\Gm(a+2\eps)} 
            \nonumber \\
    &&   + \sum_{r=1}^\infty \fr{(r+1)\Gm(r)\Gm(r+1)}{(r+3)! \Gm(r+a)}  \biggr]
            \nonumber \\
    &&   - 2 \Gm(a+\eps) \sum_{r=0}^\infty \fr{(r+1)\Gm(r+1)\Gm(r+2)}{(r+3)! \Gm(r+a+2)}  \Biggr\} 
\end{eqnarray*}
and the explicit values are
\begin{eqnarray*}
    L \left[ \bR_{1;2} \right]  &=& z^{-4\eps} \left( - \fr{3}{2 \bar{\eps}} \right), \qquad
    L \left[ x \bR_{1;2} \right]   = z^{-4\eps} \left( \fr{1}{\bar{12 \eps}^2} + \fr{7}{72 \bar{\eps}} \right), 
               \nonumber \\
    L \left[ x^2 \bR_{1;2} \right]  &=& z^{-4\eps} \fr{1}{12 \bar{\eps}} , \quad
    L \left[ x^3 \bR_{1;2} \right]   = z^{-4\eps} \fr{1}{24 \bar{\eps}} , \quad
    L \left[ x^4 \bR_{1;2} \right]   = z^{-4\eps} \fr{1}{36 \bar{\eps}} .
\end{eqnarray*}

\paragraph{\underline{$\b=2$, $m=2$}}

This case has UV divergences at the $s=0$ part only, which is given by
\begin{eqnarray*}
    (z^2)^{-2\eps} (4\pi)^{2\eps} \fr{3}{4} \Gm(a+\eps) 
    \sum_{r=0}^\infty \fr{(r+1)\Gm(r+2\eps)\Gm(r+2+\eps)}{(r+3)! \Gm(r+a+2+2\eps)} 
\end{eqnarray*}
and thus we obtain
\begin{eqnarray*}
    L\left[x^a \bR_{2;2} \right]
    = (z^2)^{-2\eps} (4\pi)^{2\eps} \fr{3}{4} \Gm(a+\eps) \biggl[ \fr{1}{6} \fr{\Gm(2\eps)\Gm(2+\eps)}{\Gm(a+2+2\eps)}
       + \sum_{r=1}^\infty \fr{(r+1)\Gm(r)\Gm(r+2)}{(r+3)! \Gm(r+a+2)}  \biggr] .
\end{eqnarray*}
The explicit values are given by
\begin{eqnarray*}
    L \left[ \bR_{2;2} \right]  &=& z^{-4\eps} \left( \fr{1}{16 \bar{\eps}^2} + \fr{1}{24 \bar{\eps}} \right), \qquad
    L \left[ x \bR_{2;2} \right]   = z^{-4\eps} \fr{1}{32 \bar{\eps}} ,
               \nonumber \\
    L \left[ x^2 \bR_{2;2} \right]  &=& z^{-4\eps} \fr{1}{96 \bar{\eps}} , \qquad
    L \left[ x^3 \bR_{2;2} \right]   = z^{-4\eps} \fr{1}{192 \bar{\eps}} .
\end{eqnarray*}

\subsubsection{Two-Loop $\Lam$-Integrals}

Using the integral formulas derived above, we can calculate UV divergences of the two-loop $\Lam$-integrals (\ref{definition of two-loop Lambda integral}).

The explicit values of the $\Lam$-integral for $I^{(n)}_\a$ are given by
\begin{eqnarray*}
    \Lam \left[ l^8 I^{(0)}_0 \right]  
         &=& \fr{z^{-4\eps}}{(4\pi)^4} \fr{1}{\bar{\eps}^2}  ,
                \qquad
    \Lam \left[ l^6 I^{(0)}_1 \right]  
         = \fr{z^{-4\eps}}{(4\pi)^4} \left( \fr{1}{2 \bar{\eps}^2} + \fr{1}{2 \bar{\eps}}  \right) ,
                \nonumber \\
    \Lam \left[ l^4 I^{(0)}_2 \right] 
         &=& \fr{z^{-4\eps}}{(4\pi)^4} \left( \fr{1}{\bar{\eps}^2} - \fr{5}{6 \bar{\eps}}  \right) ,
                \qquad
    \Lam \left[ l^2 I^{(0)}_3 \right]  
         = \fr{z^{-4\eps}}{(4\pi)^4} \left( \fr{1}{\bar{\eps}^2} - \fr{5}{6 \bar{\eps}}  \right) ,
                 \nonumber \\
    \Lam \left[  I^{(0)}_4 \right]  
         &=& \fr{z^{-4\eps}}{(4\pi)^4} \left( \fr{1}{\bar{\eps}^2} - \fr{5}{6 \bar{\eps}}  \right) ,
                 \nonumber \\
    \Lam \left[ l^6 I^{(1)}_0 \right] 
         &=& \fr{z^{-4\eps}}{(4\pi)^4} \fr{1}{2 \bar{\eps}^2} ,
                 \qquad
    \Lam \left[ l^4 I^{(1)}_1 \right] 
         = \fr{z^{-4\eps}}{(4\pi)^4} \fr{1}{2 \bar{\eps}^2} ,
                 \nonumber \\
    \Lam \left[ l^2 I^{(1)}_2 \right] 
         &=& \fr{z^{-4\eps}}{(4\pi)^4} \left( \fr{1}{\bar{\eps}^2} - \fr{5}{6 \bar{\eps}}  \right) ,
                 \qquad
    \Lam \left[ I^{(1)}_3 \right] 
         = \fr{z^{-4\eps}}{(4\pi)^4} \left( \fr{1}{\bar{\eps}^2} - \fr{5}{6 \bar{\eps}}  \right) ,
                 \nonumber \\
    \Lam \left[ l^4 I^{(2)}_0 \right] 
         &=& \fr{z^{-4\eps}}{(4\pi)^4} \left( \fr{1}{2 \bar{\eps}^2} - \fr{1}{4 \bar{\eps}}  \right) ,
                 \qquad
    \Lam \left[ l^2 I^{(2)}_1 \right] 
         = \fr{z^{-4\eps}}{(4\pi)^4} \left( \fr{5}{8 \bar{\eps}^2} - \fr{1}{3 \bar{\eps}}  \right) ,
                 \nonumber \\
    \Lam \left[ I^{(2)}_2 \right] 
         &=& \fr{z^{-4\eps}}{(4\pi)^4} \left( \fr{1}{\bar{\eps}^2} - \fr{5}{6 \bar{\eps}}  \right) ,
                 \nonumber \\
    \Lam \left[ l^2 I^{(3)}_0 \right] 
         &=& \fr{z^{-4\eps}}{(4\pi)^4} \left( \fr{1}{2 \bar{\eps}^2} - \fr{3}{8 \bar{\eps}}  \right) ,
                 \qquad
    \Lam \left[ I^{(3)}_1 \right] 
         = \fr{z^{-4\eps}}{(4\pi)^4} \left( \fr{3}{4 \bar{\eps}^2} - \fr{13}{24 \bar{\eps}}  \right) ,
                 \nonumber \\
    \Lam \left[ I^{(4)}_0 \right] 
         &=& \fr{z^{-4\eps}}{(4\pi)^4} \left( \fr{9}{16 \bar{\eps}^2} - \fr{23}{48 \bar{\eps}}  \right) .
\end{eqnarray*}

The explicit values of the $\Lam$-integrals for $J^{(n)}_\b$ are given by
\begin{eqnarray*}
   \Lam \left[ l^{10} J^{(0)}_1 \right]  &=& \fr{z^{-4\eps}}{(4\pi)^4}  \left( \fr{1}{\bar{\eps}^2} -\fr{6}{\bar{\eps}} \right)   ,
          \qquad
   \Lam \left[ l^8 J^{(1)}_1 \right]  = \fr{z^{-4\eps}}{(4\pi)^4} \fr{1}{\bar{\eps}^2}   ,
          \nonumber \\
   \Lam \left[ l^6 J^{(2)}_1 \right]  &=& \fr{z^{-4\eps}}{(4\pi)^4}  \left( \fr{5}{8 \bar{\eps}^2} -\fr{1}{8 \bar{\eps}} \right)  ,
          \qquad
   \Lam \left[ l^4 J^{(3)}_1 \right] = \fr{z^{-4\eps}}{(4\pi)^4}  \left( \fr{1}{2 \bar{\eps}^2} -\fr{3}{8 \bar{\eps}} \right)  ,
          \nonumber \\
   \Lam \left[ l^2 J^{(4)}_1 \right] &=& \fr{z^{-4\eps}}{(4\pi)^4}  \left( \fr{1}{2 \bar{\eps}^2} -\fr{17}{32 \bar{\eps}} \right) ,
          \qquad
   \Lam \left[  J^{(5)}_1 \right] = \fr{z^{-4\eps}}{(4\pi)^4}  \left( \fr{1}{2 \bar{\eps}^2} -\fr{19}{32 \bar{\eps}} \right) ,
          \nonumber \\
   \Lam \left[ l^{-2} J^{(6)}_1 \right] &=& \fr{z^{-4\eps}}{(4\pi)^4}  \left( \fr{69}{128 \bar{\eps}^2} -\fr{241}{384 \bar{\eps}} \right)  
\end{eqnarray*}
and
\begin{eqnarray*}
   \Lam \left[ l^{10} J^{(1)}_2 \right]  = \fr{z^{-4\eps}}{(4\pi)^4} \left( \fr{1}{\bar{\eps}^2} - \fr{7}{\bar{\eps}} \right)   ,
          \qquad
   \Lam \left[ l^8 J^{(2)}_2 \right]  &=& \fr{z^{-4\eps}}{(4\pi)^4}  \left( \fr{1}{\bar{\eps}^2} -\fr{3}{2 \bar{\eps}} \right)  ,
          \nonumber \\
   \Lam \left[ l^6 J^{(3)}_2 \right] = \fr{z^{-4\eps}}{(4\pi)^4}  \left( \fr{3}{4 \bar{\eps}^2} -\fr{1}{8 \bar{\eps}} \right)  ,
          \qquad
   \Lam \left[ l^4 J^{(4)}_2 \right] &=& \fr{z^{-4\eps}}{(4\pi)^4}  \left( \fr{9}{16 \bar{\eps}^2} -\fr{3}{8 \bar{\eps}} \right) ,
          \nonumber \\
   \Lam \left[  l^2 J^{(5)}_2 \right] = \fr{z^{-4\eps}}{(4\pi)^4}  \left( \fr{1}{2 \bar{\eps}^2} -\fr{19}{32 \bar{\eps}} \right) ,
          \qquad
   \Lam \left[ J^{(6)}_2 \right] &=& \fr{z^{-4\eps}}{(4\pi)^4}  \left( \fr{1}{2 \bar{\eps}^2} -\fr{17}{24 \bar{\eps}} \right)  .
\end{eqnarray*}
Here, note that $\Lam \left[ l^{12} J^{(0)}_2 \right]$ is not defined, but it is not necessary in our calculations.

\section{Expressions of $\Gm^{\rm R}_1, \cdots, \Gm^{\rm R}_5$}
\setcounter{equation}{0}

Here, we summarize the integral expressions of $\Gm^{\rm R}_i ~(i=1, \cdots 5)$ obtained from the Feynman diagrams (1) to (5) in Fig.\ref{bt^2 correction in Landau gauge}, which are calculated using MAXIMA software.

The diagram (1) is calculated using the interaction (\ref{3-point vertex bt V^3}) as
\begin{eqnarray*}
  \Gm^{\rm R}_1 &=& - \half \fr{b t^2}{(4\pi)^{D/2}} \fr{D-2}{2(D-3)} \int \fr{d^D p}{(2\pi)^D} \phi(p) \phi(-p) 
             \nonumber \\
    && \times
        \int \fr{d^D q}{(2\pi)^D} \fr{1}{q_z^4 (q-p)_z^4} I^{(\zeta)}_{\mu\nu,\lam\s}(q) 
        V^3_{\mu\nu}(q-p,p) V^3_{\lam\s} (q-p,p) 
             \nonumber \\
    &=& - \fr{b t^2}{(4\pi)^{D/2}} \fr{D-2}{4(D-3)} \int \fr{d^D p}{(2\pi)^D} \phi(p) \phi(-p) 
        \int \fr{d^D q}{(2\pi)^D} \fr{F_1 (p,q)}{q_z^4 (q-p)_z^4}  ,
\end{eqnarray*}
where 
\begin{eqnarray}
     && F_1(p,q) = \fr{16(D-2)}{D-1} \Biggl[  p^8 + 2 p^6 q^2 + p^4 q^4 - 2 (p^4 q^2 + p^6 ) (p \cdot q)
             \nonumber \\
     &&   - \left( 2 \fr{p^6}{q^2} + 3 p^4 + 2 p^2 q^2 \right) (p \cdot q)^2 
          + 4 \left( \fr{p^4}{q^2} + p^2 \right) (p \cdot q)^3 
          + \left( \fr{p^4}{q^4} + 1 \right) (p \cdot q)^4  
                      \nonumber \\
     &&   - 2 \left( \fr{p^2}{q^4} + \fr{1}{q^2} \right) (p \cdot q)^5  
          + \fr{(p \cdot q)^6}{q^4} \Biggr]
          + \zeta \Biggl[  \fr{16}{D(D-1)} p^8  + \fr{8(3D^2-7D+16)}{3D(D-1)} p^6 q^2 
              \nonumber \\
     &&   + \fr{8(11D^2-25D+32)}{9D(D-1)} p^4 q^4 + 2 p^2 q^6
          - \Biggl( \fr{32(D^2-2D+2)}{D(D-1)} p^6 
              \nonumber \\
     &&   + \fr{16(11D^2-23D+18)}{3D(D-1)} p^4 q^2  
          - \fr{16(11D-8)}{9D} p^2 q^4 \Biggr) (p \cdot q)
          + \Biggl( \fr{32(D-2)}{D-1} \fr{p^6}{q^2}  
               \nonumber \\
     &&   + \fr{8(43D^2-85D+24)}{3D(D-1)} p^4  + \fr{8(23D^2-43D+8)}{3D(D-1)} p^2 q^2
          - \fr{2(D+8)}{9D} q^4 \Biggr) (p \cdot q)^2
               \nonumber \\
     &&   + \left( - \fr{64(D-2)}{D-1} \fr{p^4}{q^2}  - \fr{16(13D-25)}{3(D-1)} p^2  + \fr{16}{3} q^2 \right) (p \cdot q)^3
          + \Biggl( - \fr{16(D-2)}{D-1} \fr{p^4}{q^4} 
              \nonumber \\
     &&   - \fr{16(4D-7)}{3(D-1)} \Biggr) (p \cdot q)^4
          + \left( \fr{32(D-2)}{D-1} \fr{1}{q^2} + \fr{32(D-2)}{D-1} \fr{p^2}{q^4} \right) (p \cdot q)^5
              \nonumber \\
     &&   - \fr{16(D-2)}{D-1} \fr{(p \cdot q)^6}{q^4}  \Biggr] .
           \label{F_1}
\end{eqnarray}

The diagram (2) is calculated using the interaction (\ref{4-point vertex bt^2 V^4}) as
\begin{eqnarray*}
  \Gm^{\rm R}_2 &=& \fr{bt^2}{(4\pi)^{D/2}} \fr{D-2}{2(D-3)} \int \fr{d^D p}{(2\pi)^D} \phi(p) \phi(-p) 
         \int \fr{d^D q}{(2\pi)^D} \fr{1}{q_z^4} I^{(\zeta)}_{\mu\nu,\lam\s}(q) V^4_{\mu\nu,\lam\s}(q,-q;-p)
               \nonumber \\
    &=& \fr{b t^2}{(4\pi)^{D/2}} \fr{D-2}{2(D-3)} \int \fr{d^D p}{(2\pi)^D} \phi(p) \phi(-p) 
        \int \fr{d^D q}{(2\pi)^D} \fr{1}{q_z^4} F_2 (p,q) ,
\end{eqnarray*}
where
\begin{eqnarray}
   && F_2(p,q) = \fr{(D-2)(D+3)}{D-1} p^4  - \fr{(D-7)(D-2)(D+1)}{6(D-1)} p^2 q^2 
                \nonumber \\
   &&  + \left( - \fr{(D-2)(D+5)}{D-1} \fr{p^2}{q^2} + \fr{(D-3)(D-2)(D+1)}{2(D-1)} \right) (p \cdot q)^2
                \nonumber \\
   &&  + \fr{2(D-2)}{D-1} \fr{(p \cdot q)^4}{q^4}
       + \zeta \Biggl[ \fr{D^2-D+4}{D(D-1)} p^4   + \fr{2(2D^2 -5D +6)}{3D(D-1)} p^2 q^2 
                \nonumber \\
   &&  + \left( \fr{(D-2)(D+5)}{D-1} \fr{p^2}{q^2} - \fr{2(D^2 -4D +6)}{3D(D-1)} \right) (p \cdot q)^2 
       - \fr{2(D-2)}{D-1} \fr{(p \cdot q)^4}{q^4} \Biggr] .    
                \nonumber \\
   &&    \label{F_2}
\end{eqnarray}

The diagram (3) is calculated using the interactions (\ref{3-point vertex from Weyl action}) and (\ref{3-point vertex bt^2 S^3}) as
\begin{eqnarray*}
  \Gm^{\rm R}_3 &=& - (D-4) 2 \fr{bt^2}{(4\pi)^{D/2}} \left( \fr{(D-2)}{2(D-3)} \right)^2 \int \fr{d^D p}{(2\pi)^D} \phi(p) \phi(-p) 
             \nonumber \\
    && \times
        \int \fr{d^D q}{(2\pi)^D} \fr{1}{q_z^4 (q-p)_z^4} I^{(\zeta)}_{\mu\nu,\gm\dl}(p-q) I^{(\zeta)}_{\lam\s,\a\b}(q)
        W^3_{\mu\nu,\lam\s}(q-p,-q) S^3_{\a\b,\gm\dl} (q,p-q)
             \nonumber \\
    &=& - \fr{b t^2}{(4\pi)^{D/2}} \fr{(D-4)(D-2)^2}{2(D-3)^2} \int \fr{d^D p}{(2\pi)^D} \phi(p) \phi(-p) 
        \int \fr{d^D q}{(2\pi)^D} \fr{1}{q_z^4 (q-p)_z^4} F_3 (p,q) ,
\end{eqnarray*}
where 
\begin{eqnarray}
   && F_3(p,q) = \fr{(D-3)(7D+19)}{12(D-1)} p^4 q^4  - \fr{(D-3)(D+1)(7D-19)}{12(D-1)} p^2 q^6 
              \nonumber \\
   &&    + \Biggl(  \fr{(D-3)(D+1)(5D-13)}{4(D-1)} p^2 q^4   - \fr{(D-3)(D+1)}{12(D-1)} p^4 q^2 \Biggr) (p \cdot q)
              \nonumber \\
   &&    + \Biggl( \fr{(D-3)^2(D+1)}{2(D-1)} q^4  - \fr{(D-3)(9D^2+D+16)}{12(D-1)} p^2 q^2 \Biggr) (p \cdot q)^2
              \nonumber \\
   &&    + \Biggl( - \fr{(D-3)^2(D+1)}{D-1} q^2  +  \fr{(D-3)D(D+1)}{12(D-1)} p^2 \Biggr) (p \cdot q)^3
              \nonumber \\
   &&    + \fr{D(D-3)}{2} (p \cdot q)^4 .
          \label{F_3}
\end{eqnarray}

The diagram (4) is calculated using the interactions (\ref{3-point vertex bt V^3}) and (\ref{3-point vertex (D-4)bt T^3}) as
\begin{eqnarray*}
  \Gm^{\rm R}_4 &=& - (D-4) \fr{b t^2}{(4\pi)^{D/2}} \fr{(D-2)}{2(D-3)} \int \fr{d^D p}{(2\pi)^D} \phi(p) \phi(-p) 
             \nonumber \\
    && \times  \int \fr{d^D q}{(2\pi)^D} \fr{1}{q_z^4 (q-p)_z^4} I^{(\zeta)}_{\mu\nu,\lam\s}(q) V^3_{\mu\nu}(p,q-p) T^3_{\lam\s}(p,q-p)
             \nonumber \\
    &=& - \fr{b t^2}{(4\pi)^{D/2}} \fr{(D-4)(D-2)}{2(D-3)} \int \fr{d^D p}{(2\pi)^D} \phi(p) \phi(-p) 
        \int \fr{d^D q}{(2\pi)^D} \fr{F_4 (p,q)}{q_z^4 (q-p)_z^4}  ,
\end{eqnarray*}
where 
\begin{eqnarray}
   && F_4(p,q) = \fr{D-2}{D-1} \Biggl[ 24 p^8  + 44 p^6 q^2  + 20 p^4 q^4  - (48 p^6 + 44 p^4 q^2 )(p \cdot q)
             \nonumber \\
   &&  - \Biggl( \fr{48 p^6}{q^2} + 64 p^4 + 40 p^2 q^2 \Biggr) (p \cdot q)^2  
       + \Biggl(  \fr{96 p^4}{q^2} + 88 p^2 \Biggr) (p \cdot q)^3 
             \nonumber \\
   &&  + \Biggl(  \fr{24 p^4}{q^4} - \fr{4 p^2}{q^2} + 20 \Biggr) (p \cdot q)^4
       - \Biggl( \fr{48 p^2}{q^4} + \fr{44}{q^2} \Biggr) (p \cdot q)^5  + 24 \fr{(p \cdot q)^6}{q^4} \Biggr]
             \nonumber \\
   &&  + \zeta \Biggl[ 
        \fr{24}{D(D-1)} p^8  + \fr{4(27D^2-59D+131)}{9D(D-1)} p^6 q^2  + \fr{20(19D^2-41D+49)}{27D(D-1)} p^4 q^4
             \nonumber \\
   &&  + \fr{31D-16}{9D} p^2 q^6  
       - \Biggl( \fr{48(D^2-2D+2)}{D(D-1)} p^6 + \fr{4(194D^2-398D+303)}{9D(D-1)} p^4 q^2  
             \nonumber \\
   &&  + \fr{52(16D-13)}{27D} p^2 q^4   + \fr{4(D-1)}{9D} q^6 \Biggr) (p \cdot q)
       + \Biggl(  \fr{48(D-2)}{D-1} \fr{p^6}{q^2}  
             \nonumber \\
   &&  + \fr{4(393D^2-761D+234)}{9D(D-1)} p^4 
       + \fr{8(107D^2-205D+53)}{9D(D-1)} p^2 q^2  + \fr{47D-92}{27D} q^4 \Biggr) (p \cdot q)^2
             \nonumber \\
   &&  - \Biggl( \fr{96(D-2)}{D-1} \fr{p^4}{q^2} + \fr{8(122D^2-239D+18)}{9D(D-1)} p^2 - \fr{4(7D+6)}{9D} q^2 \Biggr) (p \cdot q)^3
             \nonumber \\
   &&  - \Biggl( \fr{24(D-2)}{D-1} \fr{p^4}{q^4} - \fr{4(D-2)}{D-1} \fr{p^2}{q^2} + \fr{4(58D-103)}{9D(D-1)} \Biggr) (p \cdot q)^4
             \nonumber \\
   &&  + \Biggl(  \fr{48(D-2)}{D-1} \fr{p^2}{q^4}  + \fr{44(D-2)}{D-1} \fr{1}{q^2} \Biggr) (p \cdot q)^5 
       - \fr{24(D-2)}{D-1} \fr{(p \cdot q)^6}{q^4} \Biggr] .
         \label{F_4}
\end{eqnarray}

The diagram (5) is calculated using the interaction (\ref{4-point vertex (D-4)bt^2 T^4}) as
\begin{eqnarray*}
  \Gm^{\rm R}_5 &=&  (D-4) \fr{b t^2}{(4\pi)^{D/2}} \fr{(D-2)}{2(D-3)} \int \fr{d^D p}{(2\pi)^D} \phi(p) \phi(-p) 
              \nonumber \\
     && \times   \int \fr{d^D q}{(2\pi)^D} \fr{1}{q_z^4} I^{(\zeta)}_{\mu\nu,\lam\s}(q) T^4_{\mu\nu\lam\s}(q,-q;-p)
             \nonumber \\
    &=& \fr{b t^2}{(4\pi)^{D/2}} \fr{(D-4)(D-2)}{2(D-3)} \int \fr{d^D p}{(2\pi)^D} \phi(p) \phi(-p) 
        \int \fr{d^D q}{(2\pi)^D} \fr{1}{q_z^4} F_5 (p,q) ,
\end{eqnarray*}
where 
\begin{eqnarray}
   && F_5(p,q) = \fr{3(D-2)(D+3)}{2(D-1)} p^4  - \fr{(7D-43)(D-2)(D+1)}{36(D-1)} p^2 q^2 
             \nonumber \\
   &&   + \Biggl( - \fr{3(D-2)(D+5)}{2(D-1)} \fr{p^2}{q^2}  + \fr{(D-3)(D-2)(D+1)}{2(D-1)} \Biggr) (p \cdot q)^2
             \nonumber \\
   &&   + \fr{3(D-2)}{D-1} \fr{(p \cdot q)^4}{q^4}
        + \zeta \Biggl[   \fr{3(D^2-D+4)}{2D(D-1)} p^4  +  \fr{17D^2-38D+39}{9D(D-1)} p^2 q^2  
             \nonumber \\
   &&   + \Biggl( \fr{3(D-2)(D+5)}{2(D-1)} \fr{p^2}{q^2} + \fr{D^2+17D-54}{18D(D-1)} \Biggr) (p \cdot q)^2
        - \fr{3(D-2)}{D-1} \fr{(p \cdot q)^4}{q^4}  \Biggr] .
             \nonumber \\
   &&     \label{F_5}
\end{eqnarray}


\end{document}